\documentclass[10pt]{revtex4}
\usepackage{graphicx}
\usepackage[american]{babel}
\usepackage{amssymb}
\usepackage{amsfonts}
\usepackage{color}
\usepackage[normalem]{ulem}

\begin{document}
\newtheorem{corollary}{Corollary}[section]
\newtheorem{remark}{Remark}[section]
\newtheorem{definition}{Definition}[section]
\newtheorem{theorem}{Theorem}[section]
\newtheorem{proposition}{Proposition}[section]
\newtheorem{lemma}{Lemma}[section]
\newtheorem{help1}{Example}[section]
\renewcommand{\theequation}{\arabic{section}.\arabic{equation}}

\title{Escape Dynamics in the Discrete Repulsive $\phi^4$-Model}
\author{V. Achilleos$^1$, A. \'Alvarez$^2$, J. Cuevas$^3$, D. J. Frantzeskakis$^1$,
N. I. Karachalios$^4$, P. G. Kevrekidis$^5$, B. S\'{a}nchez-Rey$^3$}
\affiliation{$^1$Department of Physics, University of Athens, Panepistimiopolis, Zografos, Athens 15784, Greece\\
$^2$Grupo de F{\'i}sica No Lineal. Universidad de Sevilla. \'Area de F\'{\i}sica Te\'orica.
Facultad de F\'{\i}sica, Avda. Reina Mercedes, s/n, 41012 Sevilla, Spain\\
$^3$Grupo de F\'{\i}sica No Lineal. Universidad de Sevilla, Departamento de F\'{\i}sica Aplicada I, Escuela
 Polit\'{e}nica Superior, C/ Virgen de Africa, 7, 41011 Sevilla, Spain\\
$^4$Department of Mathematics, University of the Aegean, Karlovassi, 83200 Samos, Greece\\
$^5$Department of Mathematics and Statistics, University of Massachusetts, Amherst MA 01003-4515, USA}
\date{\today}

\begin{abstract}
We study deterministic escape dynamics of the discrete Klein-Gordon model
%
%
with a repulsive quartic on-site potential.
%
%
Using a combination of
analytical techniques, based on differential and algebraic inequalities and
selected numerical illustrations, we
first derive
conditions for collapse of an initially excited single-site unit,
for both the Hamiltonian and the linearly damped versions of the system and
showcase different
potential fates of the single-site excitation, such as the possibility
to be ``pulled back'' from outside the well or to ``drive over'' the barrier
some of its neighbors.
%
%
Next, we study
%
%
the evolution of a uniform (small) segment of the chain
and, in turn, consider the conditions that support its escape
and collapse of the chain. Finally, our path from one to the few
and finally to
the many excited sites is completed by
a modulational stability analysis and the exploration of its connection to
the escape process for plane wave initial data.
This reveals the existence of three distinct regimes, namely
modulational stability, modulational instability without escape and,
finally, modulational instability accompanied by escape. These are
corroborated by direct numerical simulations.
In each of the above cases, the variations of the relevant model
parameters enable a consideration of the interplay of discreteness and
nonlinearity within the observed phenomenology.

\end{abstract}

\maketitle

\section{Introduction}
\paragraph{ The escape problem for systems of coupled objects.} The
escape problem for
single particles of coupled degrees of freedom or for small
chains of coupled objects, concerns their exit
from the domain of attraction of a locally stable state \cite{Hanggi0, Hanggi}.
Its investigation is crucial for the understanding of evolutionary processes in natural sciences, representing the transition from a stable state to another, or to extinction and collapse.

Transition state theory deals exactly with this kind of problem by
referring to the case where the considered objects, initiating from the domain of attraction
of a locally stable state, escape to a neighboring stable state
crossing a separating energy barrier. A natural example of this sort
involves a bistable potential, where the barrier corresponds to the
height of
the
saddle point separating the two minima of the potential. This mechanism has been
widely explored in numerous physical settings, including (but not
limited to) phase transitions, chemical kinetics, pattern formation and the
kinematics of biological waves --
see, e.g., Refs.~\cite{PF,KMI,YN,AS} and references therein. Depending on the
geometry of the underlying potential energy landscape
which may possess multiple minima,
topological excitations with a ``kink shape'' (which can be thought of
as instantons) interpolating between the adjacent minima may exist and
play a critical role in the study of transition probabilities.

Another example of such an escape process is the one whereby the considered objects, initiating within the basin of attraction of a locally stable state, escape
towards infinity.
This process is often termed collapse and it is associated with the
emergence of finite time singularities for the corresponding dynamical system.
Collapse or blow-up in finite time is quite widespread as an instability
mechanism in diffusion processes and wave phenomena in
nonlinear media \cite{bb,souplet,bs,susu,ws}.
Generally, the collapse phenomenon features the domination of
nonlinearity when the initial data are sufficiently large; however, the escape process leading to collapse may arise
in more subtle cases where the initial data
may be contained even deeply within the domain of attraction of the stable
state. This process indicates the existence of interesting energy
exchange and transfer mechanisms between the interacting objects.
We remark that, generally, collapse does not occur in integrable systems related
to solitons or intrinsic localized modes \cite{Chrisbook}. As concerns discrete systems,
the most prototypical among the known 
integrable differential-difference equations, namely the Ablowitz-Ladik
and the Toda lattices, also do not exhibit collapse behavior~\cite{mark,toda}.
In both the continuum and the discrete cases this feature can be associated
with the presence of an infinity of conservation laws, which seems highly
restrictive towards the possibility of collapse-type events.

A seminal work examining the underlying energy mechanisms supporting
escape is the work of Kramers \cite{NVK,Kramers}, implying that external,
mainly stochastic, energy sources generate energy dissipation and fluctuation
enabling the escape process. These mechanisms have been investigated
in many variants of damped and driven systems and the theory of thermally activated escape has been generalized to systems with many degrees of freedom, with numerous physical applications \cite{d03,Hanggi}.

\vspace{+6pt}
\paragraph{Deterministic escape for nonlinear Hamiltonian systems of coupled oscillators.}

The possibility of an unforced, purely deterministic Hamiltonian system
realizing an
escape process from the region of phase space sustaining oscillations
around a stable fixed point
will be our focal theme herein.
This nonlinear dynamics scenario has been investigated
in Refs.~\cite{Dirk1,Dirk2,Dirk3} for
a system of
coupled harmonic oscillators
subjected to a nonlinear external potential, namely:
%
\begin{eqnarray}
\label{eq0}
\ddot{U}_n-(U_{n+1}-2U_n+U_{n-1})+W'(U_n)=0.
\end{eqnarray}
The system (\ref{eq0}), known as the Discrete Klein-Gordon (DKG)
equation, is a fundamental
model used in different contexts, including
crystals and metamaterials, ferroelectric and ferromagnetic domain
walls, Josephson junctions, nonlinear optics, complex
electromechanical devices,
the dynamics of DNA, and so on
\cite{Chrisbook,reviewsA,reviewsB,reviewsC,reviewsD}.

In Refs.~\cite{Dirk1,Dirk2} the case of a cubic potential of the form:
%
\begin{eqnarray}
\label{CP}
W(U)=\frac{\omega_d^2}{2}U^2-\frac{\beta}{3}U^3,\;\;\beta>0,
\end{eqnarray}
is considered.
This
is a prototypical example of potentials which may support collapse, since it possesses a metastable equilibrium and becomes negatively unbounded after crossing its maximum. {\em  Importantly, it may support escape dynamics associated to collapse.} Here, collapse will be taken to mean that one or more of
the oscillators acquire an amplitude $U \rightarrow \infty$.
For instance, it is established
\cite{Dirk1,Dirk2} that
an initially almost uniformly supplied energy,
associated with an almost flat mode, may progressively be
redistributed internally, creating unstable growing nonlinear modes,
and eventually concentrated within a few units forming a localized mode
that may overcome in amplitude the height of the potential well.
More precisely, the whole chain is initialized by an appropriate
randomization to an almost homogeneous state enabling non-vanishing
interactions and the exchange of energy among the units. The total energy of
this initial state is greater than the potential energy of the saddle which
defines the depth of the well; however, the position of each of
the oscillators within this state is close
to the bottom of the respective well and thus the energy of each unit is
significantly below the energy barrier of the saddle. Due to the emergence of
modulational instability \cite{KP92}, localized excitations are formed
and the total energy becomes concentrated in confined segments of the chain.
This energy exchange mechanism
results in the emergence of
a critical mode in which
one of the lattice units overcomes the saddle point barrier.

\vspace{+6pt}
\paragraph{Model and principal aims.}
In the present work, the collapse instability, as well as the above
escape scenario will be examined
by both analytical
(based mainly on analysis techniques involving
differential and algebraic inequalities for suitable norms of
the system) methods and selected numerical experiments that will be used to
complement/corroborate the analytical results.
Our model will be the following Klein-Gordon chain, characterized by a
quartic potential and a dissipative term:
\begin{eqnarray}
\label{eq1}
\ddot{U}_n +\gamma\dot{U}_n-(U_{n+1}-2U_n+U_{n-1})+\omega^2_d(U_n-\beta U_n^3)=0,\;\;\beta>0,\;\;\gamma\geq 0.
\end{eqnarray}
%
The lattice
is assumed to be infinite ($n \in \mathbb{Z}$) and
we consider initial conditions:
\begin{eqnarray}
\label{eq2}
U_n(0)\;\;\mbox{and}\;\; \dot{U}_n(0)\in\ell^2,
\end{eqnarray}
thus the results refer to spatially localized solutions (as the definition of the phase space $\ell^2$ directly implies). In some cases, especially in numerical simulations, we will consider Dirichlet or periodic boundary conditions.
When the dissipation parameter $\gamma=0$, the DKG system (\ref{eq1}) describes the equations of motion derived by the Hamiltonian:
%
\begin{eqnarray}
\label{hamp}
\mathcal{H}(t)=\frac{1}{2}\sum_{n=-\infty}^{+\infty}\dot{U}_n^2
+\frac{1}{2}\sum_{n=-\infty}^{+\infty}(U_{n+1}-U_n)^2
+\frac{\omega_d^2}{2}\sum_{n=-\infty}^{+\infty}U_n^2-\frac{\beta\omega_d^2}{4}\sum_{n=-\infty}^{+\infty} U_n^4,
\end{eqnarray}
while the case $\gamma>0$ corresponds to the linearly damped analogue of
the problem.
Equation~(\ref{eq1})
implies that each individual oscillator of unit mass
evolves within the quartic on-site potential
\begin{eqnarray}
\label{qp}
W(U)=\frac{\omega_d^2}{2}U^2-\frac{\beta\omega_d^2}{4}U^4.
\end{eqnarray}
\begin{figure}
\begin{center}
    \begin{tabular}{cc}
    \includegraphics[scale=0.4]{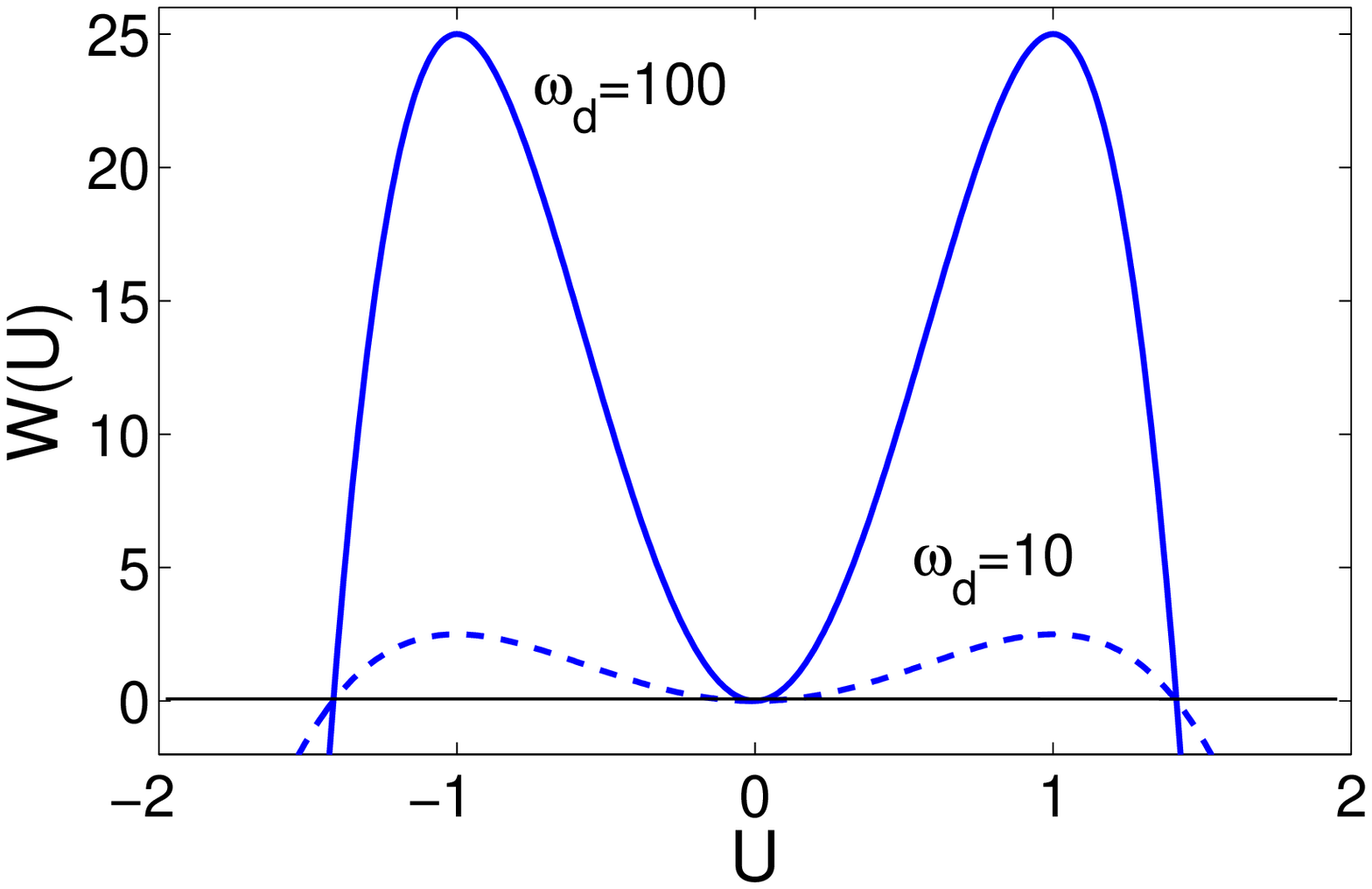} &
    \includegraphics[scale=0.4]{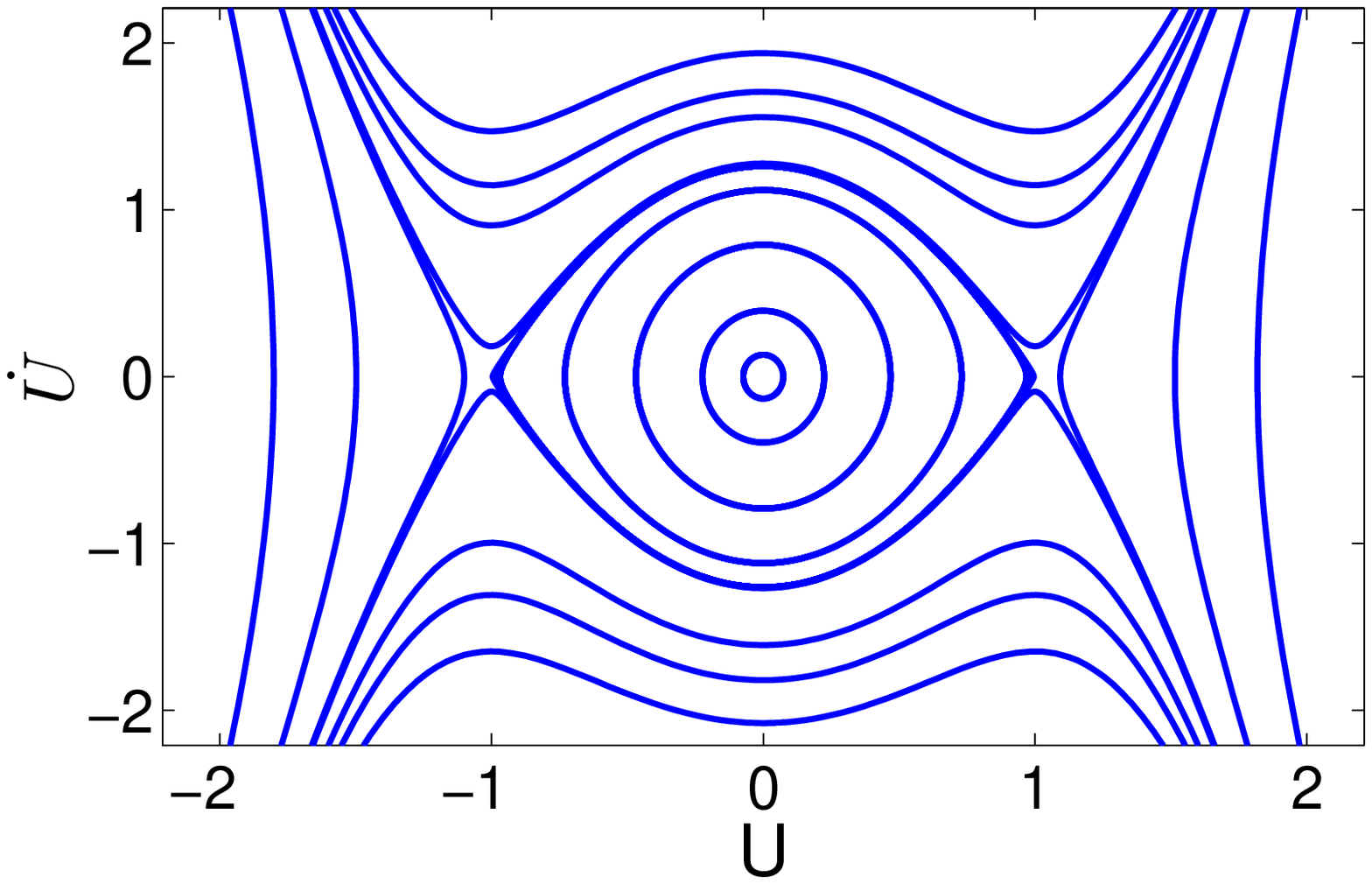}
\end{tabular}
\caption{(a) The quartic on-site potential (\ref{qp}) for $\beta=1$ and two different values of $\omega_d$ ($\omega_d^2=10$ and $\omega_d^2=100$). (b) The
phase plane for the single repulsive Duffing equation
$\ddot{U}+\omega_d^2U-\omega_d^2\beta U^3=0$ for $\beta=1$ and $\omega_d^2=3.2$. Blow-up occurs only for initial positions outside the
potential well (i.e., for all orbits beyond
the heteroclinic orbit that is connecting the two saddles).
}
\label{fig0a}
\end{center}
\end{figure}
The repulsive $\phi^4$-potential (\ref{qp}), illustrated in Fig.~\ref{fig0a}(a), has a stable minimum
at $U_{\mathrm{min}}=0$, corresponding to the rest energy $E_{\mathrm{min}}=W(0)=0$, and two unstable maxima located at $U^{\mp}_{\mathrm{max}}=\mp\frac{1}{\sqrt{\beta}}$, corresponding to $E_{\mathrm{max}}=W\left(\mp\frac{1}{\sqrt{\beta}}\right)=\frac{\omega_d^2}{4\beta}$.
%


System (\ref{eq1}) shares similar phenomenological properties with those of simplified models for DNA dynamics \cite{Peyrard1,Peyrard2,Juan}. To be more specific, the prototypical similarities are the following. The case $\beta>0$ is the physically relevant case for the study of energy-localization mechanisms in hydrogen-bonded crystals or DNA molecules, since in this type of models the potentials soften for large amplitudes \cite {KP92}. In particular, the potential (\ref{qp}) (which is also known as ``soft quartic potential'') may correspond to the hydrogen-bridge bond between nucleotides. As it is typical in the modeling of chemical bonds \cite[Section 2.1, pg. 269]{Peyrard2}, (\ref{qp}) represents qualitatively a potential with a hard repulsive part and a softer attractive part, and it diverges as $U\rightarrow\infty$. The region $[U^{-}_{\mathrm{max}},  U^{+}_{\mathrm{max}}]$, which corresponds to the potential well of depth $E_{\mathrm{max}}$, describes the region of amplitudes below the breaking of the chemical bonds.

Here, the escape problem can
simply be described as
escaping from the potential well (towards
infinite values of the field) by
crossing the saddle points
$U^{\mp}_{\mathrm{max}}$ 
in the configuration space. In order to have escape of particles over the
energy barrier $E_{\mathrm{max}}$ to the region $U>U_{\max}$, a sufficient
amount of energy has to be supplied.
Regarding the large-time behavior of solutions, the escape leads to
{\em blow-up of solutions in finite time}.

\vspace{+6pt}

\paragraph{Structure of the presentation and main findings.}
The paper organization and the main results can be described as follows.
Our aim is to progressively examine the potential blow-up of
initial conditions involving more oscillators.

First, in Section~II, we study the collapse problem for the
Hamiltonian
system ($\gamma=0$), i.e.,
the formation of finite time singularities, and find the following.

\begin{itemize}

\item
Since the system can be seen to belong  in the class of
second-order evolution equations of divergent structure, it can be
naturally treated, with respect to global non-existence,
by the energy-type methods of \cite{GP}. In particular, using
differential and algebraic inequalities we derive conditions under
which suitable norms (such as the $\ell^2$ norm) will diverge in
finite time.

\item Focusing on
initial conditions
of a {\it single} excited unit, with
the other units
initially located at the minimum $U_{\mathrm{min}}=0$, we derive an analytical
value for the initial position of the single unit outside the potential
well, which serves as a sufficient condition (threshold) for global
non-existence. This value
%
%
%
depends strongly on
the parameter $\omega_d^2$ controlling the interplay
of discreteness and nonlinearity.
Our numerical simulations demonstrate that, indeed,
%
%
blow up occurs when the
analytical prediction is satisfied and a
concerted escape may follow when we are in a regime where
the coupling effect is significant. This is the ``drive over''
phenomenon, where a particular oscillator drives its neighbors
over the collapse barrier. A reverse
phenomenon that
we also identify numerically
and explore herein is the
``pull back'' effect,
where
the attraction of an initially escaped unit by the neighbors
back to the stability domain, leads
to solutions which exist globally in time.

\item A detailed  numerical analysis of the single excited site
initial data reveals the existence of a ``true'' threshold for the
position of the excited unit outside of the well, which acts
as a separatrix for the above two different dynamical behaviors. For this
threshold, the $\omega_d^2$-dependent
analytical prediction that
we analytically identify (for collapse) serves as an upper bound; the
relevant comparison of the
analytical and true (numerically obtained)
threshold
is analyzed in the different limits of the discreteness
parameter.

\item
We also examine
the global existence result of \cite[Lemma 2.1, pg. 455]{GP} in view of the Hamiltonian system (\ref{eq1}) ($\gamma=0$).
We find the following: assuming negative initial time derivative of
the $\ell^2$ norm along with a non-positive initial Hamiltonian energy
{\it fails} to ensure global existence. The numerical results seem to
fully support this concern.

\end{itemize}


Next, Section III deals with the collapse problem for the linearly damped system (\ref{eq1}) for $\gamma>0$.
Using the abstract energy methods developed in Ref.~\cite{PS98} (having the advantage that they take into account the geometry of the
potential energy), we find the following.

\begin{itemize}
\item
We derive a $\gamma$-independent prediction for the position threshold outside the
potential well for the single excited unit.
%
%
The numerical experiments of section \ref{S2A} verify the
relevance of this criterion and that, even for significantly large
values of $\gamma>0$, the induced friction does {\it not} modify the scenarios found in the conservative case. In particular, friction does {\it not} affect the ``pull back'' effect by increasing the threshold value
(as might intuitively be expected).

\item
The $\gamma$-independence of the threshold criterion, illustrates that only the strength of the binding forces are responsible for the ``pull back'' or ``drive over'' effects. To this end, the criterion is also applied successfully to the case where $\gamma=0$, yielding an alternative and improved upper bound to the threshold for collapse or stability. This is also
corroborated by our numerical simulations.


\end{itemize}


In Section IV, we continue our analysis by studying
the escape dynamics for {\it multi-site} excitations of the Hamiltonian system ($\gamma=0$).
Here, we extend our arguments for a single-site excitation to the case
of a few and ultimately to many site excitations, but this time, positioned inside the potential well.
Our results in this setting are as follows.

\begin{itemize}

\item We derive an analytical threshold for collapse based on
the initial ``energy'' of a short-length lattice-segment,
depending on the discreteness parameter $\omega_d^2$. The numerical
results (based on a three excited units initial configuration) validate the
theoretical expectations. The threshold separates the remaining of the segment in the potential well from its escape which, in turn, is leading to collapse.

\item Combining  a violation of a small initial data global existence result (which can be  proved by implementing the methods of  Ref.~\cite{cazh}), with the
modulational instability analysis (based on the discrete nonlinear Schr\"{o}dinger (DNLS) approximation \cite{KP92}), we
reveal the existence of three different regimes for the amplitude of a plane wave initial configuration, distinguishing between the following dynamical behaviors:
modulational stability, modulational instability without escape and
finally to modulational instability combined with escape.
We also examine how  the violation of a small data global existence
criterion connects to the numerical amplitude value
beyond which modulational instability leads to escape.

\item For the linearly damped DKG chain, we perform numerical simulations which reveal a transient modulation
instability regime for small values of damping. Nevertheless, the instability is completely suppressed eventually,
in accordance with the predictions of the damped DNLS approximation.

\end{itemize}

Finally, Section~V offers a discussion and a summary of our results. The complementary appendices section contains the proofs of the  global nonexistence and global existence results which have been used in this paper for the analysis of the escape dynamics.

\vspace{+6pt}

\paragraph{Notation.}
We shall use when convenient, the short-hand notation $\{U_n(t)\}_{n\in\mathbb{Z}}=U(t)\in\ell^2$, $\{U_n(0)\}_{n\in\mathbb{Z}}=U(0)\in\ell^2$ and  $\{\dot{U}_n(0)\}_{n\in\mathbb{Z}}=\dot{U}(0)\in\ell^2$, for the solution for $t>0$ and the initial conditions at $t=0$ respectively. In this notation, $\Delta_2$ stands for the
one-dimensional discrete Laplacian
\begin{eqnarray*}
\left\{\Delta_2U\right\}_{n\in\mathbb{Z}}=U_{n+1}-2U_n+U_{n-1},
\end{eqnarray*}
defined on $\ell^2$.
 We shall also denote by
\begin{eqnarray*}
(U,Y)_{\ell^2}=\sum_{n=-\infty}^{+\infty}U_nY_n,\;\;||U||^2_{\ell^2}=\sum_{n=-\infty}^{+\infty}U_n^2,
\end{eqnarray*}
the squared-$\ell^2$ inner product and norm respectively.

\section{The case of the Hamiltonian system}
\label{SEC2}
\setcounter{equation}{0}
We may start discussing the conditions for the existence of finite-time singularities, and their relevance to the problem of escape dynamics. To this end, we have reviewed and implement in the discrete case the method of Ref.~\cite{GP}, on the derivation of a differential inequality for the norm $x(t):=||U(t)||_{\ell^2}^2$, and verify that it blows-up in finite time under appropriate sign conditions on the initial Hamiltonian
and the time derivative of the norm.
\begin{theorem}
\label{GP}
We assume that the initial data (\ref{eq2}) for the system (\ref{eq1}) are such that $\mathcal{H}(0)\leq 0$ and $(U(0), \dot{U}(0))_{\ell^2}>0$. Then $T_{\mathrm{max}}<\infty$. More precisely, the solution blows up (in the sense that $||U(t)||^2_{\ell^2}$ becomes unbounded) on the finite interval $(0,T_{\mathrm{b}})$, with
\begin{eqnarray}
\label{estbl}
T_{\mathrm{b}}=2\frac{||U(0)||^2}{(U(0), \dot{U}(0))_{\ell^2}},\;\;\mbox{and}\;\;T_{\mathrm{max}}\leq T_{\mathrm{b}}.
\end{eqnarray}
\end{theorem}
\textbf{Proof:} See Appendix A.\ \ $\Box$
\subsection{Numerical Study 1: Connection of the results of Theorem \ref{GP} with escape dynamics}
\label{IIA}
Initiating here our numerical studies, we intend to discuss the possible connections of Theorem \ref{GP} with the problem of escape dynamics. The numerical studies will consider the system (\ref{eq1}) in
a form involving the variable discretization parameter $\epsilon>0$, namely:
\begin{eqnarray}
\label{eq1h}
\ddot{U}_n-\epsilon(U_{n+1}-2U_n+U_{n-1})+\omega^2_d(U_n-\beta U_n^3)=0,\;\;\beta>0,\;\;\epsilon=\frac{1}{h^2},
\end{eqnarray}
on the interval $\left[-L, L\right]$, supplemented with Dirichlet
boundary conditions. The solution of (\ref{eq1h}) is described by
the vector $U\in\mathbb{R}^{K+2}$
\begin{eqnarray}
\label{eq1hs}
U(t)=\left(U_0(t),U_1(t),\ldots U_{K+1}(t)\right),\;\;U_0(t)=U_{K+1}(t)=0,
\end{eqnarray}
where $U_n(t):=U(x_n,t)$, $x_n=-\frac{L}{2}+nh$, $n=0,\ldots, K+1$, for all $t\geq 0$. With the change of variable $t\rightarrow\frac{1}{h}t$, we rewrite the initial-boundary value problem for (\ref{eq1h}) in the form
\begin{eqnarray}
\label{eq1ha}
\ddot{U}_n-(U_{n+1}-2U_n+U_{n-1})&+&\Omega_d^2(U_n-\beta U_n^3)=0,\;t>0\;\;n=1,\ldots,K,\;\;\Omega_d^2=h^2\omega_d^2,\\
\label{eq1hb}
&&U(0),\;\dot{U}(0)\in\mathbb{R}^{K+2},\\
\label{eq1hc}
&&U_0=U_{K+1}=0,\;\;t\geq 0.
\end{eqnarray}
Our rescaling of the lattice parameter out
of the problem makes it important that we add here a comment about
the nature of the continuous and the so-called anti-continuous limit.
In the latter, we need to consider within Eq.~(\ref{eq1}) the
limit of $\omega_d^2 \rightarrow \infty$. There, the nonlinearity
dominates, and we are essentially in the regime of individual (uncoupled)
oscillators. On the other hand, for small values of $\omega_d^2$,
the linear coupling term becomes important. However, due to the
nature of our model, we are reaching the asymptotically linear
limit (rather than the continuum limit) as $\omega_d^2 \rightarrow 0$.

We will perform numerical simulations for the simplest case of initial data
\begin{eqnarray}
\label{in1}
U_n(0)&=&A\delta_{n,0}, \\
\label{in2}
\dot{U}_n(0)&=&B\delta_{n,0},
\end{eqnarray}
where $\delta_{n,0}=1$ at site $n=0$ and zero elsewhere (Kronecker
$\delta$), for $A,B\in\mathbb{R}$.
For the initial data (\ref{in1})-(\ref{in2}), the condition for the initial Hamiltonian
$\mathcal{H}(0)\leq 0$ reads:
\begin{eqnarray}
\label{hamin}
\frac{B^2}{2}+\left(1+\frac{\Omega_d^2}{2}\right)A^2\leq \frac{\beta\Omega_d^2}{4}A^4,\;\;\Omega_d^2=h^2\omega_d^2.
\end{eqnarray}
Then, it can readily be observed that the quartic equation
\begin{eqnarray}
\label{hamineq}
\frac{\beta\Omega_d^2}{4}A^4-\left(1+\frac{\Omega_d^2}{2}\right)A^2-\frac{B^2}{2}=0,
\end{eqnarray}
%
possesses
the unique positive root (for $A^2$):
\begin{eqnarray}
\label{root1}
A^2_*=2\left[\frac{\left(1+\frac{\Omega_d^2}{2}\right)}{\beta\Omega_d^2}
+\sqrt{\frac{\left(1+\frac{\Omega_d^2}{2}\right)^2}{\beta^2\Omega_d^4}
+\frac{B^2}{2\beta\Omega_d^2}}\right],\;\;\Omega_d^2=h^2\omega_d^2.
\end{eqnarray}
Therefore, for the initial data (\ref{in1})-(\ref{in2}), the condition (\ref{hamin}) is satisfied if
\begin{eqnarray}
\label{haminAnew}
A<-{A_*}\;\;\mbox{or}\;\;A>{A_*}.
\end{eqnarray}
It can be easily checked from (\ref{root1}) that
\begin{eqnarray*}
-A_*< U^{-}_{\mathrm{max}}=-\frac{1}{\sqrt{\beta}}\;\;\mbox{and}\;\;A_*>U^{+}_{\mathrm{max}}=\frac{1}{\sqrt{\beta}},\;\;\mbox{for all}\;\;\omega_d^2, h,\beta>0,
\end{eqnarray*}
recalling that $U^{\mp}_{\mathrm{max}}$ denote the location of the saddle points of the on-site potential. The condition on the sign of the time derivative
of the $\ell^2$ norm for the initial data (\ref{in1})-(\ref{in2}) reads
\begin{eqnarray}
\label{haminAnew3}
(U_n(0), \dot{U}_n(0))_{\ell^2}=AB>0.
\end{eqnarray}
%
First, we remark that condition (\ref{haminAnew})
is {\it not} directly related to the core question of escape dynamics, which is the escape from the potential-well of the on-site potential for initial configurations of the chain inside the well:
assuming initially only one excited unit, the conditions of Theorem \ref{GP} imply that the initial position of this unit should be located outside the interval $(U^{-}_{\mathrm{max}},  U^{+}_{\mathrm{max}})$ defining the well.

On the other hand, exciting initially only one unit
outside the well, with
all the others
located at
the minimum
$U_{\mathrm{min}}=0$, {\em is indeed} of strong physical relevance in connection
to
another central question regarding the escape dynamics
\cite[pg. 041110-4]{Dirk1}:
does this initially escaped unit
continue its excursion beyond the saddle-point barrier or can it,
in fact, be pulled back into the bound chain configuration $U_n\in
(U^{-}_{\mathrm{max}},  U^{+}_{\mathrm{max}})$ by the restoring
binding forces exerted by the neighbors? Alternatively, can the unit which
is initially located outside the well, drag neighboring ones
closer to or, in a more extreme scenario,
over the barrier? The answer to these
important questions should critically depend on the strength of the
interactions between the oscillating units, imposed by the linear
coupling.

In
view of the above questions, the conditions (\ref{haminAnew})-(\ref{haminAnew3}) seem to be
quite relevant, in the sense that they determine the behavior of the whole chain regarding the escape dynamics if the 
analytical threshold
(\ref{haminAnew})
is crossed by at least one unit: in the {\em anti-continuum limit} $\epsilon\rightarrow 0$ [or $\Omega_d^2\rightarrow\infty$ since, from our scaling $\Omega_d^2=O\left(\frac{1}{\epsilon}\right)=O(h^2)$], according to Theorem \ref{GP}, we expect that the initially escaped unit will continue its
motion beyond the saddle-point barrier, while the other units should remain in the well. This is because in this case, the escaped unit is not interacting with the other units of the chain. The fact that $||U(t)||_{\ell^2}$
approaches infinity
at finite time is due to the unboundedness of the energy of this unit in finite time; while the other units remain in the well, the excursion of the escaped unit implies actually the unboundedness of the $\mathrm{sup}$-norm $||U(t)||_{\infty}=\sup_{n\in\mathbb{Z}}|U_n(t)|$ in finite time and, in turn, the unboundedness of the $||U(t)||_{\ell^2}$-energy due to the inequality:
\begin{eqnarray}
\label{expl1}
||U(t)||_{\infty}\leq ||U(t)||_{\ell^2}.
\end{eqnarray}
%
Note that in the infinite lattice only this side of the inequality holds
in contrast with the finite lattice, where the equivalence of norms
\begin{eqnarray}
\label{expl1a}
||U(t)||_{\infty}\leq ||U(t)||_{\ell^2}\leq \sqrt{N}||U(t)||_{\infty},
\end{eqnarray}
is valid in the $N$-dimensional space.

In the {\it discrete regime} $\epsilon=O(1)$ (i.e., for moderate values of $\Omega_d^2$, as
e.g., $\Omega_d^2=10$ used in Ref.~\cite{Peyrard})
and in the {\it continuum limit} $\epsilon\rightarrow\infty$,
it is expected from Theorem \ref{GP} and (\ref{expl1})
that a concerted escape of the whole chain will take place,
while the time of escape of the whole chain should depend on
the coupling strength.

The above analytical expectations and questions have been tested for $\beta=1$, $h=0.5$ and varying the parameter $\omega_d^2$. The initial speed is
taken to be $B=0.001$, i.e., a small initial velocity of the one-excited
unit is provided in order to satisfy the condition (\ref{haminAnew3}).

The first observation of the numerical studies  is {\em the justification of (\ref{haminAnew})-(\ref{haminAnew3}) as sufficient conditions for blow-up and
for the continuation of the excursion of the one-excited unit beyond the
saddle point barrier when $A>A_*$.}

The second question
examined numerically,
concerns the dynamics of
the chain when the one-excited unit is located in the region
$1<A<A_*$, i.e., beyond the location of the saddle point, but below
the analytical
critical value $A_*$.
There, our numerical simulations {revealed that blow-up occurs for initial
positions of the unit $A<A_*$;
hence, conditions
(\ref{haminAnew})-(\ref{haminAnew3}), although sufficient, are not
strictly necessary} for blow-up, which
was found to
occur for the one-excited unit even when its location is below the analytical
value $A_*$.

\begin{figure}
\begin{center}
    \begin{tabular}{cc}
    \includegraphics[scale=0.31]{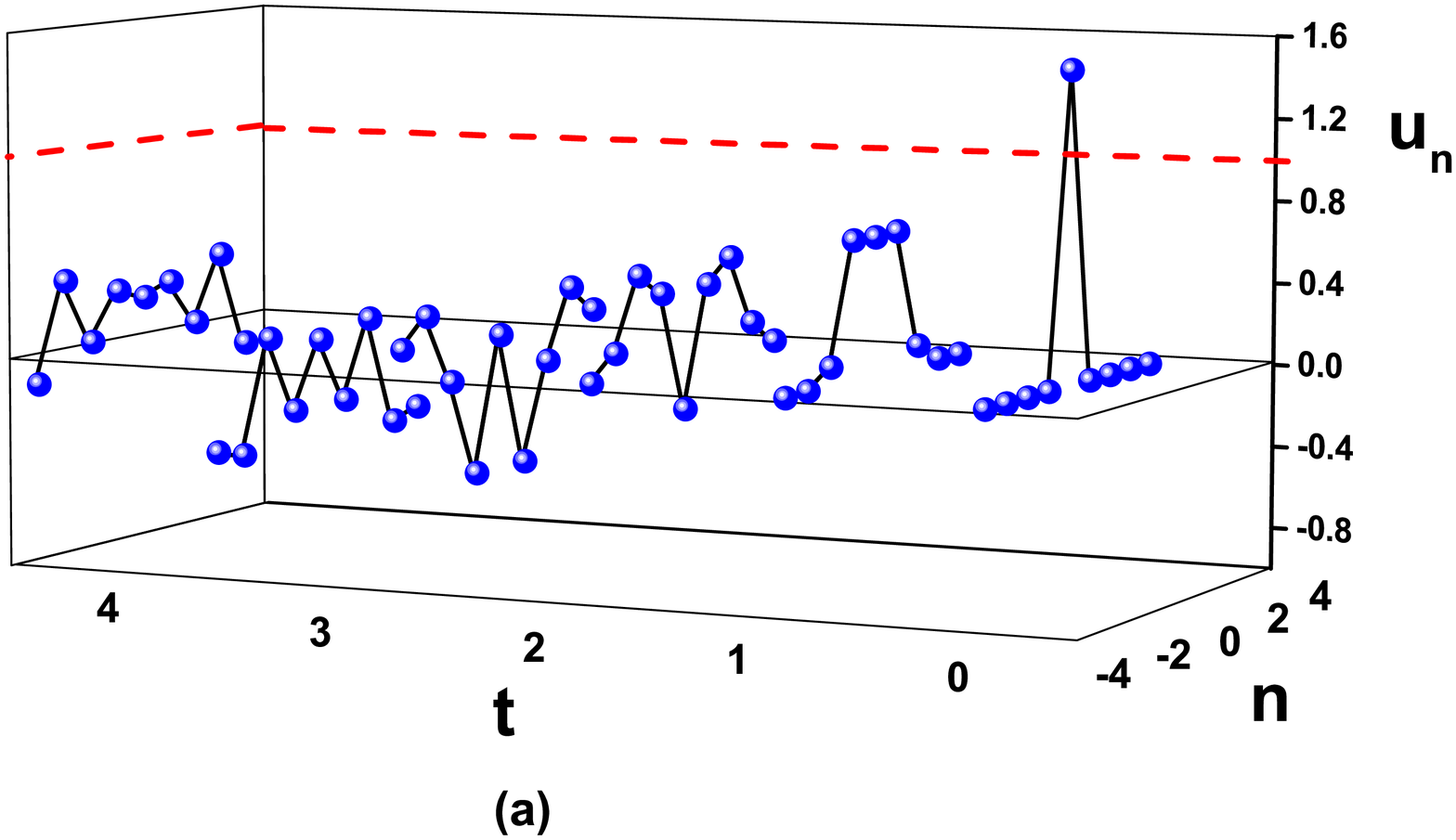} &
    \includegraphics[scale=0.31]{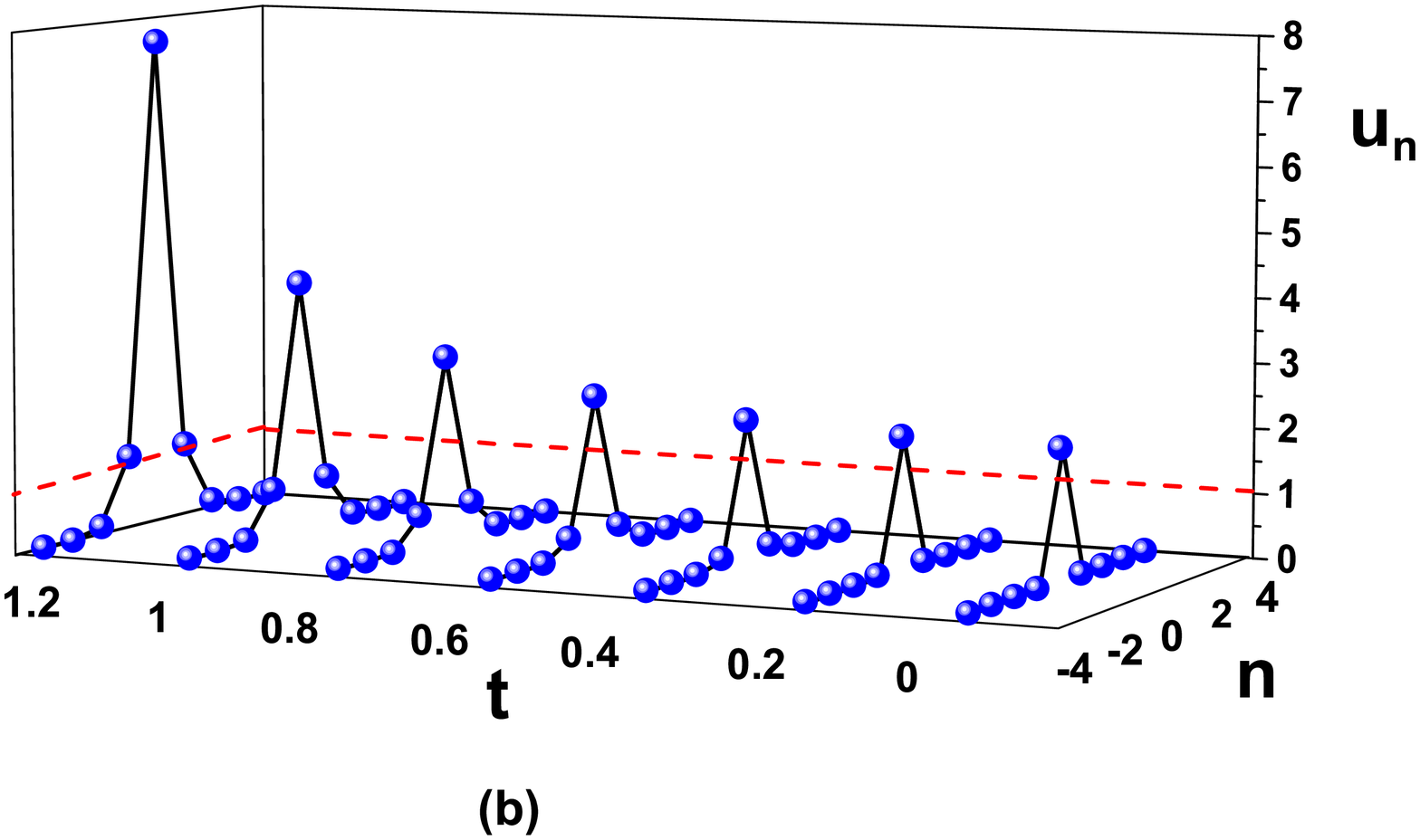}
\end{tabular}
\caption{(a)
Evolution of
a single-excited unit initially located at $A=1.2$, satisfying
$U^{+}_{\mathrm{max}}=1<A<A_*=2.45$; the value of $U^{+}_{\mathrm{max}}$ is indicated by the dashed (red) line.
Although the unit
is initially beyond the saddle point
(but below both the numerical threshold $\mathcal{U}_{\mathrm{thresh}}\sim 1.8$
and the
analytical value $A_*$),
it is pulled back in the potential well, leading to globally
existing dynamics. Lattice parameters are $\beta=1$, $\Omega_d^2=1$, for
$h=0.5$ and $\omega_d^2=4$.
(b)
In this case, the single-excited unit, initially at $A=2$, is
pulling the adjacent neighbors towards escape from the potential well.
Here, $A_*=4.24$ and $\mathcal{U}_{\mathrm{thresh}} \sim 1.9$.
The lattice parameters are: $\beta=1$, $\Omega_d^2=0.25$, for $h=0.5$ and $\omega_d^2=1$.
}
\label{fig1a}
\end{center}
\end{figure}

However, and more importantly, the numerical studies revealed the
existence of a new barrier (threshold) $\mathcal{U}_{\mathrm{thresh}}$, located
beyond the position of the saddle point $U^{+}_{\mathrm{max}}=1$,
with an important dynamical property. Although
$U^{+}_{\mathrm{max}}<\mathcal{U}_{\mathrm{thresh}}$, when the
excited unit is initially located at position $A$ satisfying
$U^{+}_{\mathrm{max}}<A<\mathcal{U}_{\mathrm{thresh}}\leq A_*$,
it may be pulled back to the
potential well and the solution exists globally.
As mentioned above,
the position of the barrier $\mathcal{U}_{\mathrm{thresh}}$, satisfying
$U^{+}_{\mathrm{max}}<\mathcal{U}_{\mathrm{thresh}}\leq A_*$, is determined by the restoring binding forces. This
fact is demonstrated in Fig.~\ref{fig1a}(a),
showing the evolution of the chain
in strong coupling
regime ($\Omega_d^2=1$). A single-excited unit, of initial amplitude
$A=1.2$ (satisfying $U^{+}_{\mathrm{max}}=1<A<A_*$),
%
is being pulled back inside the well, due to the
sufficiently strong coupling with its neighbor units. This verifies
our
analytical prediction, that restoring forces (depending on the discreteness)
can prevent the escape of elements of the chain
initialized outside the potential well. This is what we will
refer to as the ``pull back'' effect.
After this effect takes place for
the single-excited unit and it gets ``retracted''
inside the potential well $(U^{-}_{\mathrm{max}},
U^{+}_{\mathrm{max}})$, the whole
chain performs globally existing wave motions.

The fact that blow-up
occurs for $A>A_*=2.44$, as well as the result of Fig.~\ref{fig1a}(a),
confirm the existence of
the threshold $\mathcal{U}_{\mathrm{thresh}}$ and motivate a further
numerical
investigation to determine its exact location.
The corresponding  numerical
value is found to be  $\mathcal{U}_{\mathrm{thresh}}\sim 1.8<A_*$ for the near-continuum case ($\Omega_d^2=1$).
Concluding, we have shown that:
\begin{enumerate}
\item blow-up occurs when $A>\mathcal{U}_{\mathrm{thresh}}$;
\item a single-excited unit is
pulled back to the potential well when $U^{+}_{\mathrm{max}} \leq A\leq\mathcal{U}_{\mathrm{thresh}}$;
\item finally and in connection to our analytical result,
$A_*$ provides an upper bound for the exact value of
$\mathcal{U}_{\mathrm{thresh}}\in(U^{+}_{\mathrm{max}}, A_*)$
and blow-up always takes place, as predicted for $A>A_*$.
\end{enumerate}

\begin{figure}
\begin{center}
    \begin{tabular}{cc}
    \includegraphics[scale=0.4]{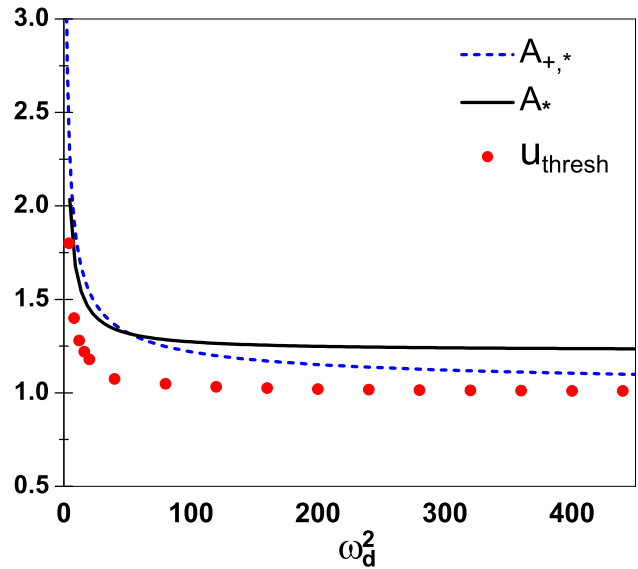}
\end{tabular}
\caption{Theoretical predictions for the blow-up amplitude
for a single site excitation, as a function of $\omega_d^2$,
compared to the actual numerical value
$\mathcal{U}_{\mathrm{thresh}}$.
The solid (black) curve corresponds to the analytical
upper bound $A_*$ [cf.~Eq.~(\ref{root1})] and the dashed (blue) curve to the
analytical value $A_{+,*}$ [cf.~Eq.~(\ref{roots12})]. Bullets (in red)
denote the numerical values $\mathcal{U}_{\mathrm{thresh}}$.
An asymptotic discrepancy
appears for $A_*$ as
$\omega_d^2\rightarrow\infty$
while the accuracy of $A_{+,*}$ increases as the anti-continuum limit
of uncoupled sites is approached-- see discussion in the text.
The parameter values are: $h=0.5$, $\beta=1$.} \label{fig2a}
\end{center}
\end{figure}

The dependence of the analytical blow-up threshold prediction $A_*$ on the discreteness parameter
$\omega_d^2$
(for fixed values of $\beta=1$ and $h=0.5$), is depicted by the solid (black) line in Fig.~\ref{fig2a}.
The respective plot for the numerically obtained value, $\mathcal{U}_{\mathrm{thresh}}$ is also depicted --in the same plot-- by (red) circles.
Comparing the two results, we conclude that
$A_*$ tends to be more
accurate as an upper bound for the ``real''
barrier $U_{\mathrm{thresh}}$, for small to moderate values of $\omega_d^2$.
On the other hand, a constant --but reasonable-- discrepancy
between the analytical and the
numerical value
is found in the intermediate regime between moderate and
large values of $\omega_d^2$ (the latter
corresponding to the near anti-continuum regime of
essentially uncoupled nonlinear
sites).
This
discrepancy can be explained
as follows: approaching
the anti-continuum limit as $\omega_d^2\rightarrow\infty$, it is
naturally expected that the
threshold value
$\mathcal{U}_{\mathrm{thresh}}\rightarrow
U^{+}_{\mathrm{max}}=\frac{1}{\sqrt{\beta}}$ since, due to the
increasingly weaker
interactions
between the oscillating units, the
system (\ref{eq1h})
is
asymptotically reduced to a single repulsive
Duffing equation for the single-excited unit
--see Fig.~\ref{fig0a}(b). On the other hand, it follows from
(\ref{root1}) that $A_*\rightarrow \frac{\sqrt{2}}{\sqrt{\beta}}$ as
$\omega_d^2\rightarrow\infty$. Thus, a constant
difference between the threshold values is found in this regime, namely
$A_*-\mathcal{U}_{\mathrm{thresh}}\sim
\frac{\sqrt{2}-1}{\sqrt{\beta}}$ (for large values of $\omega_d^2$),
which is
observed in Fig.~\ref{fig2a}.

In the asymptotically linear
limit of $\omega_d^2\rightarrow 0\;(\Omega_d^2\rightarrow 0)$,
Eq.~(\ref{root1}) predicts that $A^*\rightarrow\infty$. This prediction
can also be justified, given the absence (in that limit)
of the nonlinearity that
is responsible for the blow-up.
In the corresponding limit, we observe numerically
that  $\mathcal{U}_{\mathrm{thresh}}\rightarrow\infty$,
in accordance with the analytical prediction $A_*\rightarrow\infty$. The
limits
$A_*,\,\mathcal{U}_{\mathrm{thresh}}\rightarrow\infty$ as
$\omega_d^2\rightarrow 0$ are shown in Fig.~\ref{fig2a}.

Furthermore, as the above limit is approached and
due to the increasingly more significant role of the
interactions
between the units, if the excited amplitude has
crossed the barrier $\mathcal{U}_{\mathrm{thresh}}$ for blow-up, it will
enforce the other units to escape from the potential well. This is
``drive over'' the barrier scenario,
clearly demonstrated in Fig.~\ref{fig1a}(b), which shows
the evolution of the chain for lattice parameters $\Omega_d^2=0.25$,
i.e., $\omega_d^2=1$ for $h=0.5$.
The single-excited unit with an initial amplitude
$\mathcal{U}_{\mathrm{thresh}}\sim 1.9 <A=2<A_*$, pulls the
adjacent neighbors, enabling them
to cross the saddle point towards escape from the potential well.

The accuracy of the analytical prediction for the escape threshold
will be improved in the next
section dealing with the linearly damped version of (\ref{eq1}). The
improvement, also depicted in Fig.~\ref{fig2a} with the dashed (blue) line
will be discussed there, in detail.

\subsection{Discussion on  the sign-condition (\ref{haminAnew3}).}
We conclude this section with a discussion of \cite[Lemma 2.1, pg.~455]{GP}
for the lattice dynamical system (\ref{eq1})-(\ref{eq2}). According to the abstract results of Ref.~\cite{GP}, keeping the condition $\mathcal{H}(0)\leq 0$ and {\em violating} the sign condition $(U(0),\dot{U}(0))_{\ell^2}>0$ should lead to global existence. In terms of the initial data (\ref{in1})-(\ref{in2}), the result \cite[Lemma 2.1, pg.~455]{GP} implies that the initially escaped unit will be pulled back and return to the potential well. For example, under the assumption
\begin{eqnarray}
\label{ph1}
\dot{x}(0)=2(U(0),\dot{U}(0))_{\ell^2}<0,
\end{eqnarray}
we will reconsider the differential inequality (\ref{hamp15}) in the form
\begin{eqnarray}
\label{hamp20}
\frac{\ddot{x}(t)}{[\dot{x}(t)]^2}\geq \frac{3}{2 x(t)}.
\end{eqnarray}
\begin{figure}
\begin{center}
    \begin{tabular}{cc}
    \includegraphics[scale=0.5]{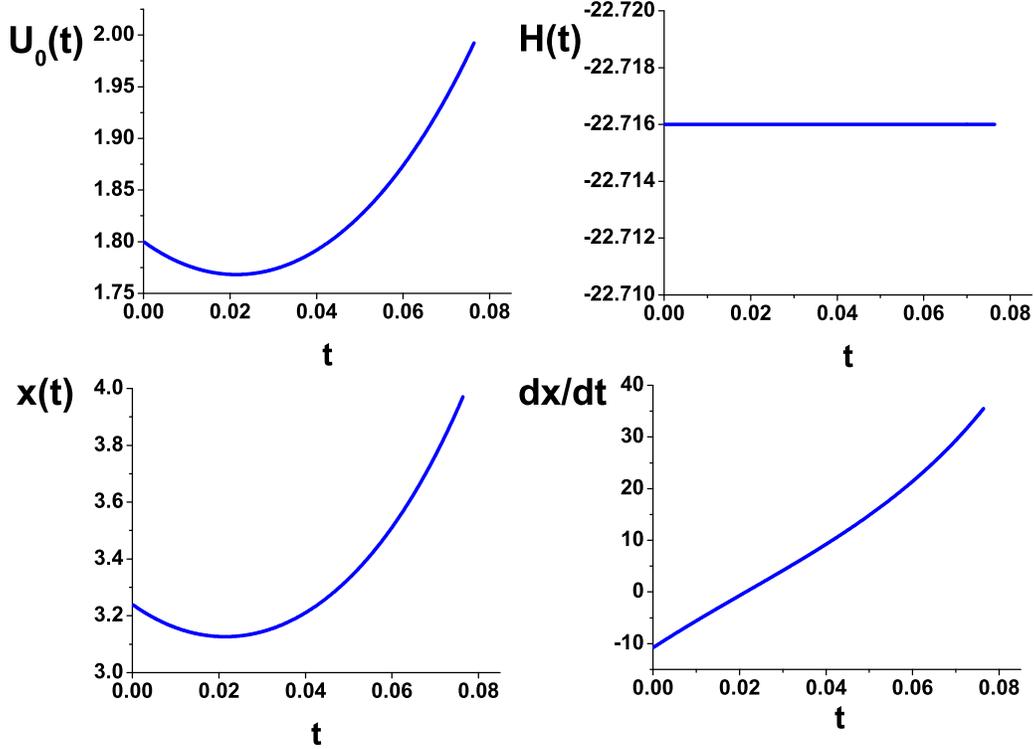}
\end{tabular}
\caption{Time evolution of the amplitude $U_0$, Hamiltonian $\mathcal{H}$ (denoted by $H(t)$ in the figure), norm $x$ and
time derivative of the norm $dx/dt$ for a single-excited unit with initial
data (\ref{in1})-(\ref{in2}) with $A=1.8$ and $B=-5.5$; the lattice parameters
are $\Omega_d^2=10$ ($\omega_d^2=40$, $h=0.5$), $\beta=1$, corresponding
to a highly nonlinear discrete regime.
%
%
The top left panel shows the excursion to
infinity for the one-excited unit; the top right panel depicts the negative
(constant) Hamiltonian
$\mathcal{H}(t)=\mathcal{H}(0)\sim 22.7$; the bottom left panel shows the
change of the slope of the norm $x(t)$ and its
increase to infinity
after time $t^*$; the bottom right panel depicts the
change of sign of $\dot{x}(t)$ after time $t^*$.
%
}
\label{fig3a}
\end{center}
\end{figure}
%
Integrating (\ref{hamp20}) in the interval $[0,t]$ for arbitrary
$t\in [0, T]$, the interval of existence, we find that
\begin{eqnarray*}
-\frac{1}{\dot{x}(t)}+\frac{1}{\dot{x}(0)}\geq\frac{3}{2}\int_{0}^{t}\frac{1}{x(s)}ds,
\end{eqnarray*}
which can be rewritten as
\begin{eqnarray}
\label{hamp21}
\frac{1}{\dot{x}(t)}\leq \frac{1}{\dot{x}(0)}-\frac{3}{2}\int_{0}^{t}\frac{1}{x(s)}ds.
\end{eqnarray}
Then, the assumption $\dot{x}(0)=2(U(0),\dot{U}(0))<0$,
as well as the
positivity of the norm function $x(t)=||U(t)||^2_{\ell^2}\geq 0$ for
all $t\in [0,T]$ in the interval of existence $[0, T]$ should imply
that $\frac{1}{\dot{x}(t)}\leq 0$, i.e.,
\begin{eqnarray}
\label{hamp21a}
\dot{x}(t)\leq 0,\;\;\mbox{{\emph{for all}}}\;\; t\in [0,T].
\end{eqnarray}
Integrating (\ref{hamp21}) once more in $[0,T]$, we see that
\begin{eqnarray}
\label{hamp22}
x(t)\leq x(0),\;\;\mbox{for all}\;\; t\in [0,T].
\end{eqnarray}
Letting $t\rightarrow\infty$ in (\ref{hamp22}) implies that $T_{\mathrm{max}}=\infty$ and
\begin{eqnarray}
\label{hamp23}
\limsup_{t\rightarrow\infty}||U(t)||^2_{\ell^2}\leq ||U(0)||_{\ell^2}^2.
\end{eqnarray}
Hence, the solution $U(t)=\left\{U_n(t)\right\}_{n\in\mathbb{Z}}$
should be defined in $[0,\infty)$ and be uniformly bounded.

However, the validity of (\ref{hamp21}) for all $t\in [0,T]$ --the
interval of existence-- {\em should be put under question:} Although
the continuity of $\dot{x}(t)=2(U(t),\dot{U}(t))_{\ell^2}<0$ in
$[0,T]$, the assumption $\dot{x}(0)=2(U(0),\dot{U}(0))_{\ell^2}<0$
and (\ref{hamp21}) guarantee the existence of $t_1>0$, such that
$\dot{x}(t)\leq 0$ for all $t\in[0,t_1]$, this assumption and
(\ref{hamp21}) do not guarantee that $t_1=T$. Recall that the
inequality (\ref{hamp10}) implies that $\ddot{x}(t)>0$ for all
$t\in[0,T]$, hence the function $\dot{x}(t)$ is increasing in the
interval $[0,T]$. Due to this fact, the existence of a $t_2>t_1>0$,
such that $\dot{x}(t)>0$ for all $t\in [t_2,T]$, cannot be excluded.

Our
concerns on the arguments of \cite[Lemma 2.1, pg. 455]{GP} have been tested numerically for
the
initial data (\ref{in1})-(\ref{in2}). If the above arguments were valid, the condition $\mathcal{H}(0)\leq 0$ on the initial Hamiltonian implying
(\ref{haminAnew}) on the position $A$ of the
single-excited unit, together with (\ref{ph1}) which for the initial data (\ref{in1})-(\ref{in2}) reads as
\begin{eqnarray}
\label{hamp24}
(U(0),\dot{U}(0))_{\ell^2}=AB<0,
\end{eqnarray}
should imply the pull-back of the initially escaped unit inside the potential well $(U^{-}_{\mathrm{max}},  U^{+}_{\mathrm{max}})$.

In Fig.~\ref{fig3a}, we present the results obtained, after  numerically integrating Eq.~(\ref{eq1}), with initial conditions
of the form of Eqs.~(\ref{in1})-(\ref{in2}),
with $A=1.8$ and $B=-5.5$, thus satisfying (\ref{haminAnew}), and the condition (\ref{hamp24}), since $(U(0),\dot{U}(0))_{\ell^2}=AB=-9.9<0$.
The lattice parameters are $\Omega_d^2=10$ ($\omega_d^2=40$ and $h=0.5$), $\beta=1$,
while the initial Hamiltonian
is negative $\mathcal{H}(0)=-22.7$ as required, and remains negative and constant in the interval of existence, due to conservation of energy, as shown in top right panel of Fig.~\ref{fig3a}.
In the bottom right panel of Fig.~\ref{fig3a}, we show the time evolution of the function $\dot{x}(t)=2(U(t),\dot{U}(t))_{\ell^2}$. The initial value of $\dot{x}(t)$ is negative,
as required by \cite[Lemma 2.1, pg. 455]{GP}, and
remains negative up to some finite time, say $t^*$. On the other hand, at $t=t^*$ (where
$\dot{x}(t^*)=0$), {\em the function $\dot{x}(t)$ becomes positive,
and remains positive for all $t>t^*$, justifying the contradiction with
the arguments of Ref.~\cite{GP} for global existence}. Also, since $\ddot{x}(t)>0$ for all $t\in [0,T]$, $\dot{x}(t)$ is strictly increasing, and the norm $x(t)$ is concave up for all $t\in [0,T]$, as shown in bottom left panel of Fig.~\ref{fig3a}. Choosing a time $t_2>t^*$ for which $\dot{x}(t)>0$ for all $t\geq t_2>t^*>0$, the results of Theorem \ref{GP} come into play again: since the system (\ref{eq1})-(\ref{eq2}) is autonomous, using the time $t_2$ as an initial time, and $(U(t_2),\dot{U}(t_2))_{\ell^2}$ as initial data, we may repeat the arguments of Theorem \ref{GP} {\em establishing that the initially escaped unit can never return in the bound-chain configuration}, as observed in the top left panel of Fig.~\ref{fig3a}.

\section{The case of the linearly damped system}
\label{S2} \setcounter{equation}{0} In this section, we will consider the conditions
for collapse in the case of
the linearly damped analogue of Eq.~(\ref{eq1}), namely,
\begin{eqnarray}
\label{eq1d}
\ddot{U}_n+\gamma\dot{U}_n-(U_{n+1}-2U_n+U_{n-1})+\omega^2_d(U_n-\beta U_n^3)=0,\;\;\;\gamma>0,\;\; \beta>0,
\end{eqnarray}
with the initial conditions:
\begin{eqnarray}
\label{eq2d}
U_n(0)\;\;\mbox{and}\;\; \dot{U}_n(0)\in\ell^2.
\end{eqnarray}
In the damped system (\ref{eq1d}), we may consider the abstract methods of Ref.~\cite{PS98} and prove global non-existence by replacing the sign condition on $\mathcal{H}(0)$ with an appropriate smallness condition. Added to the assumption of the possibly positive Hamiltonian, a condition
for a sufficiently large initial size of the quartic term of the Hamiltonian  is derived, replacing
the one --imposed in the Hamiltonian case--
concerning the sign of the time derivative of the norm of the initial
data. Of primary interest here, will be the discussion of the
effect of the damping in the behavior of the whole chain, regarding its
single node or possibly concerted escape from the potential well.

The global non-existence result of the section is stated as follows.
\begin{theorem}
\label{bud}
We assume that the initial Hamiltonian satisfies
\begin{eqnarray}
\label{eq3d}
\mathcal{H}(0)<\frac{\omega_d^2}{4\beta},
\end{eqnarray}
and that,
initially, the quartic term is sufficiently large, in the sense that the $\ell^4$-norm of the initial position satisfies:
\begin{eqnarray}
\label{eq4d}
||U(0)||_{\ell^4}=\left(\sum_{n=-\infty}^{+\infty}U_n(0)^4\right)^{\frac{1}{4}}>\frac{1}{\sqrt{\beta}}.
\end{eqnarray}
Then, the solution of (\ref{eq1d})-(\ref{eq2d}) does not exist globally in
time.
\end{theorem}
\textbf{Proof:}. See Appendix B. \ \ $\Box$
\subsection{Numerical Study 2: Connection of the results of Theorem \ref{bud}
with escape dynamics}
\label{S2A}
As in section \ref{IIA}, in the numerical
simulations we will consider the linearly damped version of (\ref{eq1h}) involving the linear coupling parameter $\epsilon>0$
\begin{eqnarray}
\label{eq1dh}
\ddot{U}_n+\gamma\dot{U}_n-\epsilon(U_{n+1}-2U_n+U_{n-1})+\omega^2_d(U_n-\beta U_n^3)=0,\;\;\beta>0,\;\;\gamma>0,\;\;\epsilon=\frac{1}{h^2},
\end{eqnarray}
on the interval $\left[-L, L\right]$, supplemented with Dirichlet boundary conditions. The initial-boundary value problem for the damped system (\ref{eq1dh}) can be written with the change of variable $t\rightarrow\frac{1}{h}t$ as
\begin{eqnarray}
\label{eq1dha}
\ddot{U}_n+\gamma h\dot{U}_n-(U_{n+1}-2U_n+U_{n-1})&+&\Omega_d^2(U_n-\beta U_n^3)=0,\;t>0\;\;n=1,\ldots,K,\;\;\Omega_d^2=h^2\omega_d^2,\\
\label{eq1dhb}
&&U(0),\;\dot{U}(0)\in\mathbb{R}^{K+2},\\
\label{eq1dhc}
&&U_0=U_{K+1}=0,\;\;t\geq 0.
\end{eqnarray}
In the numerical simulations we will again consider
the
single-unit initial excitation (\ref{in1})-(\ref{in2}). For the initial data (\ref{in1})-(\ref{in2}), conditions (\ref{eq3d})-(\ref{eq4d}) are implemented as:
\begin{eqnarray}
\label{eq53d}
&&\frac{B^2}{2}+\left(1+\frac{\Omega_d^2}{2}\right)A^2
-\frac{\beta\Omega_d^2}{4}A^4<\frac{\Omega_d^2}{4\beta},\;\;\mbox{and}\\
\label{eq54d}
&&A>\frac{1}{\sqrt{\beta}}.
\end{eqnarray}
Here, the quartic equation
\begin{eqnarray}
\label{hamineqd}
\frac{\beta\Omega_d^2}{4}A^4-\left(1+\frac{\Omega_d^2}{2}\right)A^2-\frac{B^2}{2}+\frac{\Omega_d^2}{4\beta}=0,
\end{eqnarray}
has in terms of $A^2$ the two roots:
\begin{eqnarray}
\label{roots12}
A^2_{\pm,*}=2\left[\frac{\left(1+\frac{\Omega_d^2}{2}\right)}{\beta\Omega_d^2}
\pm\sqrt{\frac{\left(1+\frac{\Omega_d^2}{2}\right)^2}{\beta^2\Omega_d^4}
+\frac{B^2}{2\beta\Omega_d^2}-\frac{1}{4\beta^2}}\right],\;\;\Omega_d^2=h^2\omega_d^2.
\end{eqnarray}
Note that $A^2_{+,*}>0$, $A^2_{+,*}>A^2_{-,*}$ and $A^2_{-,*}>0$ if and only if $B\in\left(-\frac{\Omega_d}{\sqrt{2\beta}},\frac{\Omega_d}{\sqrt{2\beta}}\right)$. Furthermore, it can be seen that
\begin{eqnarray}
\label{eq54e}
A^2_{+,*}>\frac{1}{\beta},\;\;\;\;\mbox{for all}\;\;\omega_d^2, h,\beta>0.
\end{eqnarray}
Then, both conditions (\ref{eq53d}) and (\ref{eq54d}) are satisfied, as required, if
\begin{eqnarray}
\label{eq55d}
A<-A_{+,*}\;\;\mbox{or}\;\;A>A_{+,*}.
\end{eqnarray}
As in the Hamiltonian case, conditions (\ref{eq55d}) imply that the initial position of the excited unit should be located outside the interval $(U^{-}_{\mathrm{max}},  U^{+}_{\mathrm{max}})$ defining the well,
due to (\ref{eq54e}).
We point out that
this condition of blow-up bears no signature of the damping parameter $\gamma$ and,
hence, can be used for the Hamiltonian case as well, a point
evident in Fig.~\ref{fig2a} and one to which we return below.

However, the question
of whether the single-unit, initially placed outside
the saddle-point barrier, will escape  or
whether it will be pulled back into the well
by its neighboring units, becomes
even more interesting due to the
presence of the damping:
the question is if
damping has an additional effect in increasing the value of the threshold between ``pull back'' and collapse.
Since the threshold is increased as the strength of the coupling forces is increased (cf. Fig.~\ref{fig2a}),
one might expect that that the same would happen with the damping force.

\begin{figure}
\begin{center}
    \begin{tabular}{cc}
    \includegraphics[scale=0.4]{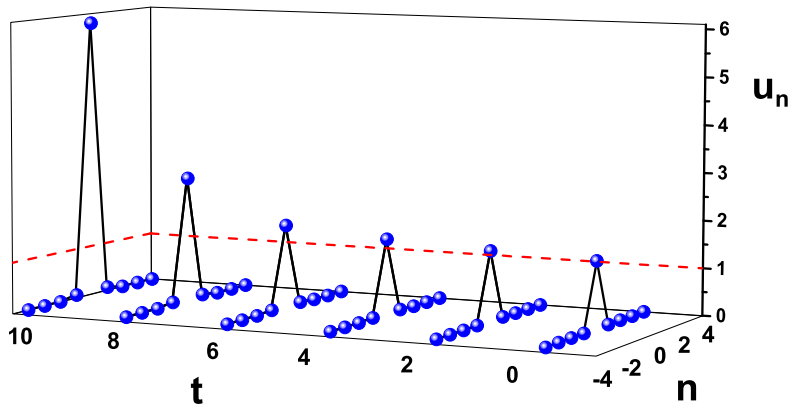}
\end{tabular}
\caption{Evolution of the damped chain for lattice parameters $\Omega_d^2=10$ ($\omega_d^2=40$, $h=0.5$) and
$\beta=1$. The central excited unit $n=0$ has initial
amplitude $A=1.4$ and initial velocity $B=0$. The large damping, characterized by the parameter $\gamma=2000$,
{ \it does}  not affect the threshold value for the ``pull back'' effect.
}
\label{fig1b}
\end{center}
\end{figure}

The time evolution of the damped chain, for lattice parameters $\Omega_d^2=10$
($\omega_d^2=40$, $h=0.5$ --corresponding to a moderately
discrete regime) and $\beta=1$, is shown in Fig.~\ref{fig1b}. For this set of parameters the analytical threshold for the amplitude of the single-excited unit, calculated by (\ref{roots12}), is $A_{+,*}=1.36$. To test this prediction, we consider an initial amplitude of
$A=1.4$ and an initial speed $B=0$ as
per the Theorem \ref{bud}. In order to investigate
the friction-induced effects,
we consider a
relatively large value
of the damping parameter, i.e., $\gamma=2000$. Figure~{\ref{fig1b} shows the escape
of the excited unit to infinity.
On the other hand, Fig.~\ref{fig2b}(a) visualizes the rapid increase of the kinetic energy of the single unit $n=0$,  despite of the fact that the total
energy is dissipated according to the identity (\ref{eq7d}) as long as the solution exists.
The
evolution of the kinetic energy is also shown in Fig.~\ref{fig2b}(b),
but for  lattice parameters closer to the asymptotically linear
 regime, i.e., for $\Omega_d^2=1$ (and
$\beta=1$ as well).
Comparing the two panels of Fig.~\ref{fig2b},
it is evident that the initially excited
unit at $n=0$ does not affect its neighboring units
in the discrete case (a), in the sense that the energy remains localized at this site; on the other hand,
in
 case (b), due to the stronger coupling, the central excited
unit
pulls the adjacent units at sites $n=\pm1$ towards the saddle point,
and their kinetic energy is
increased
compared to case (a).
Numerical
simulations have been also been performed
for
other values of $\gamma$, and they all have justified the prediction of Theorem \ref{bud}:
when $A>A_{+,*}$, the collapse
does not depend on the value of the damping parameter;
furthermore, the true threshold $\mathcal{U}_{\mathrm{thresh}}$
is also $\gamma$-independent. {\em Therefore, the ``drive over'' and ``pull back'' effects are chiefly controlled by the strength of the coupling forces and not from the damping}.
The failure of the intuitive expectation that dissipation should
decay away the motion of the oscillators and hence be less conducive
to collapse is evident here. This is
predominantly due to the principal role of dissipation in decreasing the
overall energy of the oscillator chain; yet, the latter scenario can be
achieved by large amplitudes due to the nonlinearity, hence the
concerted effect of energy decrease due to dissipation and its
achievement by increased (squared) amplitudes due to nonlinearity
gives rise to the finite time blow-up analyzed above.


Another interesting feature is the following.
The analytical value $A_{+,*}$
provided by (\ref{roots12}),
shows an increased accuracy
--in the parameter regime ranging from the moderately
to the highly discrete (and the anti-continuum)--
as an estimate of the
numerical
threshold $\mathcal{U}_{\mathrm{thresh}}$,
below which the single unit will be pulled back by the exerted
forces of the neighbors, and above which the chain configuration
collapses. This improvement is due to the fact that the proof of
Theorem \ref{bud} takes into more detailed
account the geometry of the potential energy $W(U)$.
Since the proof of Theorem \ref{bud} is independent of the damping
parameter, and is also valid
in the Hamiltonian
case, we return to
Fig.~\ref{fig2a}, where the dashed (blue) curve depicts
$A_{+,*}$ as a function of $\omega_d^2$
(for fixed $B=0$, $h=0.5$ and $\beta=1$).
The increased accuracy of $A_{+,*}$ [dashed (blue) line] towards the anti-continuum regime
is explained directly from (\ref{roots12}) since
$A_{+,*}\rightarrow U^{+}_{\mathrm{max}}=\frac{1}{\sqrt{\beta}}$ as
$\omega_d^2\rightarrow\infty$,
and the constant discrepancy
for $A_*$ asymptotically vanishes.
Also, in the asymptotically linear case of
$\omega_d^2\rightarrow 0$,
$A_{+,*}\rightarrow\infty$, thus it correctly predicts the
asymptotic behavior of $\mathcal{U}_{\mathrm{thresh}}$ in this
limit. For moderate values of $\omega_d^2$,
a comparison
of (\ref{root1}) and (\ref{roots12})
indicates that
$A_*$ may be preferable to $A_{+,*}$ in this intermediate regime.
Nevertheless, $A_{+,*}$ accurately captures both asymptotic limits
and in the intermediate regime both $A_*$ and $A_{+,*}$ are expected
to be least accurate due to the relevance of the terms omitted
in the derived (algebraic and differential) inequalities.
\begin{figure}
\begin{center}
    \begin{tabular}{cc}
    \includegraphics[scale=0.3]{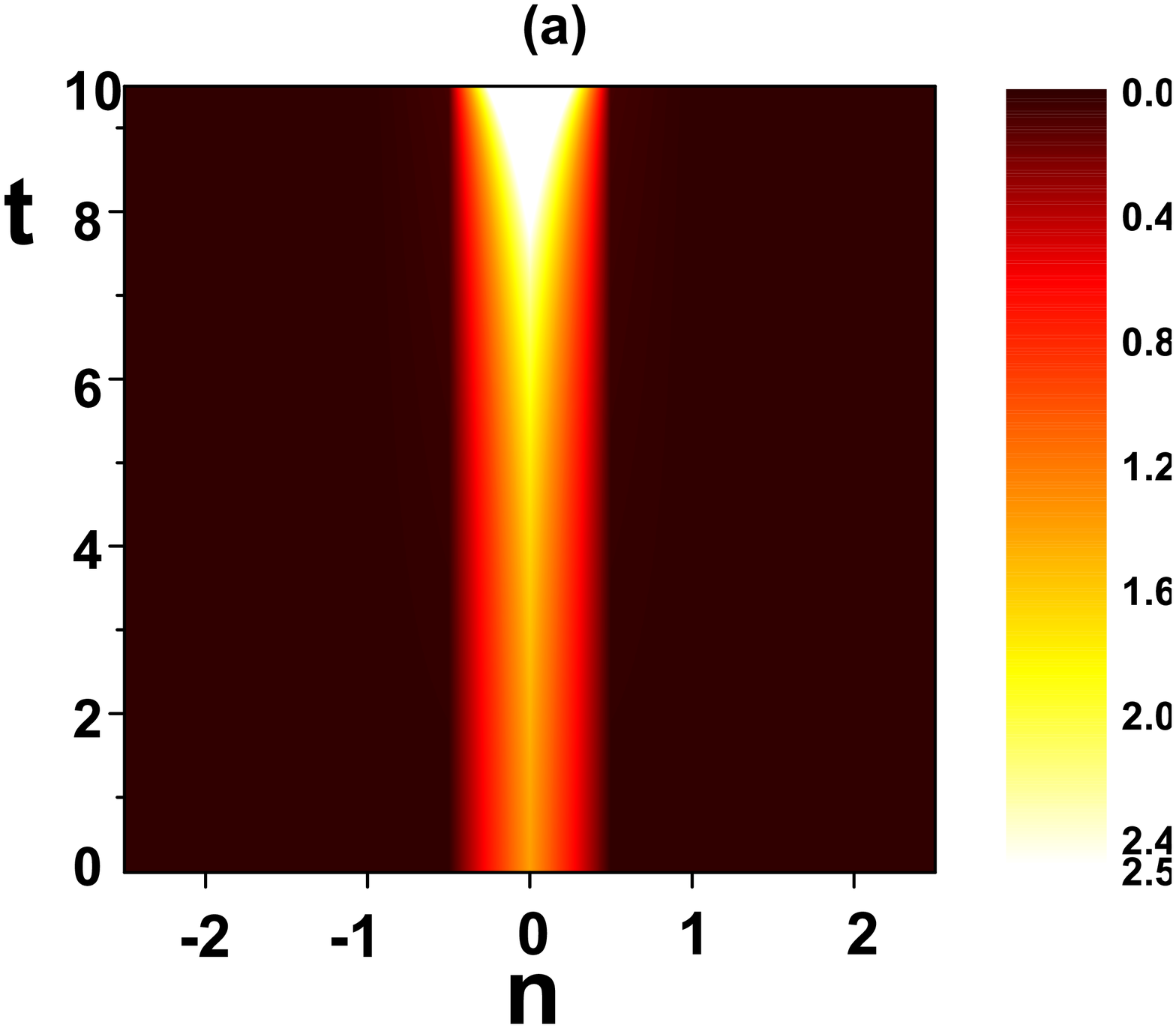} &
    \includegraphics[scale=0.3]{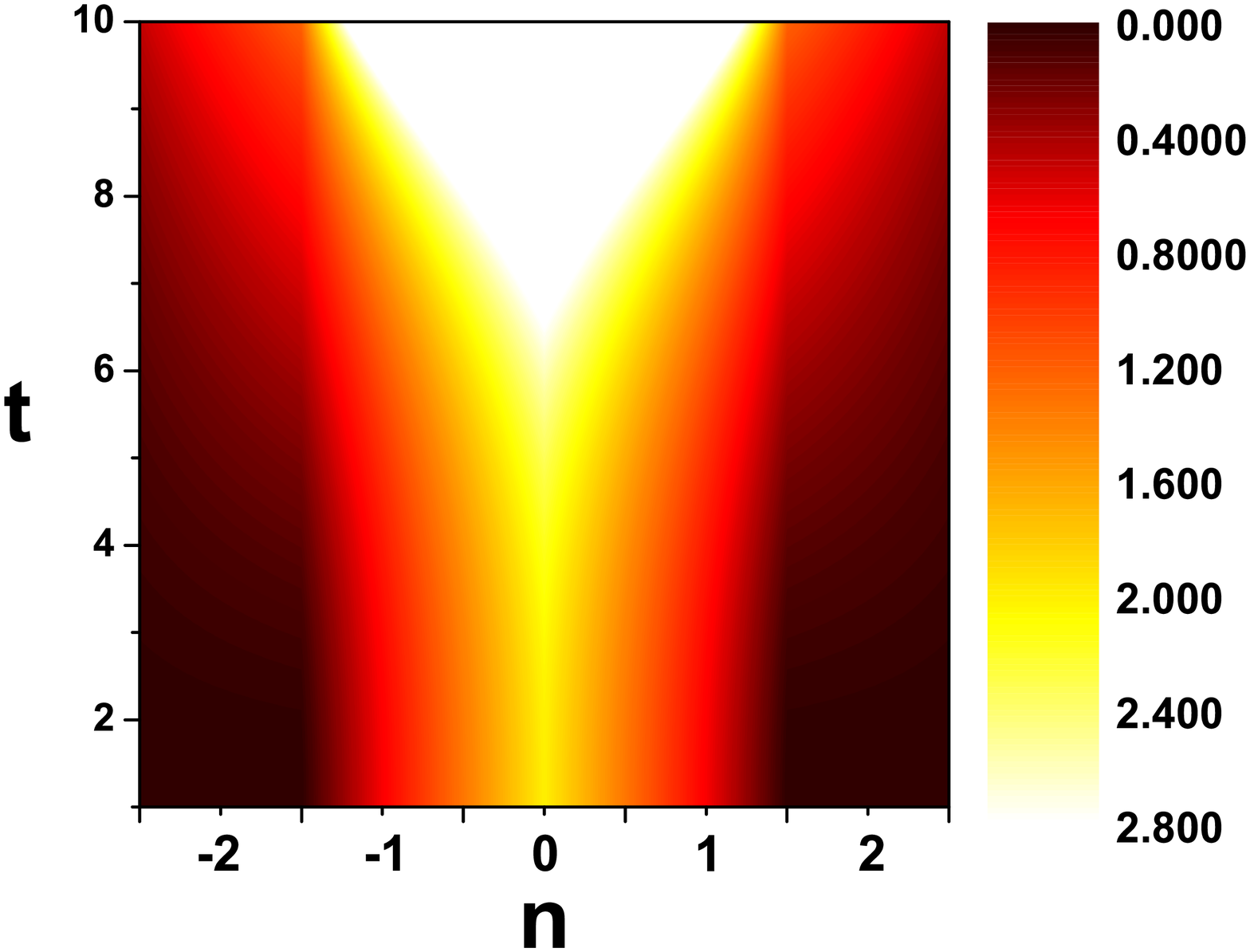}
\end{tabular}
\caption{Contour plots showing the evolution of the kinetic energy in the damped lattice
$\gamma=2000$, $\beta=1$. The central excited unit $n=0$ at initial
position $A>A_{+,*}$ and velocity $B=0$. (a) $\Omega_d^2=10$
($\omega_d^2=40$, $h=0.5$)--this is within the moderately
discrete regime. (b) $\Omega_d^2=1$
($\omega_d^2=4$, $h=0.5$)--this is in the vicinity of
the asymptotically linear regime.
} \label{fig2b}
\end{center}
\end{figure}

\section{Scenarios for escape for multi-site excitations}
\label{S5} \setcounter{equation}{0}


We now move progressively away from the single site excitation
scenario and toward settings where the energy of the chain
becomes concentrated within a (wider than one site, yet confined)
region~\cite{Dirk1}.
Even if initially the energy is equally shared among all units, after
a certain time, the dynamics may lead to its redistribution,
so that at least one of the involved units concentrates sufficient energy,
to overcome the barrier defined by the saddle point of the potential.
In this section, we will first consider such an escape scenario,
focusing on the evolution of a small lattice segment. More specifically, we will consider
the evolution of a few-site excitation, by means of the energy
methods developed in the previous sections.

Then, we will study escape dynamics of the chain,
induced by the modulation instability mechanism \cite{Dirk1,KP92}.
In the latter case, the initial excitation will involve a plane wave extending throughout the chain.
We shall review the conditions for modulation instability of plane waves for (\ref{eq1h}) and how this instability may be related to a potential
escape of the modulationally unstable excitation.

Finally, deriving conditions for the global existence of initial data of sufficiently small energy, we shall investigate how the possible violation of the global existence assumptions may be connected to the escape mechanism, and
formulate conditions for escape dynamics. This approach will consider solely
an initial condition in the form of a plane wave.

\subsection{Energy methods on a lattice segment and escape dynamics}
\label{S5A}
In view of the energy methods discussed in the previous sections and motivated by the first localization scenario discussed above, we focus on the study of the evolution of a fixed segment of the lattice. In particular, we seek
appropriate conditions for the ``initial energy'' of this segment, which may lead to an escape process.

We consider again system (\ref{eq1h}) on the interval $[-L, L]$ supplemented with Dirichlet boundary conditions. Here, we will study the evolution of a lattice segment occupying the unit subinterval $\left[-\frac{1}{2},\frac{1}{2}\right]$ $\subset$ $[-L, L]$ together with the first neighbors adjacent to the points $-\frac{1}{2}$ and $\frac{1}{2}$.

The number of oscillators located outside the piece of the chain of unit length
$I=\left[-\frac{1}{2},\frac{1}{2}\right]$ is
\begin{eqnarray}
\label{br2}
\theta=2\left\lceil\frac{\left(\frac{L}{2}-\frac{1}{2}\right)(K+1)}{L}\right\rceil,
\end{eqnarray}
where $\lceil x\rceil =\min\left\{n\in\mathbb{Z}\,|\,n\geq x\right\}$, $x\in\mathbb{R}$. Then the number of oscillators included in the unit interval $I$ is
\begin{eqnarray}
\label{br3}
m=K+2-\theta.
\end{eqnarray}
We distinguish between two different possibilities for the endpoints of $I$ --cf. cases $\mathrm{A}$ and $\mathrm{B}$ below.
\newline

$\mathrm{Case~A}.$  The simplest case is when the endpoints of $I$ are occupied by oscillators. In this case,
the lattice spacing satisfies
\begin{eqnarray}
\label{br4}
(m+1)h=1,
\end{eqnarray}
and the unit interval $I$ consists of $\lambda=m+2$ oscillators, the $m$ included in $I$ together with the two endpoints.

We denote by $I'$ the interval which comprises $I$ and the two adjacent ones. The interval $I'$ consists of
$\lambda+2=m+4$ oscillators. The length of $I'$ is, due to (\ref{br4}),
\begin{eqnarray}
\label{br6}
L'=1+2h=\frac{m+3}{m+1}.
\end{eqnarray}
We also assume that the endpoints of $I'$ occupied by the neighbors adjacent to the endpoints of $I$, are located at the sites $k$ and $k+\lambda+1$. Then, we may decompose the solution configuration vector (\ref{eq1hs}) as $U=U^{L\setminus I}+U^{I}$, where
\begin{eqnarray}
\label{br7}
U^{L\setminus I}&=&\left(U_0,U_1,\ldots U_{k-1},U_{k},0,\ldots,0,U_{k+\lambda+1},\ldots,U_{K+2}\right),\\
\label{br8}
U^{I}&=&\left(0,\ldots,0,U_{k+1},U_{k+2},\ldots, U_{k+\lambda},0\ldots,0\right).
\end{eqnarray}
The initial conditions $U(0),\dot{U}(0)\in\mathbb{R}^{K+2}$ are decomposed similarly. Since the decomposition is linear, it follows from (\ref{eq1h}) that the elements $U^{L\setminus I}$ and $U^I$ satisfy, respectively, the equations:
\begin{eqnarray*}
&&\ddot{U}_n^I-\epsilon\Delta_dU^I_n+\omega_d^2U^I_n
-\beta\omega_d^2U^I_n~U_{n}^2=0,\;\;n=0,\ldots,K+2,\\
&&\ddot{U}_n^{L\setminus I}-\epsilon\Delta_dU^{L\setminus I}_n+\omega_d^2U^{L\setminus I}_n
-\beta\omega_d^2U^{L\setminus I}_n~U_{n}^2=0\;\;n=0,\ldots,K+2.
\end{eqnarray*}
%
However, taking into account the form of $U^{L\setminus I}$ given in (\ref{br8}), the equation for $U^I$ can be written as %
\begin{eqnarray}
\label{br9}
\ddot{U}_n^I-\epsilon\Delta_dU^I_n&+&\omega_d^2U^I_n-\beta\omega_d^2U^I_n~U_{n}^2=0,\;\;n=k+1,\ldots,k+\lambda,\\
\label{br10}
U_k^I&=&U_{k+\lambda+1}=0.
\end{eqnarray}
Relabeling for convenience, the system (\ref{br9})-(\ref{br10}) can be considered on the interval $I'$ of the $\lambda+2$ oscillators $j=0,\ldots,\lambda+1$ as
\begin{eqnarray}
\label{br11}
&&\ddot{Z}_j-\epsilon\Delta_d Z_j+\omega_d^2 Z_j-\beta\omega_d^2 Z_j^3=0,\;t>0\;\;j=1,\ldots,\lambda,\\
\label{br11in}
&&Z(0),\dot{Z}(0)\in\mathbb{R}^{\lambda+2},\\
\label{br12}
&&Z_0=Z_{\lambda+1}=0,\;\;t\geq 0.
\end{eqnarray}

For  brevity of notation, we
indicate the initial data and the solution of (\ref{br11})-(\ref{br12}) as elements of the set
\begin{eqnarray*}
\mathcal{S}_{I'}=\left\{Z\in\mathbb{R}^{\lambda+2}\;:\;Z_0=Z_{\lambda+1}=0\right\}.
\end{eqnarray*}
The system (\ref{br11})-(\ref{br12}) on the segment $I'$ conserves the Hamiltonian
on $I'$, i.e.,
\begin{eqnarray}
\label{br13}
\mathcal{H}_{I'}(t)=\frac{1}{2}\sum_{j=0}^{\lambda+1}\dot{Z}_j^2+\frac{1}{2}\sum_{j=0}^{\lambda+1}(Z_{j+1}-Z_j)^2
+\frac{\omega_d^2}{2}\sum_{j=0}^{\lambda+1}Z_j^2-\frac{\beta\omega_d^2}{4}\sum_{j=0}^{j=\lambda+1} Z_j^4=\mathcal{H}_{I'}(0),
\end{eqnarray}
for all $t\in [0,T_{\mathrm{max}}]$. Let us recall that the following inequality holds
\begin{eqnarray}
\label{br14}
\epsilon\sum_{j=0}^{\lambda+1}(Z_{j+1}-Z_j)^2\geq \mu_1\sum_{j=0}^{\lambda+1}Z_j^2,
\end{eqnarray}
for all $Z\in \mathcal{S}_{I'}$, where
\begin{eqnarray}
\label{br15}
\mu_1(I')=\frac{4}{h^2}\sin^2\left(\frac{\pi h}{2L'}\right)=\frac{4}{h^2}\sin^2\left(\frac{\pi h}{2(1+2h)}\right)
\end{eqnarray}
is the first eigenvalue of the discrete Dirichlet Laplacian on $I'$,
\begin{eqnarray*}
-\epsilon\Delta_d\Phi_j&=&\mu\Phi_j,\;\;j=0,\ldots,\lambda,\\
\Phi_0&=&\Phi_{\lambda+1}=0.
\end{eqnarray*}
\newline

$\mathrm{Case~B}.$ In the general case, the end-points of the unit interval $I$ are not occupied by oscillators.
In this case, the interval $I'$ of the cut-off procedure has length
$$
L'=\left\{
  \begin{array}{ll}
    4h, & \hbox{when $h>0.5$, } \\
    (m+1)h, & \hbox{when $h<0.5$,}
  \end{array}
\right.
$$
and the first eigenvalue of the discrete Dirichlet Laplacian is
$$
\mu_1(I')=\left\{
  \begin{array}{ll}
    \frac{4}{h^2}\sin^2\left(\frac{\pi }{8}\right), & \hbox{when $h>0.5$, } \\
    \frac{4}{h^2}\sin^2\left(\frac{\pi }{2(m+1)}\right), & \hbox{when $h<0.5$.}
  \end{array}
\right.
$$
With these preparations in hand, we shall start the investigations for conditions on the initial data $Z(0)$, $\dot{Z}(0)$ of the segment, which may lead to escape dynamics.
\begin{proposition}
\label{indic1}
Assume that the initial data  $Z(0)\in\mathcal{S}_{I'}$ of (\ref{br11})-(\ref{br12}) satisfy
\begin{eqnarray}
\label{br16}
||Z(0)||_{4}>\frac{\sqrt{\mu_1+\omega_d^2}}{\omega_d\sqrt{\beta}}.
\end{eqnarray}
Then, there exists $0<t_1\leq T_{\mathrm{max}}$, such that the solution $Z(t)\in\mathcal{S}_{I'}$ satisfies
\begin{eqnarray}
\label{br17}
|||Z(t)||_{4}>\frac{\sqrt{\mu_1+\omega_d^2}}{\omega_d\sqrt{\beta}},\;\;\mbox{for all}\;\;t\in [0, t_1].
\end{eqnarray}
\end{proposition}
\textbf{Proof:} By using (\ref{br14}) and (\ref{br15}), we may observe that the Hamiltonian $\mathcal{H}_{I'}(t)$ satisfies
\begin{eqnarray}
\label{br19}
\mathcal{H}_{I'}(t)\geq\frac{\mu_1+\omega_d^2}{2}\left(\sum_{j=0}^{\lambda}Z_j^4\right)^{\frac{1}{2}}
-\frac{\beta\omega_d^2}{4}\sum_{j=0}^{\lambda} Z_j^4,\;\;\mbox{for all}\;\;t\in[0,T_{\mathrm{max}}].
\end{eqnarray}
Working as in Theorem \ref{bud}, we now consider
the function
\begin{eqnarray}
\label{br20}
\Phi(\rho):=\frac{\mu_1+\omega_d^2}{2}\rho^2-\frac{\beta\omega_d^2}{4}\rho^4.
\end{eqnarray}
having the unique positive maximum
\begin{eqnarray}
\label{br21}
\Phi(\rho_*)=\frac{(\mu_1+\omega_d^2)^2}{4\beta\omega_d^2}
\;\;\mbox{at}\;\;\rho_*=\frac{\sqrt{\mu_1+\omega_d^2}}{\omega_d\sqrt{\beta}}.
\end{eqnarray}
Assuming  that the initial data $Z(0)\in\mathcal{S}_{I'}$ are chosen such that
\begin{eqnarray*}
||Z(0)||_{4}>\rho_*=\frac{\sqrt{\mu_1+\omega_d^2}}{\omega_d\sqrt{\beta}},
\end{eqnarray*}
the continuity of the norm $||Z(t)||_{4}$ implies that there exists $0<t_1\leq T_{\mathrm{max}}$, such that
\begin{eqnarray*}
|||Z(t)||_{4}>\rho_*=\frac{\sqrt{\mu_1+\omega_d^2}}{\omega_d\sqrt{\beta}}.,\;\;\mbox{for all}\;\;t\in [0, t_1],
\end{eqnarray*}
which is (\ref{br17}). $\diamond$

Condition (\ref{br16}) is not establishing escape by itself, since it is certainly satisfied with $t_1\equiv T_{\mathrm{max}}$, when the additional restriction on the initial Hamiltonian energy $\mathcal{H}_{I'}(Z(0))$ on $I'$,
\begin{eqnarray}
\label{br22}
\mathcal{H}_{I'}(0)<\Phi(\rho_*)=\frac{(\mu_1+\omega_d^2)^2}{4\beta\omega_d^2},
\end{eqnarray}
holds. For instance, we may repeat the proof of Theorem \ref{bud} for the excited segment on $I'$ and show that the solution $Z(t)$ of (\ref{br11})-(\ref{br12}) cannot exist globally in time.

With the aim to avoid the restriction on the initial energy (\ref{br22}) and see if the condition (\ref{br16}) suffices for escape, we consider --in the set-up of (\ref{br11})-(\ref{br12})-- the invariant region introduced in \cite{PS73} (see also Ref.~\cite{PSrev}). For instance, we consider the set
\begin{eqnarray*}
\mathcal{W}(Z):=\left\{Z\in\mathcal{S}_{I'}\;:\;\mathcal{I}(Z)>0\;\;\mbox{and}\;\; \mathcal{J}(Z)<d\right\},
\end{eqnarray*}
where the functional $\mathcal{I}$ is given by
\begin{eqnarray*}
\mathcal{I}(Z)&=&\epsilon\sum_{j=0}^{\lambda+1}(Z_{j+1}-Z_j)^2+\omega_d^2\sum_{j=0}^{\lambda+1}Z_j^2
-\beta\omega_d^2\sum_{j=0}^{\lambda+1} Z_j^4,
\end{eqnarray*}
and $d=\inf\mathcal{J}(Z)$ denotes the infimum over all $Z\in\mathcal{S}_{I'}$ of the functional
\begin{eqnarray*}
\mathcal{J}(Z)=\frac{\epsilon}{2}\sum_{j=0}^{\lambda+1}(Z_{j+1}-Z_j)^2
+\frac{\omega_d^2}{2}\sum_{j=0}^{j=\lambda+1}Z_j^2-\frac{\beta\omega_d^2}{4}\sum_{j=0}^{\lambda+1} Z_j^4.
\end{eqnarray*}
The analysis of Ref.~\cite{PS73} for the continuum limit suggests that the set $\mathcal{W}(Z)$
{\em is invariant under the flow associated to (\ref{br11})-(\ref{br12}) and solutions are global in time.}
More precisely, if the initial data $Z(0)\in\mathcal{W}(Z)$, i.e.,
\begin{eqnarray}
\label{br23}
\mathcal{I}(Z(0))&>& 0,\\
\label{br24}
\mbox{and}\;\;\mathcal{H}_{I'}(0)&<&d,
\end{eqnarray}
then the solution $Z(t)$ of (\ref{br11})-(\ref{br12}) satisfies
\begin{eqnarray}
\label{br25}
Z(t)\in\mathcal{W}(t),\;\;\mbox{for all}\;\;t\in [0,\infty),
\end{eqnarray}
i.e., $T_{\mathrm{max}}=\infty$. On the other hand, a solution which blows-up in finite time should initiate from initial data violating the condition (\ref{br23}). It is interesting to observe that {\em in the discrete case, an initial condition which violates (\ref{br23}) satisfies (\ref{br16}).}
\begin{proposition}
\label{indic2}
If the initial data $Z(0)\in\mathcal{S}_{I'}$ satisfy
\begin{eqnarray}
\label{br26}
\mathcal{I}(Z(0))<0,
\end{eqnarray}
then they satisfy (\ref{br16}).
\end{proposition}
\textbf{Proof:} Condition (\ref{br26}) reads as
\begin{eqnarray*}
\beta\omega_d^2\sum_{j=0}^{\lambda+1} Z_j^4(0)>\epsilon\sum_{j=0}^{\lambda+1}(Z_{j+1}(0)-Z_j(0))^2+\omega_d^2\sum_{j=0}^{\lambda+1}Z_j^2(0).
\end{eqnarray*}
Applying (\ref{br14}) and (\ref{embp2p4}) for $Z(0)$, we have
\begin{eqnarray*}
\epsilon\sum_{j=0}^{\lambda+1}(Z_{j+1}(0)-Z_j(0))^2+\omega_d^2\sum_{j=0}^{\lambda+1}Z_j^2(0)
\geq(\mu_1+\omega_d^2)\sum_{j=0}^{\lambda+1}Z_j^2(0)
\geq(\mu_1+\omega_d^2)\left(\sum_{j=0}^{\lambda+1}Z_j^4(0)\right)^{\frac{1}{2}}.
\end{eqnarray*}
Thus, we have derived that
\begin{eqnarray*}
\beta\omega_d^2||Z(0)||_4^4>(\mu_1+\omega_d^2)||Z(0)||_4^2,
\end{eqnarray*}
which implies that
\begin{eqnarray*}
||Z(0)||_{4}>\frac{\sqrt{\mu_1+\omega_d^2}}{\omega_d\sqrt{\beta}},
\end{eqnarray*}
i.e., (\ref{br16}) is satisfied.\ \ $\diamond$

\subsection{Numerical study 3: Evolution of the lattice segment}
\label{S4A}

Propositions \ref{indic1} and \ref{indic2} indicate that (\ref{br16}) {\em may suffice for escape dynamics}, taking into account that {\em the connection with this behavior shall be established if (\ref{br16}) may describe initial configurations of the segment $Z(0)$ inside the potential well.} The numerical investigation of this issue is the purpose of the numerical study of this section. With the change of variable $t\rightarrow\frac{1}{h}t$, we rewrite the system (\ref{br11})-(\ref{br12}) as
\begin{eqnarray}
\label{br27}
&&\ddot{Z}_j-\Delta_d Z_j+\Omega_d^2 Z_j-\beta\Omega_d^2 Z_j^3=0,\;t>0\;\;j=1,\ldots,\lambda,\;\;\Omega_d^2=h^2\omega_d^2,\\
\label{br27in}
&&Z(0),\dot{Z}(0)\in\mathbb{R}^{\lambda+2},\\
\label{br28}
&&Z_0=Z_{\lambda+1}=0,\;\;t\geq 0.
\end{eqnarray}
%
We will consider a set of parameter values which will
span the whole regime, from the asymptotically linear
to the anti-continuum limit.
For a lattice spacing of $h=0.5$ we fall in
case A, where the endpoints of the unit interval $I$ are occupied by oscillators.
In this case, the unit interval consists of $\lambda=3$ oscillators, with the $m=1$ included in $I$ [see also (\ref{br4})] together with the two endpoints. The length of the interval $I'$, which is
composed by $I$ and the two adjacent neighbors, has length $L'=2$ and consists of $\lambda+2$=5 oscillators [see also (\ref{br6})].
Since the number of points is $\lambda=3$, we will examine the evolution of the $3$-point segment, by considering the simplest case of initial data, that is zero velocities $\dot{Z}(0)=(0,0,0)$, and initial positions of the form $Z(0)=(A, A, A)$. For these initial data, the  $\ell^4$-norm  is
$$||Z(0)||_4=\left(\sum_{j=0}^{\lambda+1}Z_j^4(0)\right)^{\frac{1}{4}}
=\left(\sum_{j=1}^{\lambda}Z_j^4(0)\right)^{\frac{1}{4}}=\left(\sum_{j=1}^{3}A^4\right)^{\frac{1}{4}}=3^{\frac{1}{4}}A.$$

From (\ref{br15}), the eigenvalue $\mu_1(I')\simeq 2.34315$, and the
condition (\ref{br16}) results in
a critical value for the displacements $A$, depending on $\omega_d$ as follows:
\begin{eqnarray}
\label{br29}
A>A_{\mathrm{crit}}:=\frac{\sqrt{\mu_1+\omega_d^2}}{3^{\frac{1}{4}}\omega_d}.
\end{eqnarray}
As will be justified below, the critical value $A_{\mathrm{crit}}$ is
physically relevant for the escape dynamics. This is because
in the discrete regime, it yields initial
segment configurations inside the well and has the physically expected
behavior in the limiting cases of the
asymptotically linear and of the anticontinuous limit.

\begin{figure}
\begin{center}
    \begin{tabular}{cc}
    \includegraphics[scale=0.35]{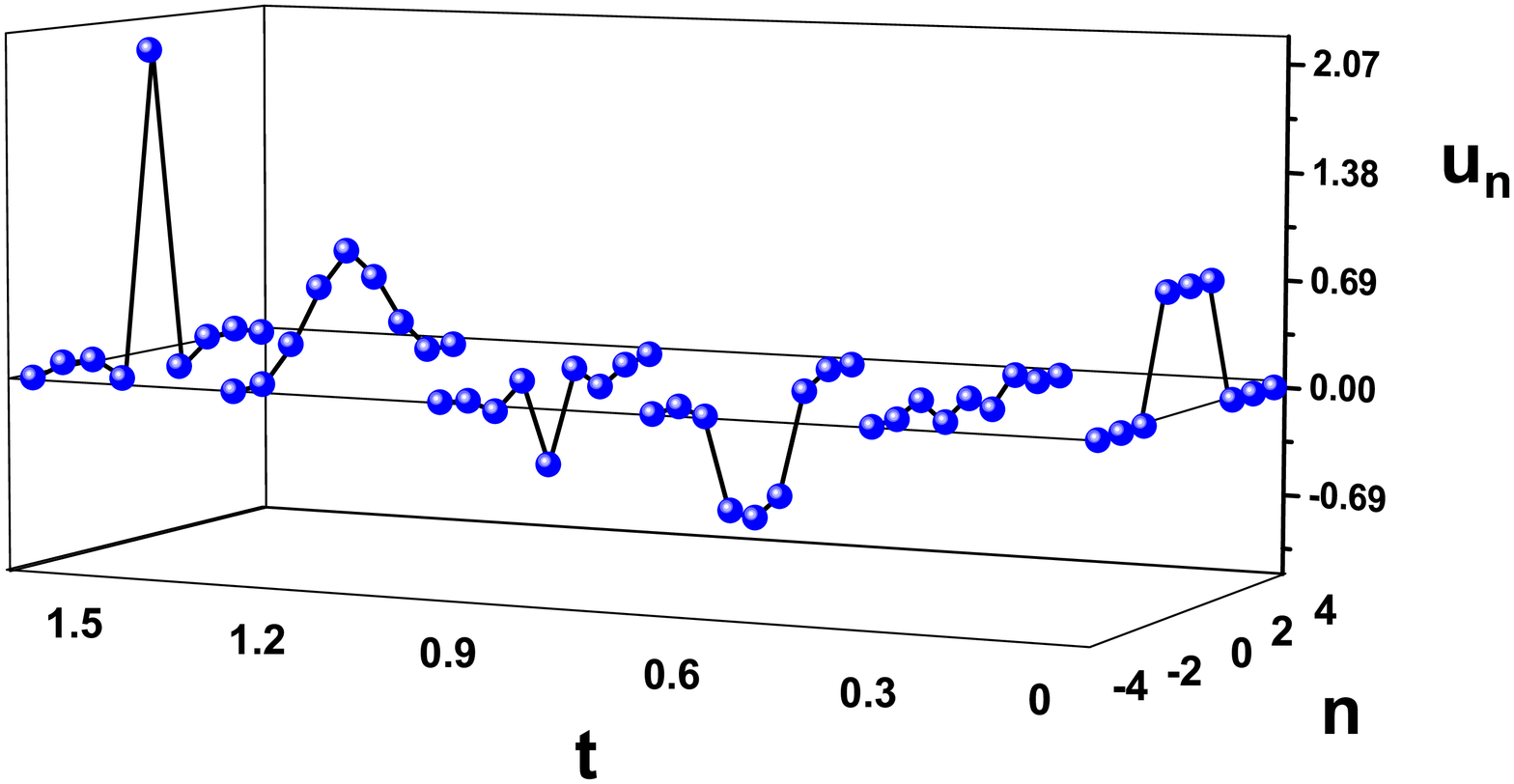} &
    \includegraphics[scale=0.3]{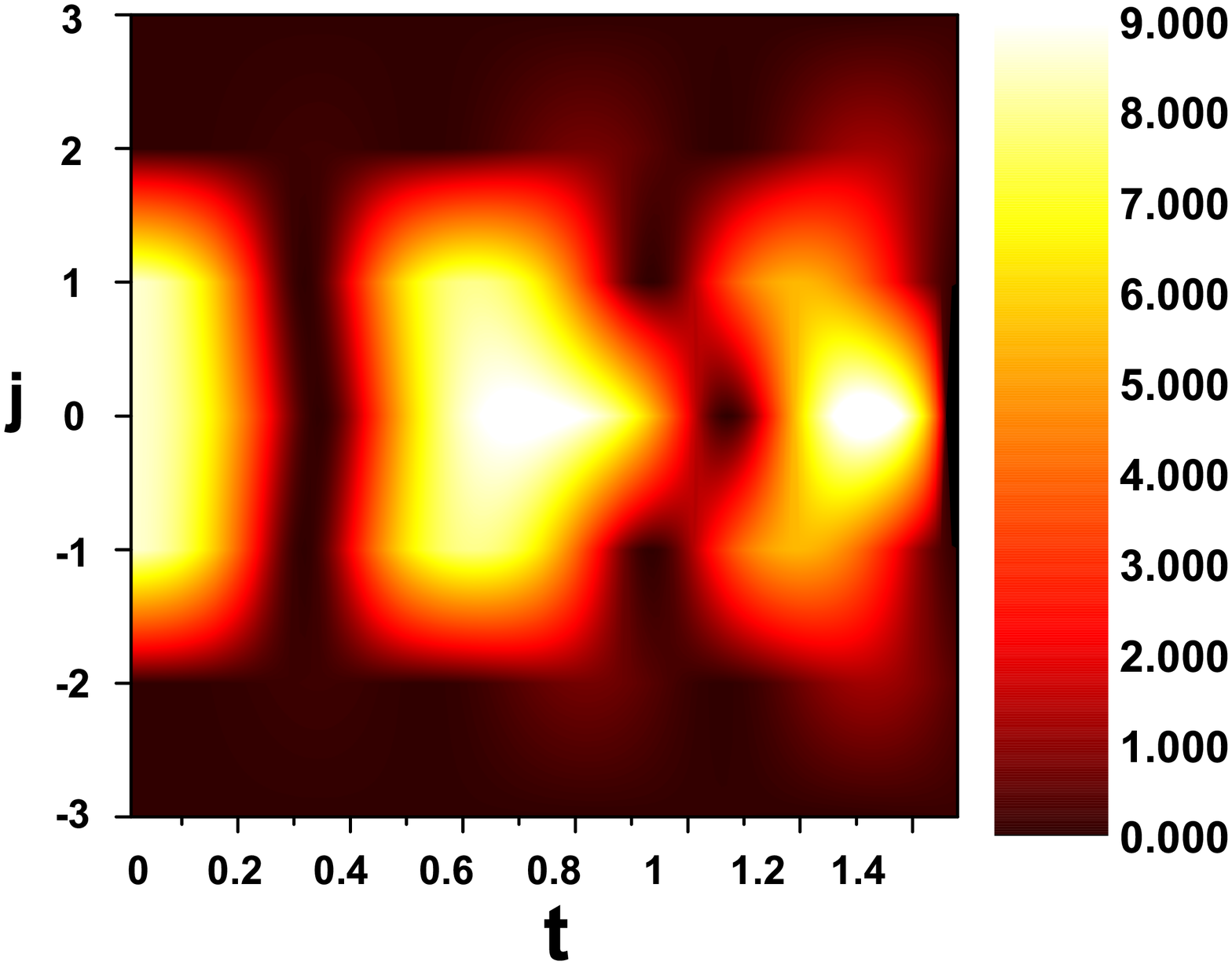}
\end{tabular}
\caption{Left panel: escape dynamics for the $3$-units segment $Z(0)=(A,A,A)$
with $A=0.79>A_{\mathrm{crit}}$;
clearly,
the central unit $j=0$ is escaping from the potential well.
Right panel: contour plot showing the evolution of the potential energy of the chain in the case of an
initially excited $3$-units segment; the plot demonstrates
the energy transfer from the adjacent units to the central unit and its localization therein.
The parameters used
are $\Omega_d^2=10$ ($\omega_d^2=40$, $h=0.5$) and $\beta=1$.}
\label{fig3b}
\end{center}
\end{figure}

In the left panel of Fig.~\ref{fig3b} we show the time  evolution
of the $3$-unit segment in the
moderately discrete regime ($\Omega_d^2=10$),
with an initial condition,
$Z(0)=(A,A,A)$ where $A=0.79>A_{\mathrm{crit}}$.
As shown in this panel,
the segment initially performs one oscillation,
but subsequently reorganizes so that
at time $t\sim 1.5$ the central unit ($j=0$) has already crossed the barrier
$\mathcal{U}_{\mathrm{thresh}}\sim 1.8$ and it will lead the chain to collapse.
In particular, it can never return to the
potential well, according to Theorem \ref{GP}, due to the autonomous
nature of the
system at hand.

The energy transfer from the adjacent units of the segment
to the central unit (resulting in growing amplitude oscillations of this unit), is visualized in the right panel of Fig.~\ref{fig3b}. The potential energy
stored in the $3$-unit segment (light colored area) is progressively
localized within the central unit. The dark area between the patterns
visualizes the passing from the bottom of the potential
well $U_{\mathrm{min}}=0$ (with a maximal
kinetic energy), as the units oscillate.

The same escape dynamics have also been
confirmed numerically, for the set of parameters
in the strongly discrete regime ($\Omega_d^2=100$),
with the initial condition $A=0.77>A_{\mathrm{crit}}$. The collapse time
in this case is
$t\simeq 6.4$, which is larger compared to the respective one
in the moderately discrete regime. This is expected
due to the weaker interaction between the units,
which results
in a longer time interval during which the energy is redistributed,
the more discrete the lattice becomes.
Besides that, in the strongly discrete regime, the depth of the potential well
is considerably increased, which has, as a result, a
longer time 
needed
for the excited units to climb over the saddle points and finally escape
(cf. Fig.~\ref{fig0a}).

The analytically obtained threshold
$A_{\mathrm{crit}}$, for which the segment $Z(0)=(A,A,A)$ will exhibit escape dynamics, is plotted
as a function of the coupling parameter $\omega_d^2$ in Fig.~\ref{fig3c} --see
the continuous (black) line;
for comparison, the numerically obtained threshold is also presented in this figure by
(red) dots.
The observation that $\lim_{\omega_d^2\rightarrow 0} A_{\mathrm{crit}}=\infty$
as $\omega_d^2 \rightarrow 0$ reflects the fact that blow-up
is not possible in the asymptotically linear regime. The numerical
simulations verified the existence of a critical value
$\omega^2_{d,\mathrm{min}}$, as we approach the
asymptotically linear
from the discrete regime, below which escape is prevented, and the
energy stored in the $3$-units segment is dispersed along the chain
leading to global existence.
For large values of $\omega_d^2$ we observe that the numerical threshold converges slowly to the
limit
$\lim_{\omega_d^2\rightarrow\infty} A_{\mathrm{crit}}\sim 0.75$.

\begin{figure}
\begin{center}
    \begin{tabular}{cc}
    \includegraphics[scale=0.3]{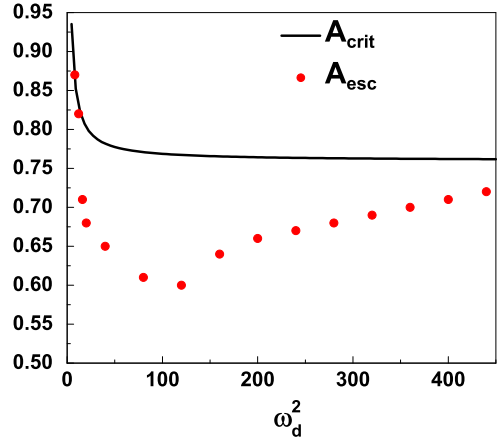}
\end{tabular}
\caption{
The analytically obtained threshold $A_{\mathrm{crit}}$, as a function of the coupling parameter $\omega_d^2$ (continuous (black) curve), against the numerical value $A_{\mathrm{esc}}$ ((red) dots) for the escape of the $3$-units segment. Other parameters used are $h=0.5$, $\beta=1$.}
\label{fig3c}
\end{center}
\end{figure}

\subsection{Modulational instability mechanism}
\label{S5MI}

In this section, we will briefly review initially
the modulational instability mechanism \cite{Dirk1}, with the subsequent
aim to investigate its connection with
escape dynamics of plane wave initial data~\cite{KP92}.
This will complete our program of examining blow-up for the chain
starting from
one site excitations and progressively passing to few and finally
to many site excitations.

The modulation instability mechanism concerns the instability of
a plane wave of the form
\begin{eqnarray}
\label{mi1}
U_n=\Phi_0e^{i(qn-\omega t)}+\hbox{c.c},
\end{eqnarray}
under small perturbations of different wave numbers.
When this
instability occurs, the exponential growth of the perturbations
creates localized excitations.
The question at hand here is if a sizeable
growth of the perturbations may create a critical localized
excitation which, in turn, may lead to escape dynamics. Let us recall
first that in the linear limit $\beta=0$ of (\ref{eq1h}), the
existence of plane waves (\ref{mi1}) is associated with the dispersion relation:
\begin{eqnarray}
\label{mi2}
\omega^2=4\epsilon\sin^2\left(\frac{q}{2}\right)+\omega_d^2.
\end{eqnarray}
For $q=0$ and $q=\pi$, (\ref{mi2}) gives $\omega^2=\omega_d^2$ and
$\omega^2=\omega_d^2+4\epsilon$ respectively. Thus, the linear
spectrum has a gap of size $\omega_d$, and is bounded by the frequency
$\omega_{\mathrm{max}}^2=\omega_d^2+4\epsilon$.

{\em In the nonlinear case} $\beta>0$, we seek for solutions of (\ref{eq1h}) of the form
\begin{eqnarray}
\label{mi3}
U_n(t)=\phi_n(t)e^{-i\omega_dt}+ \hbox{c.c},\;\;\phi_n\in\mathbb{C},
\end{eqnarray}
where
$\phi_n(t)$ denotes the varying envelope. Substitution of
(\ref{mi3}) into (\ref{eq1h}) implies that the varying envelope
$\phi_n(t)$ satisfies the equation
\begin{eqnarray}
\label{mi4}
\ddot{\phi_n}-2i\omega_d\dot{\phi_n}-\omega_d^2\phi_n=\epsilon(\phi_{n+1}-2\phi_n+\phi_{n-1})-\omega_d^2\phi_n
+\beta\omega_d^2\left(3|\phi_n|^2\phi_n+\phi_n^3e^{2i\omega_dt}\right)+\hbox{c.c}.
\end{eqnarray}
Under the assumption of the slow-variation in time of $\phi_n(t)$ with respect to the main oscillation at frequency $\omega_d$, i.e., $\dot{\phi}_n<<\omega_d\phi_n$, in the rotating-wave approximation, we may keep only the terms proportional to $e^{\pm i\omega_dt}$;
this way, (\ref{mi4}) results in the DNLS equation:
\begin{eqnarray}
\label{mi5}
2i\omega_d\dot{\phi_n}+\epsilon(\phi_{n+1}-2\phi_n+\phi_{n-1})+3\beta\omega_d^2|\phi_n|^2\phi_n=0.
\end{eqnarray}
We proceed by seeking solutions of (\ref{mi5}) of the form:
%
\begin{eqnarray}
\label{mi6}
\phi_n(t)=\phi_0e^{i\theta_n},\;\;\theta_n(t)=\,qn-\tilde{\omega} t,\;\;\phi_0\in\mathbb{R}.
\end{eqnarray}
Inserting (\ref{mi6}) in (\ref{mi5}), we obtain
\begin{eqnarray*}
2\omega_d\tilde{\omega}\phi_0e^{i\,qn}+\epsilon\phi_0\left(e^{i\,q(n+1)}-2e^{i\,qn}+e^{i\,q(n-1)}\right)
+3\beta\omega_d^2|\phi_0|^2\phi_0e^{i\,qn}=0,
\end{eqnarray*}
which leads to the following
dispersion relation
for the frequency $\tilde{\omega}$:
%
\begin{eqnarray}
\label{mi7}
2\omega_d\tilde{\omega}=4\epsilon\sin^2\left(\frac{\,q}{2}\right)-3\beta\omega_d^2|\phi_0|^2.
\end{eqnarray}
In summary, the validity of the single-frequency approximation leading to the DNLS equation (\ref{mi5})
for the varying envelope $\phi_n (t)$ of the solutions (\ref{mi3}) of the DKG equation (\ref{eq1h}),
assumes that the gap frequency $\omega_d$ is large
compared to the other frequencies of the system,
\begin{eqnarray}
\label{mi8}
4\epsilon<<\omega_d^2,
\end{eqnarray}
and
the nonlinearity is weak, in the sense:
\begin{eqnarray}
\label{mi9}
\beta\phi_0^2<<1.
\end{eqnarray}
Having the restrictions (\ref{mi8}) and (\ref{mi9}) in mind,
one can study the
modulational instability conditions of (\ref{mi6}) by considering the ansatz
%
\begin{eqnarray}
\label{mi10}
\phi_n=(\phi_0+\chi_n)e^{i(\theta_n+\psi_n)},
\end{eqnarray}
where $\chi_n$ and $\psi_n$ are small perturbations of the amplitude and phase, respectively, i.e.,
%
\begin{eqnarray*}
|\chi_n(t)|<<\phi_0,\;\;\hbox{and}\;\;|\psi_n(t)|<<|\theta_n(t)|.
\end{eqnarray*}
Then, substituting (\ref{mi10}) in the DNLS (\ref{mi5}), it turns out that
$\chi_n$ and $\psi_n$ satisfy the system of linear equations
\begin{eqnarray}
\label{mi11}
2\omega_d\dot{\chi}_n
&+&\epsilon\left[(\chi_{n+1}-\chi_{n-1})\sin\,q+\phi_0(\psi_{n+1}-2\psi_n+\psi_{n-1})\cos\,q\right]=0,\\
\label{mi12}
-2\omega_d\phi_0\dot{\psi_n}
&+&\epsilon\left[(\chi_{n+1}-2\chi_n+\chi_{n-1})\cos\,q-\phi_0(\psi_{n+1}-\psi_{n-1})\sin\,q\right]
+6\beta\omega_d^2\phi_0^2\chi_n=0.
\end{eqnarray}
Assuming that
the perturbations $\chi_n$ and $\psi_n$ have the form of plane waves, namely,
%
\begin{eqnarray}
\label{mi13}
\chi_n(t)=\chi_0e^{i(Qn-\Omega t)},\\
\label{mi14}
\psi_n(t)=\psi_0e^{i(Qn-\Omega t)},
\end{eqnarray}
where $\chi_0$ and $\psi_0$ are constant amplitudes, while $Q$ and $\Omega$ denote wavenumber and frequency respectively,
we can derive from
(\ref{mi11})-(\ref{mi12}) the following system for
$\chi_0,\psi_0$,
%
%
\begin{equation}
\label{mi15}
\left(
  \begin{array}{cc}
    -2i(\omega_d\Omega-\epsilon\sin Q\sin\,q) & -4\epsilon\phi_0\sin^2\left(\frac{Q}{2}\right)\cos\,q \\
   -4\epsilon\sin^2\left(\frac{Q}{2}\right)\cos\,q+6\beta\omega_d^2\phi_0^2  & 2i\phi_0(\omega_d\Omega-\epsilon\sin Q\sin\,q) \\
  \end{array}
\right)
\cdot
\left(
  \begin{array}{c}
    \chi_0 \\
    \psi_0 \\
  \end{array}
\right)
=\left(
   \begin{array}{c}
     0 \\
     0 \\
   \end{array}
 \right).
\end{equation}
For nontrivial solutions, we require the determinant of the matrix in (\ref{mi15}) to be zero,
which leads to the dispersion relation:
\begin{eqnarray}
\label{mi16}
(\omega_d\Omega-\epsilon\sin Q\sin\,q)^2=
\epsilon\sin^2\left(\frac{Q}{2}\right)\cos\,q
\left(4\epsilon\sin^2\left(\frac{Q}{2}\right)\cos\,q-6\beta\omega_d^2\phi_0^2\right).
\end{eqnarray}
From this,
the condition for modulation instability, i.e., for $\Omega$
with a nonvanishing imaginary part
can directly be obtained from (\ref{mi16}); this condition is expressed
(for $\cos(q)>0$; if this sign changes, so does the sign of the inequality
below) in terms of the amplitude $\phi_0$ and the wavenumber $Q$ as:
\begin{eqnarray}
\label{mi17}
6\beta\omega_d^2\phi_0^2>4\epsilon\sin^2\left(\frac{Q}{2}\right)\cos q.
\end{eqnarray}
In the case of the mode with $q=0$, the condition for
%
%
$\phi_0$ leading to modulational instability is:
%
\begin{eqnarray}
\label{mi18}
\phi_0>\sqrt{\frac{2\epsilon\sin^2\left(\frac{Q}{2}\right)}{3\beta\omega_d^2}}:=\phi_{0,\mathrm{MI}}.
\end{eqnarray}
Condition (\ref{mi18}) should be combined with restrictions (\ref{mi8}) and (\ref{mi9}) which, in our case, are:
\begin{eqnarray}
\label{mi19}
1>>\beta\phi_0^2\;\;\hbox{and}\;\;\omega_d^2>>4\epsilon.
\end{eqnarray}
We also remark that in the case of modes with $0\leq q<\frac{\pi}{2}$, since
\begin{eqnarray*}
6\beta\omega_d^2\phi_0^2-4\epsilon\sin^2\left(\frac{Q}{2}\right)\cos q\geq 6\beta\omega_d^2\phi_0^2-4\epsilon,
\end{eqnarray*}
instability of a plane wave
occurs if
\begin{eqnarray}
\label{mi20}
\phi_0^2>\frac{2}{3}\frac{\epsilon}{\beta\omega_d^2}=\frac{2}{3\beta h^2\omega_d^2}
=\frac{2}{3\beta\Omega_d^2},\;\;\Omega_d^2=h^2\omega_d^2,
\end{eqnarray}
where we have used
the scaling $t\rightarrow\frac{1}{h}t$ as before.
%

\subsection{Small data global existence conditions and their violation}
\label{S5VI}
The last scenario which will be considered is based on the violation of small data, global existence conditions, and the investigation of the possible connection of this violation with the generation of instabilities. For the derivation of the conditions for global existence, which shall guarantee non-escape dynamics, we shall consider a discrete variant of the method of Ref.~\cite{cazh}, and use the energy functional:
\begin{eqnarray}
\label{ne1}
E(t):=||\dot{U}(t)||_{\ell^2}^2+||U(t)||_{\ell^2_1}^2=  \sum_{n=-\infty}^{+\infty}\dot{U}_n^2+\sum_{n=-\infty}^{+\infty}(U_{n+1}-U_n)^2+\omega_d^2\sum_{n=-\infty}^{+\infty}U_n^2,
\end{eqnarray}
where $||U||_{\ell^2_1}^2$ denotes the ``linear coupling energy''-norm
\begin{eqnarray*}
||U||_{\ell^2_1}^2:=\sum_{n=-\infty}^{+\infty}(U_{n+1}-U_n)^2+\omega_d^2\sum_{n=-\infty}^{+\infty}U_n^2.
\end{eqnarray*}
We shall also consider the quantities
\begin{eqnarray}
\label{ne18}
E_{\pm}=\frac{\omega_d^2\left[1\pm\sqrt{1-\frac{2\beta}{\omega_d^2}E(0)
\left(1+\frac{\beta}{2\omega_d^2}\right)}\right]}{\beta},
\end{eqnarray}
which are real and positive if $E(0)$ satisfies
\begin{eqnarray}
\label{ne19}
E(0)<\frac{\omega_d^2}{2\beta\left(1+\frac{\beta}{2\omega_d^2}\right)}.
\end{eqnarray}
Then, the small data, global existence result is stated in the following Theorem.
\begin{theorem}
\label{Thne}
We consider the energy function (\ref{ne1}) and we assume that the initial data (\ref{eq2d}) have energy
\begin{eqnarray}
\label{ne27}
E(0)<\min\left\{1,\frac{\omega_d^2}{2\beta\left(1+\frac{\beta}{2\omega_d^2}\right)}\right\}.
\end{eqnarray}
Then the solution of (\ref{eq1}) exists globally in time ($T_{\mathrm{max}}=\infty$),
and satisfies for all $t\in[0,\infty)$ the energy bound
\begin{eqnarray}
\label{ne28}
E(t)<E_{-}.
\end{eqnarray}
\end{theorem}
{\bf Proof:} See Appendix C.\ \ $\Box$

On the other hand, if the smallness condition (\ref{ne19}) is violated, i.e.,
\begin{eqnarray}
\label{mi22}
E(0)>\frac{\Omega_d^2}{2\beta\left(1+\frac{\beta}{2\Omega_d^2}\right)},
\end{eqnarray}
the inequality (\ref{ne17}) is valid for $E(t)\in [0,\infty)$ and the {\em possibility of unbounded solutions cannot be excluded}.

With this observation at hand, we may investigate the dynamics of the chain for initial data satisfying (\ref{mi22}) to see if this violation condition may give an analytical value of amplitudes leading to blow-up and escape.
Let us also remark that
\begin{eqnarray}
\label{mi24}
\frac{\Omega_d^2}{2\beta\left(1+\frac{\beta}{2\Omega_d^2}\right)}>1\;\;\hbox{when}\;\;\Omega_d^2>\beta(1+\sqrt{2}).
\end{eqnarray}

\subsection{Numerical study 4: Modulational instability and violation of conditions for global existence}
\label{S5NU}
We conclude
this section with the last numerical study, investigating the modulational instability conditions (\ref{mi18})-(\ref{mi19}), and the violation of small data global existence (\ref{mi22})-(\ref{mi24}).
We will consider
lattice parameters in the strongly discrete regime, namely, $h=0.5$ and $\omega_d^2=400$ ($\Omega_d^2=\omega_d^2h^2=100$),
while for the nonlinearity parameter we will
use the value $\beta=0.1$. We intend to examine (\ref{mi20}), i.e., the conditions for modulational instability and stability for amplitudes $\phi_0>\phi_{0,\mathrm{MI}}$
and $\phi_0<\phi_{0,\mathrm{MI}}$, respectively.
For our analysis we will
focus on the region of modulation instability in the $(q,Q)$-plane defined in
\cite[Fig.~2(b), pg. 3200]{KP92}, and particularly on the line $q=0$; in this case, the corresponding domain for the
wavenumbers is $\left(0, \frac{3\pi}{4}\right)$,
and we fix $Q=\frac{\pi}{2}$.

For the above set of parameters, the critical value of the amplitude $\phi_{0,\mathrm{MI}}$ defined by the right-hand side of (\ref{mi20}) is:
\begin{eqnarray}
\label{mi26}
\phi_{0,\mathrm{MI}}=0.18.
\end{eqnarray}
The violation condition (\ref{mi22}) will also be
tested for plane wave initial data, and we will derive the relevant critical value of the amplitude $\Phi_{0,\mathrm{crit}}$, as shown below.

The plane wave (\ref{mi1}) can be rewritten as
$U_n(t)=2\phi_0\cos(qn-\omega t)$ and, thus,
at $t=0$:
\begin{eqnarray}
\label{mi26a}
U_n(0)=2\phi_0\cos qn,\;\;\mbox{and}\;\;\dot{U}_n(0)=2\omega\phi_0\sin qn,
\end{eqnarray}
In order to test the condition (\ref{mi22}) numerically we should consider a finite lattice of $K$ units.
In this case, the initial data (\ref{mi26a}) have initial energy $E(0):=E_{pw}(0)$ given by
\begin{eqnarray*}
E_{pw}(0)=4\omega^2\phi_0^2\sum_{n=1}^{K}\sin^2qn+4\phi_0^2\sum_{n=1}^{K}\left[\cos(q(n+1))-\cos qn\right]^2+4\phi_0^2\Omega_d^2\sum_{n=1}^{K}\cos^2qn.
\end{eqnarray*}
It readily follows that the mode with $q=0$
has energy
\begin{eqnarray}
\label{mi27a}
E_{pw}(0)=4\Omega_d^2\phi_0^2K,
\end{eqnarray}
and the condition (\ref{mi22}) will be satisfied for a chain of $K$ units if
\begin{eqnarray}
\label{mi27}
\phi_0>\sqrt{\frac{1}{8K\beta\left(1+\frac{\beta}{\Omega_d^2}\right)}}:=\phi_{0,\mathrm{crit}}.
\end{eqnarray}

On the other hand, to get a $K$-independent condition for the amplitude, we
first observe
that the mean value of the energy $E_{pw}(0)$  of the the $q=0$ mode denoted by $\overline{E_{pw}(0)}$ is due to (\ref{mi27a})
\begin{eqnarray*}
\overline{E_{pw}(0)}=\frac{1}{K}E_{pw}(0)=4\Omega_d^2\phi_0^2.
\end{eqnarray*}
Since $E_{pw}(0)>\overline{E_{pw}(0)}$ the condition (\ref{mi22}) will be satisfied for a chain of arbitrary size if
\begin{eqnarray*}
\overline{E_{pw}(0)}>\frac{\Omega_d^2}{2\beta\left(1+\frac{\beta}{2\Omega_d^2}\right)},
\end{eqnarray*}
implying the  inequality
\begin{eqnarray}
\label{mi2b7}
\phi_0>\sqrt{\frac{1}{8\beta\left(1+\frac{\beta}{\Omega_d^2}\right)}}:=\Phi_{0,\mathrm{crit}}.
\end{eqnarray}
For the example of lattice parameters considered above,
the critical value of the amplitude $\Phi_{0,\mathrm{crit}}$ is found to be
\begin{eqnarray}
\label{mi28}
\Phi_{0,\mathrm{crit}}=1.11.
\end{eqnarray}
Note that both critical values for the amplitudes (\ref{mi26}) and (\ref{mi28}) are physically relevant
regarding escape dynamics since in the case $\beta=0.1$ the saddle points are
located $U^{\mp}_{\mathrm{max}}=\mp\frac{1}{\sqrt{\beta}}\simeq 3.16$
(and hence the initial excitation places all nodes inside their
corresponding wells). It should be also recalled that the analytical condition (\ref{mi2b7}) rigorously does not ensure global existence or escape (blow-up), since it simply violates the global existence condition (\ref{ne27}) of Theorem \ref{Thne} for plane waves.
The numerical simulations below will examine whether data satisfying
the analytical condition (\ref{mi2b7})] may lead to escape dynamics or not.
\begin{figure}
\begin{center}
    \begin{tabular}{cc}
    \includegraphics[scale=0.3]{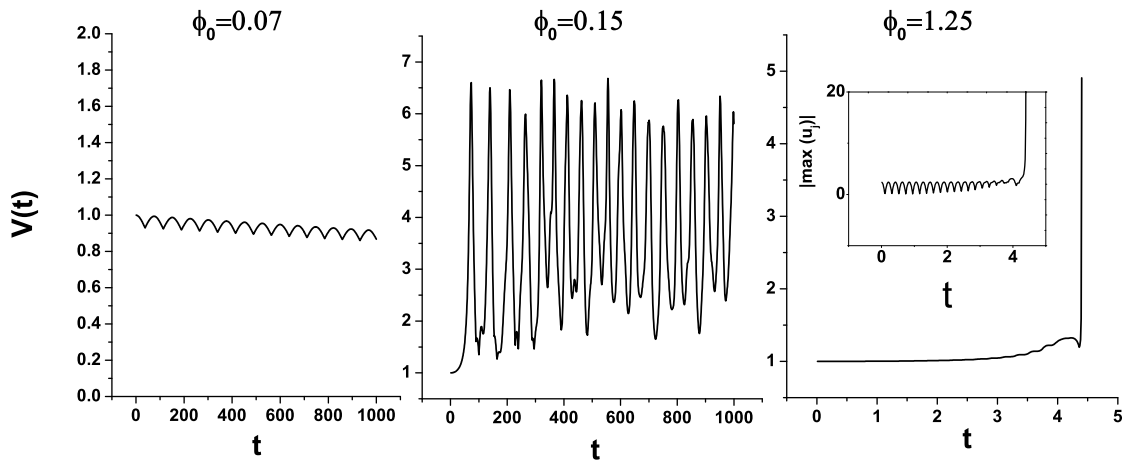}
\end{tabular}
\caption{Evolution of the maximum energy density,
normalized to the initial energy density, $V(t)$ [cf. Eq.~(\ref{mif1})],
of the lattice chain. The initial condition, in all three cases, is:
$U_n(0)=2\phi_0+\hat{\epsilon}\cos (Qn)$, with
$\hat{\epsilon}=0.005$, $Q=\frac{\pi}{2}$. The three panels correspond, respectively, to
modulational stability for $\phi_0=0.07$ (left panel), modulational instability but without collapse
for $\phi_0=0.15$ (middle panel), and  escape dynamics for $\phi_0=1.25$ (right panel). The inset in the right panel shows the evolution of the maximum
amplitude among the $U_n$,
clearly demonstrating the escape at $t \simeq 4.5$.
}
\label{fig4a}
\end{center}
\end{figure}

The initial condition used in the numerical study is
of the form [cf. Eq.~(\ref{mi26a}) for $q=0$]:
\begin{eqnarray}
\label{mpwic}
U_n(0)=2\phi_0+\hat{\epsilon}\cos (Qn),\;\;\hat{\epsilon}<<\phi_0,
\end{eqnarray}
%
where the amplitude of the small perturbation is $\hat{\epsilon}=0.005$.

Figure \ref{fig4a} demonstrates the transition from modulational
stability to escape dynamics for plane wave initial data,
by plotting the evolution of the ratio
\begin{eqnarray}
\label{mif1}
V(t)=\max_{n}\frac{\mathcal{H}_n(t)}{\mathcal{H}_n(0)},
\end{eqnarray}
%
where the quantity
\begin{eqnarray}
\label{mif2}
\mathcal{H}_n=\frac{1}{2}\dot{U}_n^2+\frac{1}{2}(U_{n+1}-U_n)^2+\frac{\omega_d^2}{2}U_n^2-\frac{\beta\omega_d^2}{4}U_n^4,
\end{eqnarray}
is the
energy density of the system,
and $\mathcal{H}_n(0)$ corresponds to the initial energy density, at $t=0$.
The figure demonstrates the transition from modulational stability (left)
to modulational instability (middle)
and finally to self-organized escape (right)
upon varying the amplitude $\phi_0$, as is explained in more detail below.

The left panel of Fig.~\ref{fig4a} shows the evolution of (\ref{mif1}) for $\phi_0=0.07<\phi_{0,\mathrm{MI}}$ and justifies  (for this choice) the
modulational stability of the plane wave;
we observe that $V(t)$ shows a small fluctuation around
its initial value $V(0)=1$, thus depicting the stability of the configuration
over the entire time interval of the numerical integration of the system ($1000$ time units).
Note that the same behavior was found for other values of the amplitude $\phi_0$, up to
$\phi_0<0.15$.

The situation changes drastically for $\phi_0=0.15$: beyond
this value,
modulational instability starts manifesting
itself, as shown in the middle panel of Fig.~\ref{fig4a}.
The onset of the instability is characterized by a significant increase of $V(t)$ from its initial value $V(0)=1$
(an order of magnitude larger than that observed in the previous case),
which implies
exponential growth of the excited mode.
We mention that $\phi_0=0.15< \phi_{0,\mathrm{MI}}=0.18$; this deviation
is expected due to the approximate description of the initial
equation (\ref{eq1h}) by the DNLS model (\ref{mi5}).
The long-time behavior of the system, in this modulationaly unstable regime, is shown in Fig.~\ref{fig4b} (the parameter values, as well as the initial condition, are the same with the ones in the middle panel of Fig.~\ref{fig4a}). The contour plot of the energy density $\mathcal{H}_n$ [cf. Eq.~(\ref{mif2})] reveals that up to $t\simeq 2000$ a periodic pattern
is formed, due to the modulational-instability-induced exponential growth of the excited mode.  This is a so-called breather lattice (see e.g.~\cite{avadh}
for a relevant
discussion of such waveforms in integrable models where they exist in
analytically tractable form).
However, after $t=2000$, and due
to the highly nontrivial dynamical interactions of such modes coming
into play,
at this stage, we observe the formation of moving breathers, each of which acquiring a small fraction of the total amount of energy. These breathers interact with each other and exchange energy; during this process, it is observed that
some ``prevailing'' breathers (the ones with the larger energy) seem to ``absorb'' the energy of the breathers that they interact with.
This process continues for a long time
leading to a gradual coarsening of the chain dynamics. 
Such processes have been debated
at considerable length for Hamiltonian systems (with and without
$\ell^2$ norm conservation properties); a recent example of such a discussion
containing an account of earlier work can be found in~\cite{iubini}.
In our long time dynamics,
one can clearly identify three strong ``eventual''
breathers; these are located far enough from each other, so that no interactions between them are observed (at least until the end of the simulation, at $t=10,000$).
A similar effect
has also been observed in Ref.~\cite{nln} (see Fig.~7 of this work and
also references
therein).

The relevant evolution also brings forth the potential for connections
of case examples such as that of Fig.~\ref{fig4b} with chaotic dynamics.
In particular, what we can see here, but also in  Figs.~2 and~7 of~\cite{nln},
is that the MI manifestation leads not just to a simple 
unstable wavenumber, but rather to a band of unstable wavenumbers. The 
dominant
one among them leads transiently to the formation of a structure resembling 
a breather
lattice \cite{avadh}, 
but as shown in both of the above examples, as well as in Figs. such
as Fig. 3 of~\cite{avadh}, such states are unstable. Thus, the excitation of 
all the additional
unstable modes eventually through an apparently chaotic evolution mixing these
modes, leads to a self-organizing end result whereby a few dominant/robust 
discrete breather
structures finally prevail.

\begin{figure}
\begin{center}
    \begin{tabular}{cc}
    \includegraphics[scale=0.5]{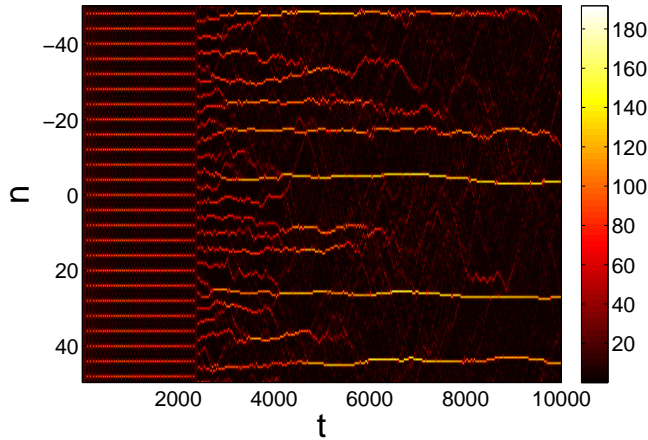}
\end{tabular}
\caption{Contour plot, showing the long time evolution of the energy density $\mathcal{H}_n$, for the same initial condition as in the middle panel of Fig.~\ref{fig4a}. Up to $t\sim 2000$, the chain follows a periodic (breather
lattice) pattern due to
the exponential growth of the initially excited mode. Beyond that point, we observe the creation of small breathers which collide and
exchange energy. At the end of the simulation ($t=10,000$),
the total energy density is chiefly
distributed among only $3$ distinct localized breathers.
}
\label{fig4b}
\end{center}
\end{figure}

%
\begin{table}[b!]
\begin{tabular}{|c|c|c|c|c|c|c|}
\hline
\multicolumn{7}{|c|}{$\Phi_{0,\mathrm{crit}}=1.11$}\\
\hline  K & $20$ & $50$ & $100$ &$200$ &$400$ &$1000$
\\
\hline $\phi_0$ & $0.98$ & $0.98$ & $0.90$ & $0.90$ & $0.90$ & $0.90$
\\
\hline
\end{tabular}
\caption{The numerically found critical amplitude $\phi_0$ for escape dynamics for various numbers $K$ of lattice sites. Here, $\Phi_{0,\mathrm{crit}}=1.11$ is the corresponding $K$-independent analytical prediction.}
\label{table1}
\end{table}

Let us now discuss the connection of the modulational instability mechanism with the emergence of escape dynamics.
Generally, modulational instability
gives rise to an exponential growth of the plane wave perturbations and, as a result,
a large-amplitude localized mode --composed by one or
few oscillators-- may be formed.
However, the modulational instability mechanism is not enough by itself to ensure escape dynamics, and modulationally unstable states corresponding to localized, even chaotic excitations, may exist globally in time.
Our numerical simulations have shown that there exists a threshold of the amplitude $\phi_0$
separating the modulational instability regime from the escape dynamics regime;
this threshold was found to be (for the set of parameters used in Fig.~\ref{fig4a}) $\phi_0=0.98$.
The right panel of Fig.~\ref{fig4a} depicts
the evolution of (\ref{mif1})
for $\phi_0=1.25$, i.e., beyond the threshold value.
It is observed that for a short time interval $V(t)$ remains close
to its initial value while, at a later time, a sudden and sharp increase appears; this is also associated with
a sharp increase of the maximum amplitude along the chain, leading to the escape from the potential well
 of the critical unit possessing this amplitude. The inset in the right panel of Fig.~\ref{fig4a} shows the evolution of
$\max_n|U_n(t)|$ crossing the saddle point barrier.

It is interesting to observe, that
the analytical value $\Phi_{0,\mathrm{crit}}$ [cf. Eq.~(\ref{mi28})] appears to
yield a value proximal to the amplitude threshold separating the modulational
 instability from the escape dynamics regime. We have confirmed numerically that an initial condition of
the form of a randomly perturbed plane wave, namely,
\begin{eqnarray}
\label{mif3}
U_n(0)=2\Phi_{0,\mathrm{crit}}+\tilde{\epsilon}\cdot\mathrm{noise},\;\;\tilde{\epsilon}=10^{-4},
\end{eqnarray}
{\em
%
leads to escape dynamics and blow-up in finite time}, and the evolution of $V(t)$
is similar to the one shown in the right panel of Fig.~\ref{fig4a}.
Additionally, it should be noted that our simulations have also confirmed the fact that $\Phi_{0,\mathrm{crit}}$ is almost independent of the number $K$ of lattice sites, as seen in Table~\ref{table1}: it is clearly observed that the numerically found critical amplitude takes the constant value $\phi_{0} = 0.90$ for chains composed by more than $K=100$ sites (and up to $K=1000$), while it takes a slightly larger value for smaller chains.

{\it The linearly damped system: Transient modulational instability within the exponential decay of solutions.} In the case of the linearly damped system (\ref{eq1dh}), seeking for solutions (\ref{mi3}) under the slow-variation in time approximation, we derive the linearly damped analogue of the DNLS (\ref{mi5}), which is
\begin{eqnarray}
\label{mi5diss}
i\omega_d\dot{\phi_n}+i\omega_d\frac{\gamma}{2}\phi_n+\epsilon(\phi_{n+1}
-2\phi_n+\phi_{n-1})+\frac{3}{2}\beta\omega_d^2|\phi_n|^2\phi_n=0.
\end{eqnarray}
Multiplying (\ref{mi5diss}) by  $\bar{\phi}_n$ and summing over $n$ (in the infinite lattice or in the finite lattice subject to Dirichlet or periodic boundary conditions), we derive the equation
\begin{eqnarray*}
\frac{1}{2}\frac{d}{dt}||\phi||^2_{\ell^2}+\frac{\gamma}{2}||\phi||^2_{\ell^2}=0,
\end{eqnarray*}
implying the decay of the $\ell^2$-norm
\begin{eqnarray}
\label{mi6diss}
||\phi(t)||^2_{\ell^2}= e^{-\gamma t}||\phi(0)||^2_{\ell^2},
\end{eqnarray}
for all solutions of the approximating DNLS equation (\ref{mi5diss}). Furthermore, (\ref{mi6diss}) and the norm relations  (\ref{expl1}) and (\ref{expl1a}) imply for both the cases of the infinite or the finite lattice, the sup-norm decay estimate
\begin{eqnarray}
\label{mi7diss}
||\phi(t)||^2_{\infty}\leq e^{-\gamma t}||\phi(0)||^2_{\ell^2}.
\end{eqnarray}
Hence $\lim_{t\rightarrow\infty}||\phi(t)||_{\infty}=0$, and all solutions of (\ref{mi5diss}) decay at an exponential rate, independently of the initial data. However, the spatially uniform decay estimate (\ref{mi7diss}) does not exclude a possible transient growth of  perturbations, and consequently, a transient modulational instability behavior within the frame of exponential decay. This transient modulational instability is especially expected to be noticeable
for small values of damping.

The above suggestions have been tested numerically, by considering the evolution of the damped DKG chain (\ref{eq1dh}) with perturbed plane wave initial data (\ref{mpwic}). The values of parameters for $h$, $\beta$, $\omega_d^2$ and $Q$, are the same as in the numerical study performed for the Hamiltonian system $\gamma=0$. The amplitude is $\phi_0=0.15$, inducing modulation instability in the Hamiltonian case. The two top panels of Fig.~\ref{fig4bD} show the contour plots of the energy density $\mathcal{H}_n$ of the dissipative system for $\gamma=2$ (left panel) and $\gamma=0.02$ (right panel), respectively. Although in both cases the solutions decay
to zero as the approximating DNLS estimate (\ref{mi7diss}) suggests, in the right panel we observe the formation of a transient breather lattice, which
survives the exponential decay up to $t\simeq 100$. The same breather lattice lives only up to $t\simeq 1$, as shown in the top left 
panel, due to the stronger damping $\gamma=2$. The two bottom panels show the contour plot of the difference
$\mathcal{H}_n\left[U_n |_{\gamma>0}\right]-\mathcal{H}_n\left[U_n |_{\gamma=0}\exp(-\gamma t)\right]$, i.e.,
of the energy density $\mathcal{H}_n\left[U_n |_{\gamma>0}\right]$
for the damped system (\ref{eq1dh}) from the energy density $\mathcal{H}_n\left[U_n |_{\gamma=0}\exp(-\gamma t)\right]$, where $U_n |_{\gamma=0}$ denotes the solution of the Hamiltonian system (\ref{eq1h}).
In the bottom right panel the time evolution of the energy density difference is shown for the damping value $\gamma=0.02$.  The peaks observed therein, within the time interval $[20,150]$, illustrate a transient modulation instability regime, where localization of energy and formation of a lattice
of localized modes occurs.
In the case of stronger damping $\gamma=2$, the transient instability regime is hardly observable.
Numerical simulations for smaller values of $\gamma$
depict that the transient modulation instability regime is increased,
as it is expected when the Hamiltonian limit $\gamma\rightarrow 0$ is approached.


\begin{figure}
\begin{center}
    \begin{tabular}{cc}
    \includegraphics[scale=0.5]{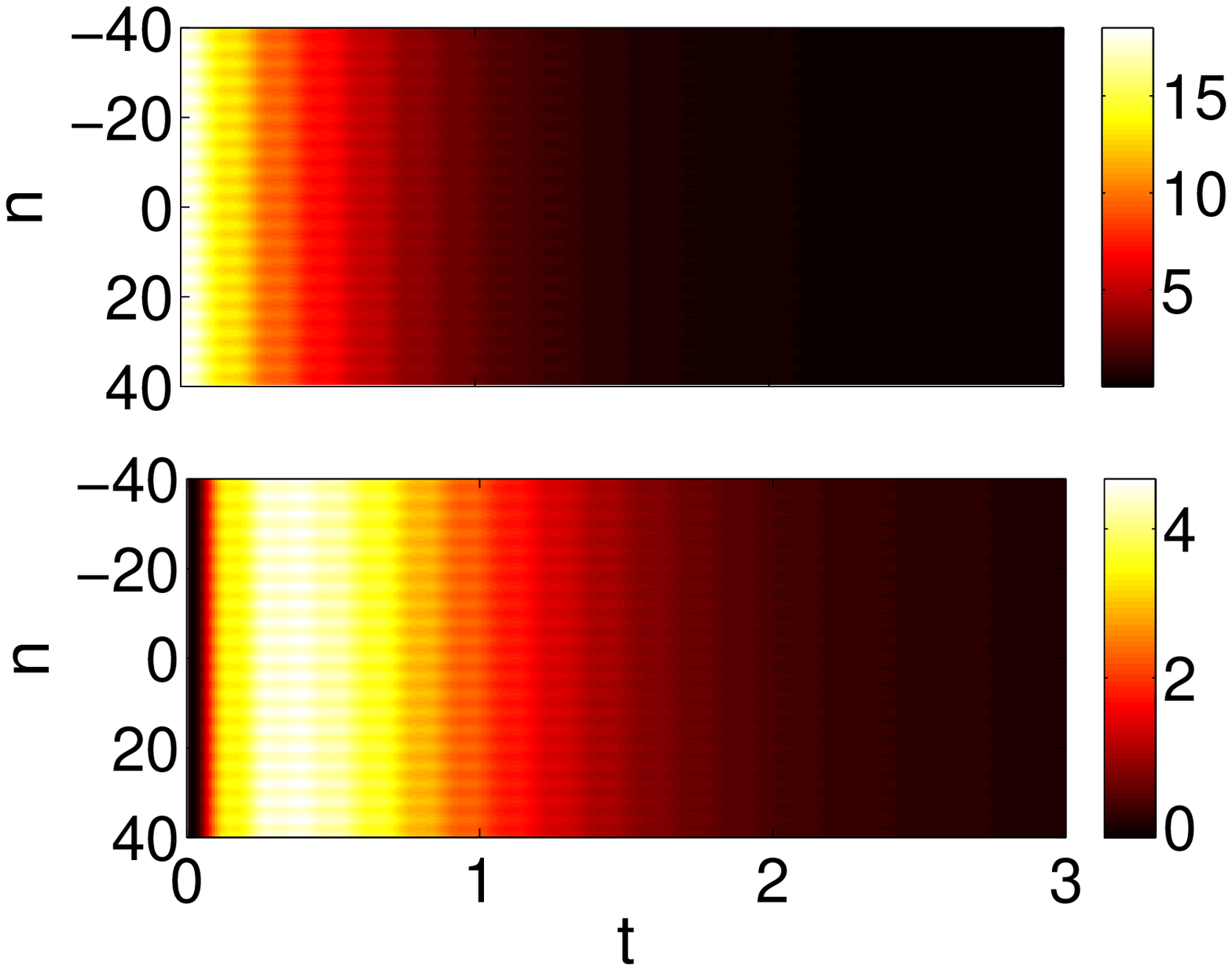}&
    \includegraphics[scale=0.52]{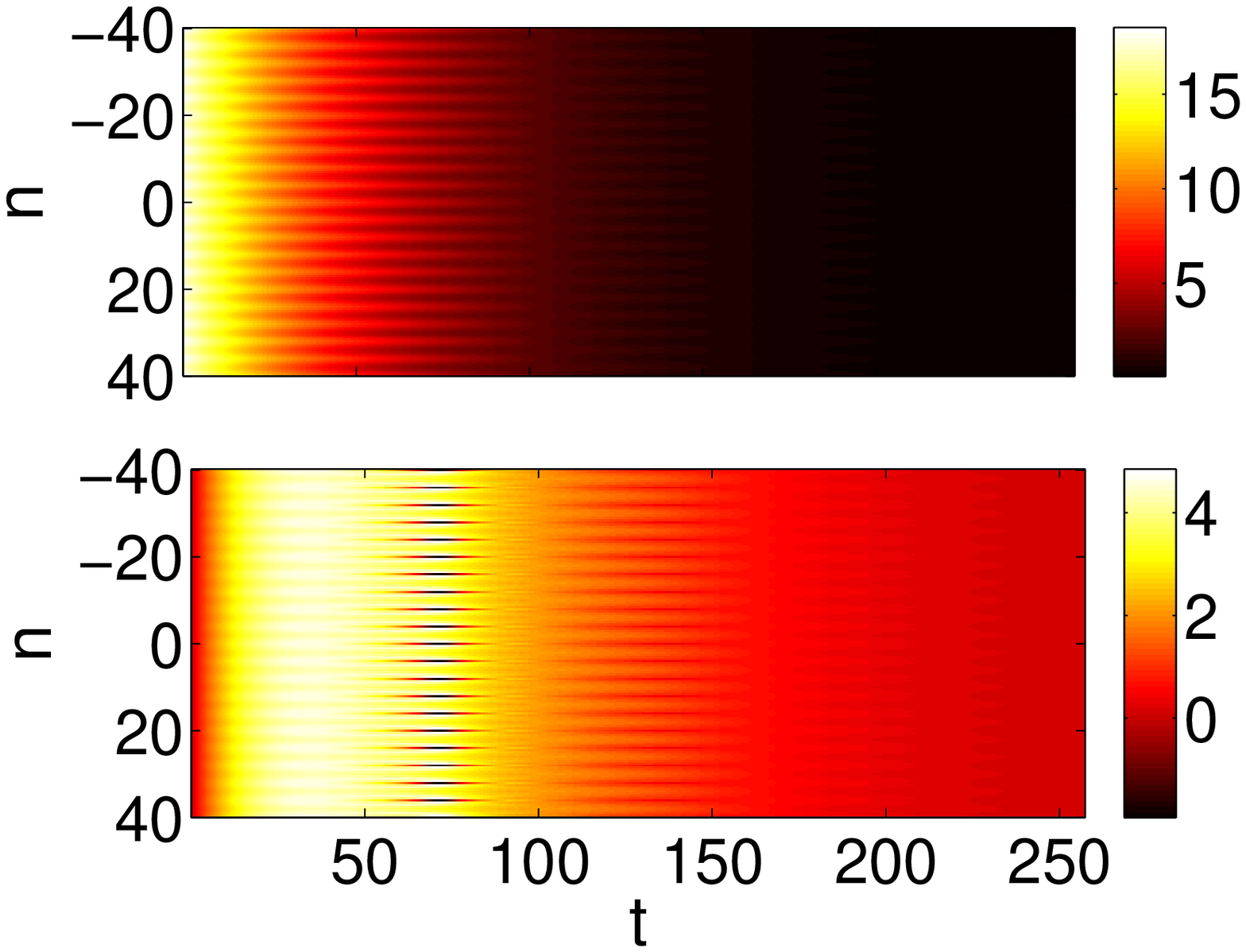}
\end{tabular}
\caption{The top panels show the long time evolution of the energy density $\mathcal{H}_n$ for the linear damped lattice $\gamma>0$. The bottom panels show the long time evolution of the difference
$\mathcal{H}_n\left[U_n |_{\gamma>0}\right]-\mathcal{H}_n\left[U_n |_{\gamma=0}\exp(-\gamma t)\right]$, i.e.,
of the energy density $\mathcal{H}_n\left[U_n |_{\gamma>0}\right]$
for the damped lattice from the
energy density $\mathcal{H}_n\left[U_n |_{\gamma=0}\exp(-\gamma t)\right]$.
The left panels are for $\gamma=2$ and the right panels for $\gamma=0.02$.
The initial condition is:
$U_n(0)=2\phi_0+\hat{\epsilon}\cos(Qn)$, with
$\hat{\epsilon}=0.005$, $Q=\frac{\pi}{2}$ and $\phi_0=0.15$. A
transient modulational instability regime within the time interval
$[20,150]$ is evident in the bottom right panel. }
\label{fig4bD}
\end{center}
\end{figure}

\section{Discussion and conclusions}

In this work, we have studied the escape problem in the discrete
repulsive $\phi^4$-model.
We have considered both the Hamiltonian ($\gamma=0$) and the dissipative ($\gamma >0$) variants of the model. We have
assumed different types of initial conditions, ranging from single-excited
units and small segments to plane waves. Our results address several relevant points related to the problem, as is briefly described below.

In the Hamiltonian case, we have discussed how the system falls within the
class of abstract second order evolution equations of divergent structure.
Such systems can be treated, with respect to global non-existence, by the
energy  methods of Ref.~\cite{GP}, involving the formulation of suitable
differential inequalities enabling the identification of
conditions for blow-up. For initial data
in the form of a single-unit excitation (all other units being initially located in the minimum $U_{\mathrm{min}}=0$), it is intuitively expected that escape occurs if the total energy of this unit is (sufficiently)
greater than the energy barrier $E_{\mathrm{max}}$, i.e., the initial
position of the unit is located outside the
potential well.
Indeed, we have derived a suitable
analytical value for the initial position of the single unit outside the
well, which serves as a sufficient condition for the absence
of global existence.
However, these single-site initial data are very relevant for two
important questions related to escape dynamics
(already posed in Refs.~\cite{Dirk1,Dirk2}):
is it possible that the neighbors coupled to the escaping unit also get
driven over the barrier and give rise to a concerted escape of the
entire chain~? As an alternative scenario,
can this escaping unit even be pulled back into
the domain of attraction by the restoring coupling forces~? We have
definitively shown that both the drive over and the pull back
effects can be realized in numerical experiments of such chains.

At this point, it should be remarked that the autonomy of the system allows to discuss the derived conditions in terms of the fate of the unit of the critical mode which, at a specific time, escapes from the well. In other words, time-invariance allows for posing as initial condition the position and momenta of the chain at an escaping snapshot, and consider as an initial time, any time referred to this barrier-crossing snapshot.

The numerical simulations performed in Section \ref{IIA} justified that the
analytical conditions capture the qualitative behavior of the chain and its dependence on the strength of the binding forces. Blow-up occurs when the
analytical prediction is satisfied and a concerted escape follows when we are
in the regime where the role of the coupling
is significant. On the contrary, if the analytical condition is not satisfied
then for suitable initial conditions, the initially out-of-well
unit may be pulled back by the neighbors inside the stability domain,
leading the initially escaping ``spike-like'' mode to give rise to
a waveform which will exist globally in time. Furthermore, the numerical
simulations revealed the existence of a ``true'' threshold for the position
of the excited unit outside of the well, acting as a separatrix between
 the above two distinct dynamical behaviors. For this numerical threshold,
the $\omega_d^2$-dependent analytical prediction serves as an upper bound,
with a justifiable error in the regime ranging from moderate to very
strong discreteness. Both theory and numerics suggest that the threshold
position moves far from the position of the saddles in the continuum limit.

In the same setting, we have also examined the global existence result of \cite[Lemma 2.1, pg. 455]{GP}, addressing the following question. Assuming negative
time derivative of the $\ell^2$ norm, while keeping the non-positivity
condition of the initial Hamiltonian energy, does global existence arise
for the solutions~? The answer we found is that
the system
under consideration [cf. Eqs.~(\ref{eq1})-(\ref{eq2})] serves as a
counter-example for the aforementioned global existence result, a feature
confirmed by our numerical simulations.

We have also studied the linearly damped system [cf. Eq.~(\ref{eq1}) with $\gamma>0$], for which the questions posed on the effect of the strength of the binding forces and their interplay with nonlinearity are additionally complicated
by the role of
damping. In that regard, we have tried to answer the
following question: can the appearance of damping
affect the position threshold which separates collapse from global existence~? 
To examine the above question, we have applied on (\ref{eq1}) suitable modifications of the energy methods
developed in Ref.~\cite{PS98}.
These methods have enabled us to provide a $\gamma$-independent prediction
for the position threshold of the single excited unit, and the numerical
experiments of section \ref{S2A} have verified this independence.
Theoretical and numerical evidence has also been provided for the
improvement that this analytical estimate yields in connection
to the true threshold.

We have also examined the escape phenomenon for multi-site excitations
inside the respective wells, trying to answer the following questions: is it
possible to describe  escape scenarios for such initial excitations~? Also,
is it possible to derive quantitative  conditions for these
excitations (e.g., initial positions, initial amplitudes) which are sufficient
for escape~? These questions were answered in
the positive by using suitable modifications of our energy arguments.
Finally, for plane wave excitations, relevant answers
were provided by the analysis of the modulational instability mechanism.

Motivated by the fact that escape is a phenomenon characterized by the concentration of energy within confined segments of the chain, in Section \ref{S5A} we have examined the evolution of a lattice segment of unit length. With an appropriate cut-off argument, we have derived a condition on the initial ``energy'' of the segment (in terms of the $\ell^4$-norm). Although this condition formally is not establishing collapse by itself, heuristic arguments based on the violation of the invariance principles of potential wells \cite{PS73,PSrev}, indicate that it is relevant for  initiating escape in the discrete regime.
The numerical results validated the theoretical expectations and enabled us to
recover the parametric regimes on which the escape process may be observed
for our finite segment of excited initial data.

Finally, we have examined escape, in terms of the instabilities of
plane wave initial conditions (Section IV). In that regard, we have reviewed
the derivation of the modulational instability condition for the amplitude
of plane waves~\cite{KP92}, based on the DNLS approximation of the
slowly-varying envelope solutions. On the other hand, as explained above,
analytical energy arguments and direct numerical simulations revealed the existence of three different
regimes for the plane wave amplitudes: modulational stability, modulational
instability without escape, and modulational instability accompanied by
escape. For the linearly damped DKG chain, the numerical results verified a transient modulation instability regime. The instabilities will be completely 
suppressed ultimately, 
in accordance with the predictions of the damped DNLS approximation.

It is crucial to remark that modulational instability analysis structurally {\em cannot} be used for the examination of the self-organized escape dynamics for the DKG chain;
the latter, is associated with blow-up in finite time (equivalent to the escape time when the initial data are in the potential well), while the modulational instability analysis relies on an approximation based on the use of
the DNLS equation (\ref{mi5}); for
the latter, solutions {always exist globally in time}, independently of the initial data and the strength of the nonlinear term \cite{KY1,Wein99}. This is a vast difference between the DNLS and DKG systems. Regarding the collapse behavior, the  DKG demonstrates some analogies with its continuous counterpart, 
i.e., the nonlinear KG partial differential equation (PDE), 
which also may exhibit collapse depending on the size of the initial data and the strength of nonlinearity. Such analogies between DNLS and NLS equations are limited only to one spatial dimension and restrictively to the case of the cubic nonlinearity; in the case of focusing, power-type nonlinearities $f(z)=|z|^{2p}z$, solutions of the one-dimensional NLS PDE may blow-up in finite time when $p\geq 2$, depending on the size and type of initial data \cite{cazh}, even in the linearly damped case \cite{tsu}. Furthermore, in view of the modulation stability analysis, it should be recalled that the DNLS approximation drastically modifies the results concerning the stability domains, which may be deduced from a continuous NLS PDE approximation. In particular, it has been illustrated in \cite{KP92}, that a stability analysis based on the continuous NLS approximation (the analogue of Eq.~(\ref{mi16}) for $q,Q <<1$) may fail, by erroneously predicting stability in regions of modulation instability which are correctly detected by the DNLS analogue. Modulation instability effects appear in the NLS PDE equation with dissipation as it was shown, e.g., in Refs.~\cite{RKDM,segur}; depending on the type and the size of the damping, modulation instability effects may be considerable. 

The results presented in this work may pave the way for future work in many interesting directions. Below, we briefly present a few such examples.
\begin{itemize}
\item A natural possibility is to extend the present
considerations to higher dimensional settings, where the role of the
coupling will be more significant due to the geometry enforcing additional
neighbors. 
\item It is also relevant to consider  via the methods and diagnostics
presented herein the phenomenology of different types of potentials
including e.g., the Morse potential, which is relevant to DNA denaturation
as modeled by the Peyrard-Bishop model~\cite{PB}. We speculate that
the techniques used herein may turn out to be more broadly relevant to 
problems involving hydrogen-bonds and, more generally, aspects
of molecular dissociation. 
\item While the present study has focused on cubic nonlinearities,
it might also be of interest to consider quintic nonlinearities
in the one-dimensional problem, or cubic ones in the two-dimensional
case. In the present case,
the DKG model is structurally closer to its continuum analog 
(regarding collapse properties) 
or to the DNLS (regarding MI features-- 
but the  DNLS model does not have collapse due to the conservation
of the $\ell^2$ norm). However, for the NLS equation in the continuum case,
it is well-known~\cite{susu} that a quintic nonlinearity is the threshold
for collapsing dynamics in 1d, while the cubic one is the corresponding
threshold for 2d. Hence, for such models comparing/contrasting the collapse
features of the lattice (DKG) model with the continuum (NLS) PDE 
problem would be
of interest in its own right. 
\item Finally, it is certainly
relevant, following similar  lines of approach as the work
of~\cite{Dirk1,Dirk2} to attempt to quantify the effects of noise
in these systems and generalize in a probabilistic way the
deterministic statements about escape presented herein for
noise realizations with different correlation properties.
\end{itemize}
Such studies are currently in progress and will be presented
in future publications.


\section*{Acknowledgments}

J.C., B.S.R and A.\'A. acknowledge financial support from the MICINN project FIS2008-04848.
The work of D.J.F. was partially supported by the Special Account for Research Grants of the University of Athens.
P.G.K. gratefully acknowledges support from the U.S.
National Science Foundation via grants NSF-DMS-0806762 and
NSF-CMMI-1000337, from the U.S. Air Force via award
FA9550-12-1-0332, as well as from the Alexander von Humboldt
Foundation, the Alexander S. Onassis Public Benefit Foundation
via grant RZG 003/2010-2011 and the Binational Science
Foundation via grant 2010239.
\section*{APPENDICES: Proofs of global non-existence and small data global existence results for the system (\ref{eq1})}
\setcounter{equation}{0}
In this complementary section, we provide the proofs of the analytical results concerning the global non-existence conditions, as well as the small data global existence conditions for the system (\ref{eq1}).

We start our considerations from the local in time existence of solutions.
Let us recall from \cite[pg.~468]{K2} that (\ref{eq1})-(\ref{eq2}) can be formulated as an abstract evolution equation in $\ell^2\times\ell^2$ (see \cite{Ball1b,cazh} for the continuous analogue). For instance, we may set $\mathbf{\hat{A}}:=\Delta_2-\omega_d^2 $, and check that the operator
\begin{equation}
\label{matrixA}
\mathbf{B}:=\left[
\begin{array}{cc}
0&1\\
\mathbf{\hat{A}}&0
\end{array}
\right],\;\;D(\mathbf{B}):=\left\{\left[
\begin{array}{l}
z\\
\omega
\end{array}
\right]\,:\,z,\omega\in \ell^2,\,\mathbf{\hat{A}}z\in\ell^2 \right\},
\end{equation}
is a skew-adjoint operator, since
$$
\mathbf{B}^*:=-\left[
\begin{array}{cc}
0&1\\
\mathbf{\hat{A}}&0
\end{array}
\right],\;\;D(\mathbf{B}^*):=\left\{\left[
\begin{array}{l}
\chi\\
\zeta
\end{array}
\right]\,:\,\chi,\zeta\in \ell^2,\,\mathbf{\hat{A}}\chi\in\ell^2 \right\},
$$
and $D(\mathbf{B})=D(\mathbf{B}^*)$.
The operator $\mathbf{B}$, is the generator of an isometry group $\mathcal{T}(t):\mathbb{R}\rightarrow\mathcal{L}(\ell^2\times\ell^2)$ --the space of linear and bounded operators of $\ell^2\times\ell^2$. Setting
$$
\mathbf{U}=\left[
\begin{array}{c}
U\\
\dot{U}
\end{array}
\right],\;\;\mathbf{F}(\mathbf{U})=\left[
\begin{array}{c}
0\\
\beta \omega_d^2U^3
\end{array}
\right],
$$
equation (\ref{eq1}) can be rewritten as
\begin{eqnarray}
\label{system}
\dot{\mathbf{U}}=\mathbf{B}\mathbf{U}+\mathbf{F}(\mathbf{U}).
\end{eqnarray}
In the above set-up, fixing $T>0$ and considering initial data $\mathbf{U}(0)=[U(0),\dot{U}(0)]^{\mathrm{T}}\in\ell^2\times\ell^2$, a function $\mathbf{U}(t):=[U(t),\dot{U}(t)]^{\mathrm{T}}\in\mathrm{C}([0,T],\ell^2\times\ell^2)$ is a solution of (\ref{system}), if and only if
\begin{eqnarray}
\label{milds}
\mathbf{U}(t)=\mathcal{T}(t)\mathbf{U}(0)+\int_{0}^{t}\mathcal{T}(t-s)\mathbf{F}(\mathbf{U}(s))ds:=S(t)\mathbf{U}(0).
\end{eqnarray}
The nonlinear term defines a locally Lipschitz map $\mathbf{F}:\ell^2\times\ell^2\rightarrow\ell^2\times\ell^2$ and
Eq.~(\ref{milds}) can be treated exactly as in \cite[Theorem 2.1, pg.~94]{KY1} in order to prove the following.
\setcounter{theorem}{0}
\begin{theorem}
\label{le}
(a)\ For all $\mathbf{U}(0)\in\ell^2\times\ell^2$, there exists $T_{\mathrm{max}}(\mathbf{U}(0))>0$  and a function $\mathbf{U}(t)\in\mathrm{C}([0,T_{\mathrm{max}}),\ell^2\times\ell^2)$ which is
for all $0<T<T_{\mathrm{max}}$, the unique solution of (\ref{system}) in $\mathrm{C}([0,T],\ell^2\times\ell^2)$ (well posedness).\vspace{.2cm}\\
(b)\  The following alternatives exist: (i) $T_{\mathrm{max}}=\infty$, or (ii) $T_{\mathrm{max}}<\infty$ and $$\lim_{T\uparrow T_{\mathrm{max}}}||\mathbf{U}(t)||_{\ell^2\times\ell^2}=\infty,\;\;\mbox{(maximality)}.$$
(c)\ The unique solution depends continuously on the initial data: If $\{\mathbf{U}_{n}(0)\}_{n\in\mathbb{N}}$ is a sequence in $\ell^2\times\ell^2$ such that $\mathbf{U}_{n}(0)\rightarrow \mathbf{U}(0)$ and if $T<T_{\mathrm{max}}$,
then $S(t)\mathbf{U}_n(0)\rightarrow S(t)\mathbf{U}(0)$ in $\mathrm{C}([0,T],\ell^2\times\ell^2)$.
\end{theorem}
\label{S1}
Let us remark that for the local existence of solutions for (\ref{eq1d})-(\ref{eq2d}) we have the same result of Theorem \ref{le}. For its proof we just have to replace the matrix operator (\ref{matrixA}) with the matrix operator:
\begin{equation}
\label{matrixAg}
\mathbf{B}_{\gamma}:=\left[
\begin{array}{cc}
0&1\\
\mathbf{\hat{A}}&-\gamma
\end{array}
\right],\;\;D(\mathbf{B}):=\left\{\left[
\begin{array}{l}
z\\
\omega
\end{array}
\right]\,:\,z,\omega\in \ell^2,\,\mathbf{\hat{A}}z\in\ell^2 \right\}.
\end{equation}
\section*{Appendix A: Proof of Theorem \ref{GP}}
\label{AppendixA}
We start by observing that the Hamiltonian can be written
in the Lagrangian form
\begin{eqnarray}
\label{hamp2}
\mathcal{H}(t)=\frac{1}{2}||\dot{U}||_{\ell^2}^2-V(U),
\end{eqnarray}
where
\begin{eqnarray}
\label{hamp3}
V(U):= -\frac{1}{2}\sum_{n=-\infty}^{+\infty}(U_{n+1}-U_n)^2-\frac{\omega_d^2}{2}\sum_{n=-\infty}^{+\infty}U_n^2+\frac{\beta\omega_d^2}{4}\sum_{n=-\infty}^{+\infty} U_n^4.
\end{eqnarray}
The functional $V:\ell^2\rightarrow\mathbb{R}$ is differentiable (see \cite[Lemma 2.3, pg. 121]{K1}), and for the derivative $V'(U)$ at $U\in\ell^2$, it holds that
\begin{eqnarray}
\label{hamp5}
(V'(U), Y)_{\ell^2}=-\sum_{n=-\infty}^{+\infty}(U_{n+1}-U_n)(Y_{n+1}-Y_n)-\omega_d^2\sum_{n=-\infty}^{+\infty}U_nY_n+\beta\omega_d^2\sum_{n=-\infty}^{+\infty}U_n^3Y_n,\;\;\mbox{for all}\;\;W\in\ell^2.
\end{eqnarray}
Furthermore, by setting $Y=U$ in (\ref{hamp5}), we get that
\begin{eqnarray}
\label{hamp5a}
(V'(U),U)_{\ell^2}=-\sum_{n=-\infty}^{+\infty}(U_{n+1}-U_n)^2-\omega_d^2\sum_{n=-\infty}^{+\infty}U_n^2+\beta\omega_d^2\sum_{n=-\infty}^{+\infty}U_n^4.
\end{eqnarray}
With these preparations we proceed in various steps.
\newline
\emph{Step 1:} {\em The inequality}
\begin{eqnarray}
\label{hamp6}
(V'(U),U)-4V(U)\geq 0,\;\;\mbox{for all}\;\;U\in\ell^2,
\end{eqnarray}
{\em holds.} Indeed, we see from (\ref{hamp3}) and (\ref{hamp5}), that
\begin{eqnarray}
\label{hamp7}
(V'(U),U)_{\ell^2}-4V(U)=\sum_{n=-\infty}^{+\infty}(U_{n+1}-U_n)^2+\omega_d^2\sum_{n=-\infty}^{+\infty}U_n^2\geq 0.
\end{eqnarray}
Thus the claim (\ref{hamp6}) is proved.
\newline
\emph{Step 2:} {\em The function $x(t)=||U(t)||^2_{\ell^2}$ satisfies the relations}
\begin{eqnarray}
\label{hamp8}
\dot{x}(t)&=&2(U(t),\dot{U}(t))_{\ell^2},\\
\label{hamp9}
\ddot{x}(t)&\geq &2||\dot{U}||_{\ell^2}^2+8V(U).
\end{eqnarray}
Equation (\ref{hamp8}) follows by direct differentiation of $x(t)=||U(t)||^2_{\ell^2}=\sum_{n=-\infty}^{+\infty}U_n(t)U_n(t)$. Note that all differentiations in the proof are justified by the local existence theorem of solutions in the interval $[0,T_{\mathrm{max}})$ of local existence.

For the inequality (\ref{hamp9}) we first differentiate (\ref{hamp8}) with respect to time and we substitute the expression $\ddot{U}$ from the system (\ref{eq1}), which will be written for brevity as $\ddot{U}=\Delta_2U-\omega_d^2U+\beta\omega_d^2U^3$.
Indeed, we have
\begin{eqnarray}
\label{hamp9aa}
\ddot{x}(t)&=&2(\dot{U},\dot{U})_{\ell^2}+2(U,\ddot{U})_{\ell^2}\nonumber\\
&=&2||\dot{U}||_{\ell^2}^2+2(U,\Delta_2U-\omega_d^2U+\beta\omega_d^2U^3)_{\ell^2}\nonumber\\
&=&2||\dot{U}||_{\ell^2}^2+2(V'(U),U)_{\ell^2}.
\end{eqnarray}
Next, by using the inequality (\ref{hamp6}) of Step 1, which implies that $(V'(U),U)_{\ell^2}\geq 4V(U)$, we estimate the second term of the right~(\ref{hamp9aa}) as
\begin{eqnarray*}
\ddot{x}(t)\geq 2||\dot{U}||_{\ell^2}^2+2(V'(U),U)_{\ell^2}\geq 2||\dot{U}||_{\ell^2}^2+8V(U).
\end{eqnarray*}
Hence, the inequality (\ref{hamp9}) is proved.
\newline
\emph{Step 3:} {\em Assume that the initial Hamiltonian
is $H(0)\leq 0$. Then it holds that}
\begin{eqnarray}
\label{hamp10}
\ddot{x}(t)\geq 6||\dot{U}||^2_{\ell^2}\geq 0.
\end{eqnarray}
To prove (\ref{hamp10}), we observe first that the equivalent expression (\ref{hamp2}) for the Hamiltonian implies that
\begin{eqnarray}
\label{hamp11}
V(U)=\frac{1}{2}||\dot{U}||^2_{\ell^2}-\mathcal{H}(t).
\end{eqnarray}
Due to the conservation of the Hamiltonian
$\mathcal{H}(t)=\mathcal{H}(0)$, (\ref{hamp11}) can be rewritten as
\begin{eqnarray}
\label{hamp12}
V(U)=\frac{1}{2}||\dot{U}||^2_{\ell^2}-\mathcal{H}(0).
\end{eqnarray}
By replacing the term $V(U)$ of the inequality (\ref{hamp9}) proved in Step 2 with the right-hand side of (\ref{hamp12}), we have that
\begin{eqnarray}
\label{hamp13}
\ddot{x}(t)\geq 2||\dot{U}||_{\ell^2}^2+8\left(\frac{1}{2}||\dot{U}||^2_{\ell^2}-\mathcal{H}(0)\right).
\end{eqnarray}
\newline
Since $\mathcal{H}(0)\leq 0$ by assumption,  from  (\ref{hamp13}) we derive the inequality
\begin{eqnarray*}
\ddot{x}(t)\geq 2||\dot{U}||_{\ell^2}^2+8\left(\frac{1}{2}||\dot{U}||^2_{\ell^2}\right)=6||\dot{U}||^2_{\ell^2}\geq 0.
\end{eqnarray*}
Hence (\ref{hamp10}) is proved.
\newline
\emph{Step 4:} Under the assumption $\mathcal{H}(0)\leq 0$, {\em the norm function $x(t)=||U(t)||_{\ell^2}^2$ satisfies the differential inequality}
\begin{eqnarray}
\label{hamp15}
\ddot{x}(t)\geq\frac{3}{2}\frac{[\dot{x}(t)]^2}{x(t)}.
\end{eqnarray}
\newline
To prove (\ref{hamp15}), we observe first that the Cauchy-Schwarz inequality implies the estimate
\begin{eqnarray*}
[\dot{x}(t)]^2=[2(U,\dot{U})_{\ell^2}]^2\leq 4||U||^2_{\ell^2}||\dot{U}||^2_{\ell^2}=4x(t) ||\dot{U}||^2_{\ell^2},
\end{eqnarray*}
which can be written as
\begin{eqnarray}
\label{hamp14}
||\dot{U}||^2_{\ell^2}\geq\frac{[\dot{x}(t)]^2}{4x(t)}.
\end{eqnarray}
Then, from a combination of (\ref{hamp10}) and (\ref{hamp14}), the inequality (\ref{hamp15}) readily follows.
\newline
\emph{Step 5 (finite-in time-singularity):} In this final step for the proof of the theorem we shall use together with the assumption for negative initial Hamiltonian
$\mathcal{H}(0)\leq 0$,  {\em the additional assumption $\dot{x}(0)=2(U(0),\dot{U}(0))_{\ell^2}>0$ (positive time derivative/increase of the $\ell^2$ norm
initially). Under these assumptions, $\dot{x}(t)>0$ in the interval $[0, T_{\mathrm{max}})$ of existence and  $x(t)=||U(t)||_{\ell^2}^2$ is strictly increasing in the interval of existence.}

Indeed, from the inequality (\ref{hamp10}) it follows that the function $\dot{x}(t)$ is increasing in the interval $[0,T]$, $T<T_{\mathrm{max}}$. Since we have assumed $\dot{x}(0)=2(U(0),\dot{U}(0)>0$ and $\dot{x}(t)$ is increasing, it readily follows  that $\dot{x}(t)>0$, on $[0,T]$. Therefore, $x(t)$ is strictly increasing on $[0,T]$.
The inequality (\ref{hamp15}), can be rewritten as
\begin{eqnarray}
\label{hamp16}
\frac{\ddot{x}(t)}{\dot{x}(t)}\geq\frac{3}{2}\frac{\dot{x}(t)}{x(t)}.
\end{eqnarray}
Note that all the quantities involved are {\em strictly positive} on $[0,T]$ as well as their initial values at $t=0$. Integrating (\ref{hamp16}) on $[0, t]$ for $t\leq T$, we get that
\begin{eqnarray*}
\ln [\dot{x}(t)]-\ln [\dot{x}(0)]\geq\frac{3}{2}\left(\ln[x(t)]-\ln [x(0)]\right),
\end{eqnarray*}
which implies that
\begin{eqnarray*}
\frac{\dot{x}(t)}{x(t)^{\frac{3}{2}}}\geq \frac{\dot{x}(0)}{x(0)^{\frac{3}{2}}}.
\end{eqnarray*}
Integrating on $[0, t]$ for $t\leq T$ once more, we get that
\begin{eqnarray}
\label{hamp17}
-\frac{2}{x(t)^{\frac{1}{2}}}+\frac{2}{x(0)^{\frac{1}{2}}}\geq \frac{\dot{x}(0)}{x(0)^{\frac{3}{2}}}t.
\end{eqnarray}
Solving (\ref{hamp17}) in terms of $x(t)$ we eventually derive the lower estimate
\begin{eqnarray}
\label{hamp18}
x(t)\geq \frac{x(0)^4}{\left(2x(0)^{\frac{3}{2}}-x(0)^{\frac{1}{2}}\dot{x}(0)t\right)^2}.
\end{eqnarray}
The right-hand side diverges (blows-up) at finite time
\begin{eqnarray*}
T_{\mathrm{b}}:=2\frac{x(0)}{\dot{x}(0)}=2\frac{||U(0)||^2}{(U(0), \dot{U}(0))_{\ell^2}},
\end{eqnarray*}
the time given in (\ref{estbl}). Therefore the left-hand side
$x(t)=||U(t)||_{\ell^2}^2$ cannot exist more than $T_{\mathrm{b}}$.
\section*{Appendix B: Proof of Theorem \ref{bud}}
\label{AppendixB}
\emph{Step 1 (Energy identities):} Multiplying (\ref{eq1d}) by $\dot{U}_n$ and summing over $\mathbb{Z}$ (and by parts in the coupling term), we derive the energy identity
\begin{eqnarray}
\label{eq5d}
\frac{1}{2}\frac{d}{dt}\left\{\frac{1}{2}\sum_{n=-\infty}^{+\infty}\dot{U}_n^2
+\frac{1}{2}\sum_{n=-\infty}^{+\infty}(U_{n+1}-U_n)^2
+\frac{\omega_d^2}{2}\sum_{n=-\infty}^{+\infty}U_n^2-\frac{\beta\omega_d^2}{4}\sum_{n=-\infty}^{+\infty} U_n^4\right\}+\gamma\sum_{n=-\infty}^{+\infty}\dot{U}_n^2=0,
\end{eqnarray}
which, using the definition
of the Hamiltonian
(\ref{hamp}), can be written as
\begin{eqnarray}
\label{eq6d}
\frac{d}{dt}\mathcal{H}(t)+\gamma\sum_{n=-\infty}^{+\infty}\dot{U}_n^2=0.
\end{eqnarray}
Integrating in $[0,t]$ for any $t\in[0,T]$, $T<T_{\mathrm{max}}$ --the maximal interval of existence, we get the energy identity
\begin{eqnarray}
\label{eq7d}
\mathcal{H}(t)+\gamma\int_{0}^t||\dot{U}(s)||^2_{\ell^2}ds=\mathcal{H}(0).
\end{eqnarray}
\newline
\emph{Step 2 (Inequality for the Hamiltonian $\mathcal{H}(t)$):} {\em For every $t\in [0, T]$, the Hamiltonian
satisfies}
\begin{eqnarray}
\label{eq8d}
\mathcal{H}(t)\geq \frac{\omega_d^2}{2}\left(\sum_{n=-\infty}^{+\infty}U_n^4\right)^{\frac{1}{2}}-\frac{\beta\omega_d^2}{4}\sum_{n=-\infty}^{+\infty}U_n^4.
\end{eqnarray}
To derive (\ref{eq8d}), we first observe
that
\begin{eqnarray}
\label{eq9d}
\frac{1}{2}\sum_{n=-\infty}^{+\infty}(U_{n+1}-U_n)^2+\frac{\omega_d^2}{2}\sum_{n=-\infty}^{+\infty}U_n^2\geq \frac{\omega_d^2}{2}\sum_{n=-\infty}^{+\infty}U_n^2.
\end{eqnarray}
Furthermore, by applying the inequality
\begin{eqnarray}
\label{discembA}
\left(\sum_{n=-\infty}^{+\infty}| U_n|^p\right)^{\frac{1}{p}}\leq \left(\sum_{n=-\infty}^{+\infty}| U_n|^q\right)^{\frac{1}{q}},\;\;\mbox{for all}\;\; 1\leq q\leq p\leq\infty,
\end{eqnarray}
in the case $p=4$ and $q=2$,  we get that
\begin{eqnarray}
\label{embp2p4}
\left(\sum_{n=-\infty}^{+\infty}U_n^4\right)^{\frac{1}{4}}\leq \left(\sum_{n=-\infty}^{+\infty}U_n^2\right)^{\frac{1}{2}}.
\end{eqnarray}
Thus, we have
 \begin{eqnarray*}
\left(\sum_{n=-\infty}^{+\infty}U_n^4\right)^{\frac{1}{2}}\leq \sum_{n=-\infty}^{+\infty}U_n^2.
\end{eqnarray*}
With the above inequality at hand, we estimate the right-hand side of (\ref{eq9d}) from below as
\begin{eqnarray}
\label{eq10d}
\frac{1}{2}\sum_{n=-\infty}^{+\infty}(U_{n+1}-U_n)^2+\frac{\omega_d^2}{2}\sum_{n=-\infty}^{+\infty}U_n^2&\geq& \frac{\omega_d^2}{2}\sum_{n=-\infty}^{+\infty}U_n^2.\nonumber\\
&\geq&\frac{\omega_d^2}{2}\left(\sum_{n=-\infty}^{+\infty}U_n^4\right)^{\frac{1}{2}}.
\end{eqnarray}
The left-hand side of (\ref{eq10d}) is the sum of the coupling and quadratic $U_n$-dependent terms of the Hamiltonian. 
Since the kinetic energy term is always $\frac{1}{2}\sum_{n=-\infty}^{+\infty}\dot{U}_n^2\geq 0$, for all $t\in [0, T]$, we have due to (\ref{eq10d}) that
\begin{eqnarray*}
\mathcal{H}(t)&=&\frac{1}{2}\sum_{n=-\infty}^{+\infty}\dot{U}_n^2+\frac{1}{2}\sum_{n=-\infty}^{+\infty}(U_{n+1}-U_n)^2
+\frac{\omega_d^2}{2}\sum_{n=-\infty}^{+\infty}U_n^2-\frac{\beta\omega_d^2}{4}\sum_{n=-\infty}^{+\infty} U_n^4\\
&&\geq +\frac{1}{2}\sum_{n=-\infty}^{+\infty}(U_{n+1}-U_n)^2+\frac{\omega_d^2}{2}\sum_{n=-\infty}^{+\infty}U_n^2
-\frac{\beta\omega_d^2}{4}\sum_{n=-\infty}^{+\infty} U_n^4\\
&&\geq\frac{\omega_d^2}{2}\left(\sum_{n=-\infty}^{+\infty}U_n^4\right)^{\frac{1}{2}}
-\frac{\beta\omega_d^2}{4}\sum_{n=-\infty}^{+\infty} U_n^4.
\end{eqnarray*}
The claim (\ref{eq8d}) is proved.
\newline
\emph{Step 3 (Derivation of the conditions (\ref{eq3d}) and (\ref{eq4d}) and their consequences):} The right-hand side of (\ref{eq8d}) defines a functional for the scalar continuous function $x(t)=||U(t)||_{\ell^4}=\left(\sum_{n=-\infty}^{+\infty}U_n^4\right)^{\frac{1}{4}}$. This functional is of the repulsive $\phi^4$-type. For instance, (\ref{eq8d}) can be written in the short-hand notation
\begin{eqnarray}
\label{eq11d}
\mathcal{H}(t)\geq \frac{\omega_d^2}{2}||U(t)||_{\ell^4}^2-\frac{\beta\omega_d^2}{4}||U(t)||^4_{\ell^4},\;\;\mbox{for all}\;\;t\in [0, T],
\end{eqnarray}
and motivated form the right-hand side of (\ref{eq11d}), we consider the function
\begin{eqnarray}
\label{eq12d}
J(x):=\frac{\omega_d^2}{2}x^2-\frac{\beta\omega_d^2}{4}x^4,\;x\geq 0.
\end{eqnarray}
Employing
$J$, we will work out the conditions for $x(t)=||U(t)||_{\ell^4}$; we are interested on its unique positive maximum
\begin{eqnarray}
\label{eq13d}
J(x_*)=\frac{\omega_d^2}{4\beta}\;\;\mbox{at}\;\;x_*=\frac{1}{\sqrt{\beta}}.
\end{eqnarray}
Now, let us assume that the initial data $U(0)$ and $\dot{U}(0)$ are chosen to satisfy
\begin{eqnarray}
\label{eq13couple}
\mathcal{H}(0)<J(x_*)=\frac{\omega_d^2}{4\beta}\;\;\mbox{and}\;\;||U(0)||_{\ell^4}>x_*=\frac{1}{\sqrt{\beta}},
\end{eqnarray}
i.e., we assume that conditions (\ref{eq3d}) and (\ref{eq4d}) are satisfied. Let us remark that the second of (\ref{eq13couple}) and the continuity of $x(t)=||U(t)||_{\ell^4}$, implies that there exists $0<t_1<T$, such that
\begin{eqnarray}
\label{eq14ad}
|||U(t)||_{\ell^4}>x_*=\frac{1}{\sqrt{\beta}},\;\;\mbox{for all}\;\;t\in [0, t_1].
\end{eqnarray}
Due to the energy identity (\ref{eq7d}), we have the bound
\begin{eqnarray}
\label{eq14d}
\mathcal{H}(t)\leq \mathcal{H}(0),\;\;\mbox{for all}\;\;t\in [0, T].
\end{eqnarray}
Then, from the first of (\ref{eq13couple}) and (\ref{eq14d}), we derive that the Hamiltonian
should be bounded by the maximum of $J$,
\begin{eqnarray}
\label{eq15d}
\mathcal{H}(t)<J(x_*)=\frac{\omega_d^2}{4\beta},\;\;\mbox{for all}\;\;t\in [0, T].
\end{eqnarray}
Then (\ref{eq14d}), together with (\ref{eq11d}), imply that
\begin{eqnarray}
\label{eq16d}
J(x(t))\leq\mathcal{H}(t)<J(x_*)=\frac{\omega_d^2}{4\beta},\;\;\mbox{for all}\;\;t\in [0, T].
\end{eqnarray}
The consequences (\ref{eq15d}) and (\ref{eq16d}) can be used as in \cite[pg. 212]{PS98}, in order to prove that in (\ref{eq14ad}) $t_1=T$, i.e.,
(\ref{eq14ad}) holds for all $t\in [0, T]$. To see this, we assume (for the contradiction) that $t_1<T$. Then, there exists some $t^*>t_1$ for which
\begin{eqnarray*}
x(t^*)=|||U(t^*)||_{\ell^4}=x_*=\frac{1}{\sqrt{\beta}}.
\end{eqnarray*}
Inserting this $t^*\in [0, T]$ in (\ref{eq16d}), the upper and lower bounds for $\mathcal{H}(t)$,
\begin{eqnarray*}
J(x_*)=J(x(t^*))\leq\mathcal{H}(t)<J(x_*)=\frac{\omega_d^2}{4\beta},\;\;\mbox{for all}\;\;t\in [0, T]
\end{eqnarray*}
follow, meaning that $J(x_*)=\mathcal{H}(t)$ for all $t\in [0, T]$. This is a contradiction with (\ref{eq15d}). Thus:
\begin{eqnarray}
\label{eq17d}
|||U(t)||_{\ell^4}>x_*=\frac{1}{\sqrt{\beta}},\;\;\mbox{for all}\;\;t\in [0, T].
\end{eqnarray}
It should be noted that the strict inequality of the assumptions (\ref{eq3d}) and (\ref{eq4d}) is crucial for the derivation of the validity of (\ref{eq15d}) and (\ref{eq17d}). The consequence (\ref{eq16d}) has the geometric implementation in view of ``Lyapunov functions'',  that under the assumptions (\ref{eq13couple}), both $J$ and $\mathcal{H}$ calculated along the orbits $\{U,\dot{U}\}\in\ell^2\times\ell^2$ are always less that the maximum $J(x_*)$ of the function $J$.
\newline
\emph{Step 4 (Definition of a ``blow-up'' functional $F(t)$ and the properties of its derivatives):} We shall derive the differential inequality (\ref{hamp15}) for the function,
\begin{eqnarray}
\label{eq18d}
F(t):&=&||U(t)||_{\ell^2}^2+\gamma\int_{0}^{t}||U(s)||_{\ell^2}^2ds+\gamma(T_0-t)||U(0)||^2_{\ell^2}
+\delta(t+t_0)^2\nonumber\\
&=&(U(t),U(t))_{\ell^2}+\gamma\int_{0}^{t}(U(s),U(s))_{\ell^2}ds+\gamma(T_0-t)||U(0)||^2_{\ell^2}+\delta(t+t_0)^2,
\end{eqnarray}
by applying the ``quadratic form argument'' of Pucci and Serrin \cite[cf. Theorem 1, pg. 206]{PS98}. This argument will be explained in detail in Step 5. For its implementation it is necessary to discuss the derivatives $\dot{F}(t)$, $\ddot{F}(t)$. In (\ref{eq18d}), the constants $t_0, T_0, \delta$ are positive and will be determined later.

The functional $F$ is well defined for all $t\in [0, T]$.
The first derivative $\dot{F}(t)$ is found to be
\begin{eqnarray*}
\dot{F}(t)&=&2(U,\dot{U})_{\ell^2}+\frac{d}{dt}\left\{\int_{0}^{t}||U(s)||_{\ell^2}^2ds\right\}-\gamma||U(0)||^2_{\ell^2}
+2\delta(t+t_0)\\
&=&2(U,\dot{U})_{\ell^2}+\gamma||U(t)||^2_{\ell^2}-\gamma||U(0)||^2_{\ell^2}+2\delta(t+t_0)\\
&=&2(U,\dot{U})_{\ell^2}+\gamma\left\{(U(t),U(t))_{\ell^2}-(U(0),U(0))_{\ell^2}\right\}+2\delta(t+t_0)\\
&=&2(U,\dot{U})_{\ell^2}+\int_{0}^t\frac{d}{ds}(U(s),U(s))_{\ell^2}ds+2\delta(t+t_0)\\
&=&2(U,\dot{U})_{\ell^2}+\gamma\int_{0}^t\left\{(U(s),\dot{U}(s))_{\ell^2}+(U(s),\dot{U}(s))_{\ell^2}\right\}ds+2\delta(t+t_0).
\end{eqnarray*}
Therefore, the derivative $\dot{F}(t)$ is
\begin{eqnarray}
\label{eq19d}
\dot{F}(t)=2(U,\dot{U})_{\ell^2}+2\gamma\int_{0}^t(U(s),\dot{U}(s))_{\ell^2}ds+2\delta(t+t_0).
\end{eqnarray}
Differentiating (\ref{eq19d}), we find a first expression for the second derivative $\ddot{F}(t)$,
\begin{eqnarray}
\label{eq20d}
\ddot{F}(t)&=&2(\dot{U},\dot{U})_{\ell^2}+2(U,\ddot{U})_{\ell^2}+2\gamma(U,\dot{U})_{\ell^2}+2\delta\nonumber\\
&=&2||\dot{U}||_{\ell^2}^2+2(U,\ddot{U})_{\ell^2}+\gamma\frac{d}{dt}||U||_{\ell^2}^2+2\delta.
\end{eqnarray}
Writing (\ref{eq1d}) in the shorthand notation $\ddot{U}=\Delta_2U-\gamma\dot{U}-\omega_d^2U+\beta\omega_d^2U^3$, and substituting the derivative $\ddot{U}$ in the second term of (\ref{eq20d}), we see that the expression for $\ddot{F}(t)$ takes the form:
\begin{eqnarray*}
\ddot{F}(t)&=&2||\dot{U}||_{\ell^2}^2+2(U,\Delta_2U)_{\ell^2}-2\gamma(U,\dot{U})_{\ell^2}
-2\omega_d^2(U,U)_{\ell^2}+2\beta\omega_d^2(U,U^3)_{\ell^2}
+\gamma\frac{d}{dt}||U||_{\ell^2}^2+2\delta\nonumber\\
&=&2||\dot{U}||_{\ell^2}^2-2\sum_{n=-\infty}^{+\infty}(U_{n+1}-U_n)^2-\gamma\frac{d}{dt}||U||_{\ell^2}^2
-2\omega_d^2\sum_{n=-\infty}^{+\infty}U_n^2
+2\beta\omega_d^2\sum_{n=-\infty}^{+\infty} U_n^4+\gamma\frac{d}{dt}||U||_{\ell^2}^2+2\delta\nonumber\\
&=&2||\dot{U}||_{\ell^2}^2-2\sum_{n=-\infty}^{+\infty}(U_{n+1}-U_n)^2-2\omega_d^2\sum_{n=-\infty}^{+\infty}U_n^2
+2\beta\omega_d^2\sum_{n=-\infty}^{+\infty} U_n^4+2\delta.
\end{eqnarray*}
Hence:
\begin{eqnarray}
\label{eq21d}
\frac{1}{2}\ddot{F}(t)=||\dot{U}||_{\ell^2}^2-\sum_{n=-\infty}^{+\infty}(U_{n+1}-U_n)^2
-\omega_d^2\sum_{n=-\infty}^{+\infty}U_n^2
+\beta\omega_d^2\sum_{n=-\infty}^{+\infty} U_n^4+\delta.
\end{eqnarray}
On the other hand, from the Hamiltonian (\ref{hamp}), we see that the quartic term equals to
\begin{eqnarray}
\label{eq22d}
\beta\omega_d^2\sum_{n=-\infty}^{+\infty} U_n^4=2\sum_{n=-\infty}^{+\infty}\dot{U}_n^2+2\sum_{n=-\infty}^{+\infty}(U_{n+1}-U_n)^2
+2\omega_d^2\sum_{n=-\infty}^{+\infty}U_n^2-4\mathcal{H}(t).
\end{eqnarray}
Then, replacing the quartic term of (\ref{eq21d}) by the right-hand side of (\ref{eq22d}), we write
$\frac{1}{2}\ddot{F}(t)$ as
\begin{eqnarray}
\label{eq23d}
\frac{1}{2}\ddot{F}(t)=3||\dot{U}||_{\ell^2}^2+\sum_{n=-\infty}^{+\infty}(U_{n+1}-U_n)^2
+\omega_d^2\sum_{n=-\infty}^{+\infty}U_n^2-4\mathcal{H}(t)+\delta.
\end{eqnarray}
Recall that by the energy identity (\ref{eq7d}),
\begin{eqnarray}
\label{eq24d}
\mathcal{H}(t)=\mathcal{H}(0)-\gamma\int_{0}^t||\dot{U}(s)||^2_{\ell^2}ds.
\end{eqnarray}
By replacing the $\mathcal{H}(t)$-term of (\ref{eq23d})
with the right-hand side of (\ref{eq24d}), $\frac{1}{2}\ddot{F}(t)$ becomes:
\begin{eqnarray}
\label{eq25d}
\frac{1}{2}\ddot{F}(t)=3||\dot{U}||_{\ell^2}^2+\sum_{n=-\infty}^{+\infty}(U_{n+1}-U_n)^2
+\omega_d^2\sum_{n=-\infty}^{+\infty}U_n^2-4\left\{\mathcal{H}(0)-\gamma\int_{0}^t||\dot{U}(s)||^2_{\ell^2}ds\right\}+\delta.
\end{eqnarray}
Furthermore, multiplying
(\ref{eq10d}) by $2$, namely,
\begin{eqnarray}
\label{eq26d}
\sum_{n=-\infty}^{+\infty}(U_{n+1}-U_n)^2+\omega_d^2\sum_{n=-\infty}^{+\infty}U_n^2
\geq \omega_d^2\left(\sum_{n=-\infty}^{+\infty}U_n^4\right)^{\frac{1}{2}},\;\;\mbox{for all}\;\;t\in [0, T],
\end{eqnarray}
and from (\ref{eq17d}), it follows that $|||U||_{\ell^4}^2$ satisfies
\begin{eqnarray}
\label{eq27d}
\omega_d^2|||U||_{\ell^4}^2=\omega_d^2\left(\sum_{n=-\infty}^{+\infty}U_n^4\right)^{\frac{1}{2}}>\omega_d^2{x_*}^2
=\frac{{\omega_d}^2}{\beta},\;\;\mbox{for all}\;\;t\in [0, T].
\end{eqnarray}
Therefore, by estimating the right-hand side of (\ref{eq26d}) as it is suggested by (\ref{eq27d}), we have
\begin{eqnarray}
\label{eq28d}
\sum_{n=-\infty}^{+\infty}(U_{n+1}-U_n)^2+\omega_d^2\sum_{n=-\infty}^{+\infty}U_n^2\geq \omega_d^2{x_*}^2,\;\;\mbox{for all}\;\;t\in [0, T].
\end{eqnarray}
Estimating further (from below) the sum $\sum_{n=-\infty}^{+\infty}(U_{n+1}-U_n)^2+\omega_d^2\sum_{n=-\infty}^{+\infty}U_n^2$ which appears in (\ref{eq25d}) by the right hand side of (\ref{eq28d}), we eventually derive the following inequality
for $\frac{1}{2}\ddot{F}(t)$,
\begin{eqnarray}
\label{eq29d}
\frac{1}{2}\ddot{F}(t)\geq 3||\dot{U}||_{\ell^2}^2+4\gamma\int_{0}^t||\dot{U}(s)||^2_{\ell^2}ds+[\omega_d^2 x_*^2-4\mathcal{H}(0)]+\delta.
\end{eqnarray}
We
require that the term $\omega_d^2 x_*^2-4\mathcal{H}(0)$
is positive. Hence, if
\begin{eqnarray*}
\omega_d^2 x_*^2-4\mathcal{H}(0)>0,
\end{eqnarray*}
in terms of the initial Hamiltonian,
the above requirement reads:
%
\begin{eqnarray*}
\mathcal{H}(0)<\frac{\omega_d^2 x_*^2}{4}=\frac{\omega_d^2}{4\beta}=J(x_*),
\end{eqnarray*}
which is exactly the assumption (\ref{eq3d}). Since this
is satisfied as an assumption on the initial data, we may select
\begin{eqnarray}
\label{eq30d}
2\delta=[C_{\omega_d} x_*^2-4\mathcal{H}(0)].
\end{eqnarray}
With the choice (\ref{eq30d}) for  $\delta$, (\ref{eq29d}) can be rewritten as
\begin{eqnarray*}
\frac{1}{2}\ddot{F}(t)\geq 3||\dot{U}||_{\ell^2}^2+4\gamma\int_{0}^t||\dot{U}(s)||^2_{\ell^2}ds+3\delta,
\end{eqnarray*}
or equivalently
\begin{eqnarray}
\label{eq31d}
\ddot{F}(t)\geq 6\left(||\dot{U}||_{\ell^2}^2+\delta\right)+8\gamma\int_{0}^t||\dot{U}(s)||^2_{\ell^2}ds.
\end{eqnarray}
\newline
\emph{Step 5 (Blow-up by non-continuation to $[0,\infty))$:} We argue by contradiction
and assume that the solution $U(t)$ of (\ref{eq1d}) exists for all $t\in [0,\infty)$, i.e., $T_{\mathrm{max}}=\infty$. Then, at first, by choosing an arbitrary $T_0\in [0,\infty)$ and since $\delta, t_0>0$ (recall that a specific range for $t_0$ will be fixed later), we have that
\begin{eqnarray}
\label{eq32d}
F(t)>0,\;\;\mbox{for all}\;\;t\in [0, T_0].
\end{eqnarray}
By the expression for the derivative $\dot{F}(t)$ given in (\ref{eq19d}), we see that at $t=0$:
\begin{eqnarray}
\label{eq33d}
\dot{F}(0)=2(U(0),\dot{U}(0))_{\ell^2}+2\delta t_0.
\end{eqnarray}
The choice of $t_0$ will be determined by the necessity to assume $\dot{F}(0)>0$. This will be guaranteed by selecting
\begin{eqnarray}
\label{eq34d}
t_0>-\frac{1}{\delta}(U(0),\dot{U}(0))_{\ell^2}.
\end{eqnarray}
Let us note that (\ref{eq34d}) is valid for any $t_0>0$ if $(U(0),\dot{U}(0))_{\ell^2}\geq 0$, while the choice (\ref{eq34d}) for $t_0$ is needed in the case where $(U(0),\dot{U}(0))_{\ell^2}< 0$.

With the above choice for $t_0$, $\dot{F}(0)>0$ and by (\ref{eq31d}), which implies that $\ddot{F}(t)>0$, we get that $\dot{F}(t)$ is strictly increasing in $[0, T_0]$. Together with the continuity of $\dot{F}(t)$, we derive that $\dot{F}(t)>0$ on $[0,T_0]$. Summarizing, we have derived that $F$ and its derivatives are positive, i.e.,
\begin{eqnarray}
\label{eq35d}
F(t), \dot{F}(t),\ddot{F}(t)>0,\;\;\mbox{for all}\;\;t\in [0, T_0]\;\;\mbox{for some}\;\;T_0\in [0,\infty).
\end{eqnarray}
With (\ref{eq31d}) at hand, {\em we may apply the ``quadratic form argument'' of \cite{PS98}}: recall that for any $(\chi,y)\in\mathbb{R}^2$, the quadratic form
\begin{eqnarray*}
\mathbf{A}\chi^2+2\mathbf{B}\chi y+\mathbf{C}y^2,\;\;\mathbf{A},\mathbf{B},\mathbf{C}\in\mathbb{R},\;\;\mathbf{A},\mathbf{C}>0,
\end{eqnarray*}
is positive definite ($\geq 0$), if and only if
\begin{eqnarray}
\label{eq36d}
\mathbf{AC}-\mathbf{B}^2\geq 0.
\end{eqnarray}
The idea is to derive the differential inequality (\ref{hamp15}) for $F(t)$ from (\ref{eq36d}), by proving that a quadratic form for suitable choices of real coefficients $\mathbf{A},\mathbf{B},\mathbf{C}$ involving $F(t), \dot{F}(t),\ddot{F}(t)$ is positive definite.

To apply this argument, we consider the quadratic form with the coefficients
\begin{eqnarray}
\label{eq37d}
\mathbf{A}&:=&(U(t),U(t))_{\ell^2}+\gamma\int_{0}^{t}(U(s),U(s))_{\ell^2}ds+\delta(t+t_0)^2\\
&=&||U||^2_{\ell^2}+\gamma\int_{0}^{t}||U(s)||^2_{\ell^2}ds+\delta(t+t_0)^2,\nonumber\\
\label{eq38d}
\mathbf{B}&:=&\frac{1}{2}\dot{F}(t)=(U(t),\dot{U}(t))_{\ell^2}
+\gamma\int_{0}^{t}(U(s),\dot{U}(s))_{\ell^2}ds+\delta(t+t_0),\\
\label{eq39d}
\mathbf{C}&:=&(\dot{U}(t),\dot{U}(t))_{\ell^2}+\gamma\int_{0}^{t}(\dot{U}(s),\dot{U}(s))_{\ell^2}ds+\delta\\
&=&||\dot{U}||^2_{\ell^2}+\gamma\int_{0}^{t}||\dot{U}(s)||^2_{\ell^2}ds+\delta.\nonumber
\end{eqnarray}
Note that $\mathbf{A},\mathbf{B}>0$ in $[0,T_0]$, the first requirement for a positive definite quadratic form. Also, $\mathbf{B}>0$ on $[0,T_0]$. With these coefficients, we have
\begin{eqnarray}
\label{eq40d}
\mathbf{A}\chi^2&=&(\chi U(t),\chi U(t))_{\ell^2}+\gamma\int_{0}^{t}(\chi U(s),\chi U(s))_{\ell^2}ds+\chi^2\delta(t+t_0)^2,\\
\label{eq41d}
2\mathbf{B}\chi y&=&\dot{F}(t)\chi y=2(\chi U(t),y\dot{U}(t))_{\ell^2}+2\gamma\int_{0}^{t}(\chi U(s),y\dot{U}(s))_{\ell^2}ds+2\chi y\delta(t+t_0),\\
\label{eq42d}
\mathbf{C}y^2&=&(y\dot{U}(t),y\dot{U}(t))_{\ell^2}+\gamma\int_{0}^{t}(y\dot{U}(s),y\dot{U}(s))_{\ell^2}ds+y^2\delta.
\end{eqnarray}
Before summing (\ref{eq40d})-(\ref{eq42d}) to recover the quadratic form $\mathbf{A}\chi^2+2\mathbf{B}\chi y+\mathbf{C}y^2$, we remark that
\begin{eqnarray}
\label{eq43d}
0\leq ||\chi U+y\dot{U}||_{\ell^2}^2&=&(\chi U+y\dot{U},\chi U+y\dot{U})_{\ell^2}\nonumber\\
&=&(\chi U,\chi U)_{\ell^2}+(\chi U,y\dot{U})_{\ell^2}+(y\dot{U},\chi U)_{\ell^2}+(y\dot{U},y\dot{U})_{\ell^2}\nonumber\\
&=&(\chi U,\chi U)_{\ell^2}+2(\chi U,y\dot{U})_{\ell^2}+(y\dot{U},y\dot{U})_{\ell^2}.
\end{eqnarray}
Summing now (\ref{eq40d})-(\ref{eq42d}), the right hand side of (\ref{eq43d}) appears as the sum of the inner product terms (outside and inside the integral). The sum of the remaining terms satisfies
\begin{eqnarray*}
\chi^2\delta(t+t_0)^2++2\chi y\delta(t+t_0)+y^2\delta=\delta\left(\chi (t+t_0)+y\right)^2.
\end{eqnarray*}
Therefore,
\begin{eqnarray*}
\mathbf{A}\chi^2+2\mathbf{B}\chi y+\mathbf{C}y^2=||\chi U+y\dot{U}||_{\ell^2}^2+\gamma\int_{0}^{t}||\chi U(s)+y\dot{U}(s)||_{\ell^2}^2ds+\delta\left(\chi (t+t_0)+y\right)^2\geq 0,
\end{eqnarray*}
and therefore (\ref{eq36d}) should hold, i.e.,
\begin{eqnarray}
\label{eq44d}
\mathbf{AC}\geq\mathbf{B}^2.
\end{eqnarray}
To derive the differential inequality (\ref{hamp15}) for $F(t)$ with the help of (\ref{eq44d}), we note the following:
by comparing the definition of $F(t)$ in (\ref{eq18d}) and the definition (\ref{eq40d}) of $\mathbf{A}$ we have:
\begin{eqnarray}
\label{eq45d}
F(t)>\mathbf{A},\;\;\mbox{for all}\;\;t\in [0, T_0].
\end{eqnarray}
Next, to compare $\ddot{F}(t)$ and $\mathbf{C}$, we use (\ref{eq31d}) and the definition of $\mathbf{C}$ (\ref{eq42d}), showing that
\begin{eqnarray*}
\ddot{F}(t)&\geq& 6||\dot{U}||_{\ell^2}^2+6\delta+8\gamma\int_{0}^t||\dot{U}(s)||^2_{\ell^2}ds\nonumber\\
&\geq&  6||\dot{U}||_{\ell^2}^2+6\delta+6\gamma\int_{0}^t||\dot{U}(s)||^2_{\ell^2}ds=6\mathbf{C}
\end{eqnarray*}
Thus, we have the differential inequality
\begin{eqnarray}
\label{eq46d}
\frac{1}{6}\ddot{F}(t)\geq\mathbf{C},\;\;\mbox{for all}\;\;t\in [0, T_0].
\end{eqnarray}
All the quantities in (\ref{eq45d}) and (\ref{eq46d}) are positive in $[0,T_0]$, hence
\begin{eqnarray}
\label{eq47d}
\frac{1}{6}F(t)\ddot{F}(t)>\mathbf{A}\mathbf{C},\;\;\mbox{for all}\;\;t\in [0, T_0].
\end{eqnarray}
By the definition of $\mathbf{B}$ in (\ref{eq38d}),
\begin{eqnarray}
\label{eq48d}
\mathbf{B}^2=\frac{1}{4}[\dot{F}(t)]^2,\;\;\mbox{for all}\;\;t\in [0, T_0].
\end{eqnarray}
Since (\ref{eq44d}) and (\ref{eq47d}), (\ref{eq48d}) hold, we have that $\frac{1}{6}F(t)\ddot{F}(t)>\frac{1}{4}[\dot{F}(t)]^2$, i.e., the differential inequality
\begin{eqnarray}
\label{eq49d}
F(t)\ddot{F}(t)>\frac{3}{2}[\dot{F}(t)]^2,\;\;\mbox{for all}\;\;t\in [0, T_0].
\end{eqnarray}
The functions $F(t),\dot{F}(t),\ddot{F}(t)>0$ in $[0,T_0]$ due to (\ref{eq35d}), and (\ref{eq49d}) can be rewritten in the form of (\ref{hamp15}),
\begin{eqnarray}
\label{eq50d}
\frac{\ddot{F}(t)}{\dot{F}(t)}>\frac{3}{2}\frac{\dot{F}(t)}{F(t)},\;\;\mbox{for all}\;\;t\in [0, T_0].
\end{eqnarray}
%
Manipulating (\ref{eq50d}) exactly as in Step 5 of the proof of Theorem \ref{GP}, we derive that the function
$F(t)$ cannot be continued further in time than $[0,T^*)$, where $T^*$ is given by
\begin{eqnarray*}
T^*=2\frac{F(0)}{\dot{F}(0)}.
\end{eqnarray*}
For instance, by choosing $T_0=T^*$, the blow-up of the function $F(t)$ in $[0, T_0)$ contradicts the fact that $T_0$ can be arbitrarily chosen in $[0,\infty)$ and that the solution $U$ of (\ref{eq1d})-(\ref{eq2d}) can be continued further than $[0, T_0)$.
\section*{Appendix C: Proof of Theorem \ref{Thne}}
\label{AppendixC}
\emph{Step 1 (Definition of the energy functional for the study of global existence):}
For convenience, we will denote the quartic term of the Hamiltonian
(\ref{hamp}) by
\begin{eqnarray}
\label{ne2}
R(U(t)):=\frac{\beta\omega_d^2}{4}\sum_{n=-\infty}^{+\infty} U_n^4.
\end{eqnarray}
Obviously,
from the conservation of
energy, we have $2\mathcal{H}(t)=2\mathcal{H}(0)$, and this can be rewritten by using (\ref{ne1}) and (\ref{ne2}) as:
\begin{eqnarray}
\label{ne3}
E(t)-2R(U(t))=E(0)-2R(U(0)),\;\;\mbox{for all}\;\;t\in [0, T_{\mathrm{max}}].
\end{eqnarray}
\newline
\emph{Step 2 (An inequality for $R(U(t))$ and $E(t)$): The energy functionals $E(t)$ and $R(U(t))$ satisfy the inequality}
\begin{eqnarray}
\label{ne4}
2R(U(t))\leq \frac{\beta}{2\omega_d^2}E^2(t),\;\;\mbox{for all}\;\;t\in [0, T_{\mathrm{max}}].
\end{eqnarray}
To prove (\ref{ne4}), we first apply
on (\ref{ne2}) the inequality (\ref{discembA}) for $p=4$ and $q=2$ to see that
\begin{eqnarray}
\label{ne5}
2R(U)=\frac{\beta\omega_d^2}{2}\sum_{n=-\infty}^{+\infty} U_n^4&=&\frac{\beta\omega_d^2}{2}||U||_{\ell^4}^4\nonumber\\
&\leq&\frac{\beta\omega_d^2}{2}||U||_{\ell^2}^4.
\end{eqnarray}
On the other hand, the energy $E(t)$ can be estimated from below as
\begin{eqnarray}
\label{ne6}
E(t)\geq \omega_d^2\sum_{n=-\infty}^{+\infty}U_n^2.
\end{eqnarray}
Recalling that
$||U||_{\ell^2}^4=\left\{\left(\sum_{n=-\infty}^{+\infty}U_n^2\right)^{\frac{1}{2}}\right\}^4
=\left(\sum_{n=-\infty}^{+\infty}U_n^2\right)^2$,
it follows from(\ref{ne6}) that $E(t)^2$ satisfies
\begin{eqnarray*}
E^2(t)\geq \omega_d^4\left(\sum_{n=-\infty}^{+\infty}U_n^2\right)^2=\omega_d^4||U||_{\ell^2}^4,
\end{eqnarray*}
which can be written equivalently as
\begin{eqnarray}
\label{ne7}
||U||_{\ell^2}^4\leq\frac{E^2(t)}{\omega_d^4}.
\end{eqnarray}
Then, estimation of the $||U|||_{\ell^2}^4$-term in the right-hand side of (\ref{ne5}) by (\ref{ne7}), results in
\begin{eqnarray*}
2R(U)\leq \frac{\beta\omega_d^2}{2}\cdot\frac{E^2(t)}{\omega_d^4}=\frac{\beta}{2\omega_d^2}E^2(t),\;\;\mbox{for all}\;\;t\in [0, T_{\mathrm{max}}],
\end{eqnarray*}
which is (\ref{ne4}).
\newline
\emph{Step 3 (Local in time estimates for $E(t)-2R(U(t))$):} Some useful local in time estimates for the energy $E(t)-2R(U(t))$ can be derived with the help of (\ref{ne4}).

By writing the conservation of energy (\ref{ne3}) as
\begin{eqnarray*}
E(t)=E(0)-2R(U(0))+2R(U(t)),\;\;\mbox{for all}\;\;t\in [0, T_{\mathrm{max}}],
\end{eqnarray*}
and estimating the $2R(U(t))$-term by using (\ref{ne4}), we derive a first
estimate, namely
\begin{eqnarray}
\label{ne9}
E(t)\leq E(0)-2R(U(0))+\frac{\beta}{2\omega_d^2}E(t)^2,\;\;\mbox{for all}\;\;t\in [0, T_{\mathrm{max}}].
\end{eqnarray}
Note that (\ref{ne4}) is valid for $t=0$. Hence
\begin{eqnarray}
\label{ne10}
2R(U(0))\leq \frac{\beta}{2\omega_d^2}E^2(0).
\end{eqnarray}
Then, using again the conservation (\ref{ne3}) and the positivity of $R(U)$, we obtain
%
\begin{eqnarray*}
E(t)-2R(U(t))=E(0)-2R(U(0))\leq E(0)-2R(U(0))+4R(U(0))=E(0)+2R(U(0)).
\end{eqnarray*}
Estimating the $2R(U(0))$-term in the right-hand side by (\ref{ne10}), leads to the second --local in time-- upper bound, in terms of $E(0)$:
%
\begin{eqnarray}
\label{ne11}
E(t)-2R(U(t))\leq E(0)+\frac{\beta}{2\omega_d^2}E^2(0),\;\;\mbox{for all}\;\;t\in [0, T_{\mathrm{max}}].
\end{eqnarray}
Consequently, since (\ref{ne11}) is valid at $t=0$, the initial data satisfy
\begin{eqnarray}
\label{ne12}
E(0)-2R(U(0))\leq E(0)+\frac{\beta}{2\omega_d^2}E^2(0).
\end{eqnarray}
\newline
\emph{Step 4 (Derivation of the condition on $E(0)$ for global existence)}.
We first assume
that the initial data are small in the sense
\begin{eqnarray}
\label{ne14}
E(0)<1.
\end{eqnarray}
Under condition (\ref{ne14}), we have that $E^2(0)<E(0)$, as well as that $E^2(0)<1$. Therefore, from the inequality (\ref{ne12}) it follows that
\begin{eqnarray*}
E(0)-2R(U(0))\leq E(0)+\frac{\beta}{2\omega_d^2}E^2(0)\leq E(0)+\frac{\beta}{2\omega_d^2}E(0),
\end{eqnarray*}
or in other words,
\begin{eqnarray}
\label{ne15}
E(0)-2R(U(0))\leq E(0)\left(1+\frac{\beta}{2\omega_d^2}\right).
\end{eqnarray}
On the other hand, the estimate (\ref{ne9}) can be written as
\begin{eqnarray}
\label{ne16}
0\leq E(0)-2R(U(0))+\frac{\beta}{2\omega_d^2}E^2(t)-E(t), \;\;\mbox{for all}\;\;t\in [0, T_{\mathrm{max}}].
\end{eqnarray}
Estimating the $E(0)-2R(U(0))$-term in the right-hand side of (\ref{ne16}) from above, by using (\ref{ne15}), yields the quadratic inequality for the derivation of conditions for small-data global existence
\begin{eqnarray}
\label{ne17}
\Theta(E(t)):=\frac{\beta}{2\omega_d^2}E^2(t)-E(t)+ E(0)\left(1+\frac{\beta}{2\omega_d^2}\right)\geq 0,\;\;\mbox{for all}\;\;t\in [0, T_{\mathrm{max}}].
\end{eqnarray}
The quadratic equation $\Theta(E)=0$ defined from (\ref{ne17}) has the two positive roots (\ref{ne18}),
if $E(0)$ satisfies the condition (\ref{ne19}). Then, if (\ref{ne19}) holds, inequality (\ref{ne17}) implies that
\begin{eqnarray}
\label{ne24}
E(t)\in [0, E_{-}]\cup [E_{+},+\infty).
\end{eqnarray}
However, the function $E(t)$ is continuous on $[0,\infty)$, therefore
\begin{eqnarray}
\label{ne25}
E(t)\in [0,E_{-}],
\end{eqnarray}
implying that solutions exist globally, and are uniformly bounded in time.

\bibliographystyle{amsplain}

\begin{thebibliography}{10}
%

\bibitem{mark} M. J. Ablowitz, B. Prinari, and A. D. Trubatch,
{\it Discrete and Continuous Nonlinear Schr{\"o}dinger Systems},
Cambridge University Press (Cambridge, 2004).

\bibitem{Juan} J. F. R. Archilla, P. L. Christiansen and Y. B. Gaididei,  Phys Rev. E \textbf{65} (2001), 016609.



\bibitem{Ball1b} J. M. Ball, Quart. J. Math. Oxford \textbf{28} (1977), 473.

\bibitem{Peyrard} O. Bang and M. Peyrard,  Phys. Rev. E \textbf{53} (1996), 4143.

\bibitem{bb} J. Bebernes and J. Eberly, {\em Mathematical Problems from Combustion Theory}
(Springer-Verlag, Berlin, 1989).

\bibitem{cazh} T. Cazenave, A. Haraux, {\em Introduction to Semilinear Evolution Equations},
Oxford Lecture Series in Mathematics and its Applications \textbf{13}, 1998.

\bibitem{nln}  I. Daumont, T. Dauxois, and M. Peyrard, Nonlinearity \textbf{10} (1997), 617.

\bibitem{Peyrard1} T. Dauxois and M. Peyrard, Phys. Rev. Lett. \textbf{70} (1993), 3935.

\bibitem{Peyrard2} T. Dauxois, M. Peyrard and C. R. Willis, Phys. D \textbf{57} (1992), 267.



\bibitem{Chrisbook} R. K. Dodd, J. C. Eilbeck, J. D. Gibbon, and H. C.  Morris,
{\em Solitons and nonlinear wave equations} (Academic Press, NY, 1982).

\bibitem{d03} J. Dunkel, W. Ebeling, L. Schimansky-Geier, and P. H{\"a}nggi,
Phys. Rev. E \textbf{67} (2003), 061118.

\bibitem{reviewsA} S. Flach and C. R. Willis, Phys. Rep. {\bf 295} (1998), 181.

\bibitem{reviewsC} S. Flach and A. V. Gorbach,
Phys. Rep. {\bf 467} (2008) 1.

\bibitem{PF} P. Fife, {\em Dynamics of internal layers and diffusive interfaces}
(Society for Industrial and Applied Mathematics, 1988).

\bibitem{Dirk3} S. Fugmann, D. Hennig, L. Schimansky-Geier, and P. H{\"a}nggi,
Phys. Rev. E \textbf{77} (2008), 061135.



\bibitem{GP} V. A. Galaktionov and S. I. Pohozaev, Nonlinear Analysis \textbf{53} (2003), 453.

\bibitem{Hanggi0} P. H{\"a}nggi,  J. Stat. Phys. \textbf{42} (1986), 105.

\bibitem{Hanggi} P. H{\"a}nggi, P. Talkner, and M. Borkovec, Rev. Mod. Phys. \textbf{62} (1990), 251.

\bibitem{Dirk1} D. Hennig, S. Fugmann, L. Schimansky-Geier, and P. H{\"a}nggi,
Phys. Rev. E \textbf{76} (2007), 041110.

\bibitem{Dirk2} D. Hennig, L. Schimansky-Geier, and P. H{\"a}nggi, Europhys. Lett. \textbf{78} (2007), 20002.


\bibitem{reviewsB} D. Hennig and G. Tsironis,
Phys. Rep. {\bf 307} (1999), 333.




\bibitem{iubini} S. Iubini, R. Franzosi, R. Livi, G.-L. Oppo, A. Politi,
arXiv:1203.4162.

\bibitem{NVK} N. van Kampen, {\em Stochastic Processes in Physics and Chemistry} (Elsevier, 2004).

\bibitem{K1} N. I. Karachalios, Proc. Edinb. Math. Soc. \textbf{49} (2006),
115.

\bibitem{K2} N. I. Karachalios, Glasg. Math. J. \textbf{48} (2006),
463

\bibitem{KY1} N. I. Karachalios and A. N. Yannacopoulos, J. Differential Equations \textbf{217} (2005),
88.

\bibitem{reviewsD} P. G. Kevrekidis, IMA J. Appl. Math. \textbf{76} (2011),
389.


\bibitem{avadh} P.G. Kevrekidis, A. Saxena and A.R. Bishop,
Phys. Rev. E {\bf 64} (2001) 026611.


\bibitem{KP92} Y. S. Kivshar and M. Peyrard, Phys. Rev. A \textbf{46} (1992), 3198.

\bibitem{Kramers} H. Kramers, Physica \textbf{7} (1940), 284.

\bibitem{KMI} K. Mischaikow, SIAM J. Math. Anal. \textbf{26} (1995),
1199.

\bibitem{YN} Y. Nishiura, {\em Far-from-equilibrium dynamics} (American Mathematical Society, 2002).

\bibitem{PS73} L. E. Payne and D.H. Sattinger, Israel J. Math. \textbf{22} (1975), 273.

\bibitem{PB} M. Peyrard and A.R. Bishop,
Phys. Rev. Lett. {\bf 62}, 2755 (1989).

\bibitem{PS98} P. Pucci and J. Serrin, J. Differential Equations \textbf{150} (1998), 203.

\bibitem{souplet} P. Quitner and P. Souplet, {\em Superlinear Parabolic Problems: Blow-up, Global Existence and Steady States} (Birkh\''{a}user, 2007).

\bibitem{RKDM} Z. Rapti, P.G Kevrekidis, D. J. Frantzeskakis and B. A. Malomed,
Physica Scripta \textbf{T113} (2004), 74.

\bibitem{AS} A. C. Scott, {\em Nonlinear Universe} (Springer-Verlag, 2007).



\bibitem{segur} H. Segur, D. Henderson, J. Carter, J. Hammack, C. M. Li, D. Pheiff, and K. Socha,
J. Fluid Mech. {\bf 539} (2005), 229.

\bibitem{bs} B. Straughan, {\em Explosive Instabilities in Mechanics} (Springer-Verlag, 1998).

\bibitem{ws} W. Strauss, {\em Nonlinear Wave Equations} (American Mathematical Society, 1990).

\bibitem {susu} C. Sulem and P. L. Sulem, {\em The Nonlinear Schr{\"o}dinger Equation. Self-Focusing
and Wave collapse} (Springer-Verlag, 1999).

\bibitem {takeno} S. Takeno and G. P. Tsironis, Phys.Lett. A 343 (2005) 274.

\bibitem{toda} M. Toda, Theory of nonlinear lattices, Springer-Verlag
(Berlin, 1989).

%
%
\bibitem{tsu} M. Tsutsumi,  SIAM J. Math. Anal. \textbf{15} (1984), 357.
%
\bibitem{Wein99} M. Weinstein, Nonlinearity \textbf{12} (1999), 673.

\bibitem{PSrev} L. Yacheng and Z. Junsheng, Nonlinear Anal. \textbf{64} (2006),
2665.
%
%
\end{thebibliography}

\end{document}